\documentclass{pinchcr}

\usepackage{amsfonts}
\usepackage{amsmath,amssymb}
\usepackage{graphicx}
\usepackage{psfrag}
\usepackage{bbm}
\usepackage{bm}
\usepackage{stmaryrd,dsfont}
\usepackage[latin1]{inputenc}
\usepackage{amsbsy}

\usepackage{listings} 

\usepackage{diff-commands}

\usepackage{bm,cite,amsmath,amssymb,amsfonts,float,graphicx,epsf,epsfig}

\title{Three Topics in Renormalization and Improvement}

\author{Anastassios Vladikas}
\author{}
\affiliation{ INFN - ``Tor Vergata",\\
c/o Department of Physics,\\
University of Rome ``Tor Vergata",\\
Via della Ricerca Scientifica 1,\\
I-00133 Rome, Italy}
\author{}
\affiliation{{\rm Summer School on} ``Modern perspectives in lattice QCD"}
\author{}
\affiliation{\rm Les Houches, August 3 - 28, 2009}
\authors{4}

\begin{document}

\maketitle

\tableofcontents

\maintext


\newpage

\chapter * {Acknowledgements}

The present set of lecture notes are based on a short course, delivered at the XCIII Les Houches Summer School (August 2009). The exciting atmosphere of the School and the students' enthusiasm have encouraged me to  write up a significantly expanded version of the original course. I thank my colleagues at Les Houches, students and lecturers alike, for their stimulating and supportive attitude. I also thank Ben Svetitsky for his constant (pure malt) spiritual support  during the preparation of my lectures at Les Houches.

I am grateful to Giancarlo Rossi, Rainer Sommer and Chris Sachrajda for carefully reading the manuscript, for their numerous useful suggestions, and for their constructive criticism. I am especially indebted to  Massimo Testa, for his patience and constant advise during several long encounters, throughout the various phases of preparation of this manuscript. If these lectures prove useful to young lattice researchers, 
it is largely thanks to the help provided by  these colleagues. Naturally, responsibility for any remaining shortcomings rests entirely with the author.

\newpage

\chapter{Introduction}

The important topic of lattice renormalization and improvement, covered by P.~Weisz in this School, is quite complicated, even for students who have considerable expertise in calculations based on continuum regularization (e.g. Pauli-Villars, dimensional regularization). The reader is advised to familiarize himself  with the basic concepts, before tackling the present chapter, which is meant to be a complement to Chapter 2 of this book~\shortcite{Weisz:2010nr}. Our aim is to present three very specific topics:
\begin{itemize}
\item
The consequences of the loss of chiral symmetry in the Wilson lattice regularization of the fermionic action and its recovery in the continuum limit. The treatment of these arguments involves lattice Ward identities.
\item
The definition and properties of mass independent renormalization schemes, which are suitable for a non-perturbative computation of various operator renormalization constants.
\item
The modification of the Wilson fermion action, by the introduction of a chirally twisted mass term (known as twisted mass QCD and abbreviated as tmQCD), which results to improved (re)normalization and scaling properties for physical quantities of interest.
\end{itemize}

Loss of chiral symmetry implies that the quark mass is not multiplicatively renormalizable, having also an additive counter-term, proportional to the ultraviolet cutoff (i.e. the inverse lattice spacing). Moreover, the no-renormalization theorem of the partially conserved vector and axial currents (PCVC and PCAC) turns out to be somewhat intricate. Wilson fermions preserve flavour (vector) symmetry, with an exactly conserved current, which is a point-split discretization of the corresponding continuum operator. The vector local current $V_\mu(x)$, familiar to us from continuum QCD, is to be normalized by a scale independent normalization factor $Z_V(g_0) \ne 1$, which is a function of the bare coupling $g_0$, tending to
unity in the continuum limit. On the other hand, axial symmetry is broken by the Wilson term. Thus  {\it any} discretization of the axial current on the lattice is accompanied by its scale independent normalization, which tends to unity in the continuum limit. For instance, the familiar axial current $A_\mu(x)$ has a normalization $Z_A(g_0) \ne 1$.
Another consequence of the loss of chiral symmetry is that operators belonging to the same chiral multiplet do not renormalize by the same renormalization factor. Although they all have identical anomalous dimensions (i.e. the same scale dependence), ratios of their renormalization constants are scale independent quantities which become unity only in the continuum limit. These properties are established with the aid of Ward identities. The very same Ward identities may also be used in practical computations for the non-perturbative determination of these scale invariant current normalizations and ratios of renormalization constants, at fixed lattice spacing.

It is a well established fact that lattice perturbation theory is characterized by poor convergence. Renormalization constants calculated perturbatively are therefore sources of rather large systematic errors in lattice predictions and postdictions of many interesting physical quantities. The second topic of these lecture notes is the discussion of a family of renormalization schemes (RI/MOM, RI/SMOM) which are used for the non-pertubative renormalization of fermionic composite operators. The two main sources of systematic error (unwanted low energy effects in the infrared and discretization effects in the ultraviolet) necessitate the existence of a ``renormalization window", in which 
these errors are controlled. The infrared problems due to a Goldstone pole in some renormalization constants is discussed in considerable detail. Although our presentation is based on Wilson fermions, these renormalization methods are equally well applied in other lattice regularizations, such as domain wall fermions.

The last topic concerns the benevolent effects of tmQCD in the renormalization and improvement properties of physical quantities. For this modified Wilson fermion action the Wilson term (responsible for a hard breaking of chiral symmetry) and the twisted mass term (which breaks the symmetry softly) are in some sense orthogonal in chiral space. This results to some fermion operators having simpler renormalization patterns in tmQCD than in the standard Wilson lattice theory. In these circumstances, the computation of their matrix elements is under better control in the tmQCD framework. Moreover, discrete symmetries in tmQCD imply that a particular tuning of the twisted mass action results to $\cO(a)$ improvement of physical quantities, without the need of Symanzik counter-terms. This is the so-called ``automatic improvement"  of Wilson fermions.

In these lecture notes we have strived to give explicit derivations of the most important results. Having opted for a detailed treatment of the theoretical issues, we have not addressed the many interesting numerical results, which have appeared in the literature over the years. Moreover, since these are meant to be paedagogical lectures addressing this advanced subject at an introductory level, we have only included the essential references which would facilitate the student's further study. This inevitably introduces some bias, for which we apologize.

\chapter{Basics}
\label{sec:defs}

Although we understand that the student is familiar with the basics of quark mass and composite operator renormalization, we recapitulate them here for completeness and in order to fix our notation. We also collect several useful definitions; although this is somewhat tedious, it is important to spell out the not-so-standard notation right from the beginning.

Concerning notation, we have preferred economy to mathematical rigour. Since the various bare quantities discussed below
are defined in the lattice regularization,  integrals like say, $\int d^4x_1 d^4 x_2$ of eq.~(\ref{eq:gp}), are really sums ($a^8 \sum_{x_1,x_2}$), which run over all lattice sites, labelled by $x_1$ and  $x_2$ etc. The occasional use of integrals instead of sums, partial derivatives instead of finite differences, and Dirac functions instead of Kronecker symbols, simplifies notation, hopefully avoiding any confusion. Moreover, a space-time function and its Fourier-transform will be indicated by the same symbol with a different argument (e.g.  $f(x)$ and $f(p)$); again mathematical rigour is being sacrificed in favour of notational economy.

\section{The Wilson lattice action and its symmetries}

We opt for the lattice regularization scheme, proposed by Wilson,
which consists of a gluonic action\footnote{Clearly renormalization schemes, constants etc. depend on the lattice actions. However, the choice of a specific gluonic action, such as that of eq.~(\ref{eq:sg}), is of no particular consequence to the topics discussed in the present lectures.}\shortcite{Wilson},    
\be   
S_G = \dfrac{6}{g_0^2}\sum_{P} \bigg[ 1 - \dfrac{1}{6}   
\Tr \bigl[ U_P + U_P^{\dagger} \bigr] \bigg]   
\label{eq:sg}   
\ee   
and a fermionic one \cite{Wilson:1975id},    
\bea   
S_F & = & a^4 \sum_x \bar \psi(x) \Big [ \dfrac{1}{2} \sum_\mu \big \{ \gamma_\mu ( \dcr^*_\mu + \dcr_\mu ) - a \dcr^*_\mu \dcr_\mu \}
+ M_0  \Big ] \psi(x) 
\nonumber \\
& = & -a^4 \sum_{x,\mu} \dfrac{1}{2a}    
 \bigg [ \bar \psi (x) (1 - \gamma_\mu) U_\mu (x) \psi (x+a\hat\mu)   
 + \bar \psi (x+a\hat\mu) (1 + \gamma_\mu) U^\dagger_\mu (x) \psi (x) \bigg]   
\nonumber \\   
&& + a^4 \sum_x \bar \psi (x) \bigg ( M_0 + \dfrac{4}{a} \bigg ) \psi (x)    \,\,\, .
\label{eq:sf} 
\eea    
In standard notation, $U_\mu(x) \equiv \exp[i g_0 a G_\mu(x)]$ is the lattice gauge link, 
depending on the gauge (gluon) field $G_\mu(x)$, $g_0$ is the bare coupling
constant, $a$ the lattice spacing (i.e. our UV cutoff) and $U_P$ the Wilson oriented plaquette.
Two discretizations of the covariant derivative are used in the above definition:
\bea
\label{eq:covder1}
&& a \dcr_\mu \psi (x) = U_\mu(x) \psi(x+a\hat\mu) - \psi(x) \\
&& a \dcr_\mu^* \psi (x) = \psi(x) - U_\mu^\dagger(x-a\hat\mu) \psi(x-a\hat\mu) \,\,\, .
\label{eq:covder2}
\eea
Later on we will also make use of the backward covariant derivative, defined as:
\be
a \bar \psi (x) \dcl_\mu = \bar \psi(x+a\hat\mu) U_\mu^\dagger(x) - \bar \psi(x) \,\,\, .
\label{eq:covderback}
\ee

The quark field  $\psi (x)$ is a vector in flavour space. Its    
components are denoted by $\psi_f$ ($f = 1,2,3, \dots , N_F$). The diagonal bare mass matrix  is denoted by $M_0$ and its elements   
by $m_{0f}$ ($f = 1, \dots , N_F$). Two continuous symmetries of particular interest are the flavour and chiral symmetries. The flavour group $SU(N_F)_V$ consists in the global vector transformations of the fermion field:
\bea   
\psi(x)  \,\,\, & \ \rightarrow & \,\,\, \psi^\prime (x) \,\,\, = \,\,\,  \exp  \left[ i \, \alpha_V^a \dfrac{\lambda^a}{2} \right] \,\,\, \psi (x) \,\,\, ,
\nonumber \\
\bar \psi (x)  \,\,\, & \rightarrow & \,\,\, \bar \psi^\prime (x) \,\,\, = \,\,\,\bar \psi (x)  \,\,\,  \exp  \left[ -i \, \alpha_V^a  \dfrac{\lambda^a}{2}  \right] \,\,\, ,
\label{eq:glob-vtransf}
\eea   
with $a = 1, \cdots , N_F^2-1$. The group generators in the fundamental representation are the flavour matrices $\lambda^a/2$
($a = 1,\dots,N_F^2-1$), satisfying:
\bea   
&& \Tr[\lambda^a \lambda^b ]  \,\,\, = \,\,\, 2 \delta^{ab} \,\,\, ,
\nonumber \\   
&& \left[ \dfrac{\lambda^a}{2}, \dfrac{\lambda^b}{2}\right]   
\,\,\, = \,\,\, i f^{abc}\dfrac{\lambda^c}{2} \,\,\, ,
\nonumber \\   
&& \left\{\dfrac{\lambda^a}{2}, \dfrac{\lambda^b}{2}\right\}   
\,\,\, = \,\,\, d^{abc}\dfrac{\lambda^c}{2}+\dfrac{\delta^{ab}}{N_F}\,I   \,\,\, ,
\label{eq:ds}   
\eea   
where $I$ represents the $N_F \times N_F$ unit matrix in flavour space; $f^{abc}$ are the $SU(N_F)$
structure constants and $d^{abc}$ are totally symmetric. For $N_F = 3$ the generators $\lambda^a$ are
the eight Gell-Mann matrices, while for $N_F = 2$ they are the three Pauli matrices. Under these transformations the Wilson lattice action~(\ref{eq:sf}) is invariant, provided all quark masses are degenerate ($m_{01} = \cdots = m_{0N_F}$). Vector (or flavour) symmetry is preserved by the Wilson lattice regularization.

Next we consider the axial transformation of the fermion field:
\bea   
\psi(x)  \,\,\, & \ \rightarrow & \,\,\, \psi^\prime (x) \,\,\, = \,\,\,  \exp  \left[ i \, \alpha_A^a \dfrac{\lambda^a}{2} \gamma_5 \right] \,\,\, \psi (x) \,\,\, ,
\nonumber \\
\bar \psi (x)  \,\,\, & \rightarrow & \,\,\, \bar \psi^\prime (x) \,\,\, = \,\,\,\bar \psi (x)  \,\,\,  \exp  \left[ i \, \alpha_A^a  \dfrac{\lambda^a}{2} \gamma_5 \right] \,\,\, ,
\label{eq:glob-atransf}
\eea   
with $a = 1, \cdots , N_F^2-1$. Even in the absence of a mass matrix ($M_0 = 0$), the presence of the Wilson term ($a \bar \psi \dcr^*_\mu \dcr_\mu \psi$) in the action~(\ref{eq:sf}) ensures that it is not invariant under these transformations. Consequently, the chiral group $SU(N_F)_L \otimes SU(N_F)_R$, related to axial and vector transformations, is not a symmetry of Wilson fermions. An interesting generalization of chiral symmetries is described in Appendix~\ref{app:spurionic}. For completeness, we also collect the discrete symmetries of the Wilson action in Appendix~\ref{app:disc-symm}.

We close this subsection with a general observation. Many important quantities we will be using, such as the quark propagator and the various correlation functions $G_O, \Lambda_O, \Gamma_O$ (to be defined below), are gauge dependent. This 
implies that some gauge fixing procedure has been applied in their calculation. The
gauge of choice is usually the Landau gauge, which on the lattice is understood as imposing a discrete version
of the condition $\partial_\mu G_\mu = 0$ on all lattice gauge fields $G_\mu $. The reader is advised to consult a
review of lattice gauge fixing for more details \cite{Giusti:2001xf}.

\section{Quark mass renormalization}

We recapitulate the general renormalization properties of the quark masses with Wilson fermions. It is convenient~\cite{Bhattacharya:2005rb}  to separate non-singlet and singlet quark mass contributions of the bare mass matrix:
\be
M_0 \,\,\, = \,\,\, {\widetilde \sum}_d \, \widetilde m^d  \lambda^d \,\, + \,\, \mav \, I \,\,\, .
\label{eq:mdecomp}
\ee
The sum runs over diagonal group generators only (e.g. $d=3$ for $SU(2)_V$ and $d=3,8$ for $SU(3)_V$) and $\mav$ is the average of the diagonal elements of $M_0$. Note that in the flavour symmetric theory, where all masses are degenerate, we have $\widetilde m^d = 0$ and $M_0 = \mav I$; cf. eq.~(\ref{eq:mass-comp}). In terms of the mass decomposition~(\ref{eq:mdecomp}), the bare mass of a specific quark of flavour
$f$ is
\be
m_{0f} \,\,\, = \,\,\, \mav \,\, + \,\, {\widetilde \sum}_d \, \widetilde m^d  \lambda^d_{ff}
\label{eq:m0f}
\ee
with $\lambda^d_{ff}$ the $f^{\rm th}$ diagonal entry of the matrix $\lambda^d$. 

As shown in Appendix~\ref{app:spurionic}, spurionic flavour symmetry implies that all components $\widetilde m^a$ transform as a multiplet in the adjoint representation of $SU(N_F)_V$, while $\mav$ is a singlet. Thus, as far as this symmetry is concerned,  members of the adjoint multiplet renormalize in the same way, while the singlet $\mav$ could renormalize differently. In the same Appendix we have also seen that spurionic axial transformations mix flavour non-singlets $\widetilde m^a$ and singlets $\mav$. If that were a symmetry of the theory, then $\widetilde m^a$ and $\mav$, being members of the same chiral multiplet would share a common renormalization constant. This is not the case, however, as the Wilson term of the lattice action breaks chiral spurionic symmetry. Moreover, flavour symmetry also suggests that the $\widetilde m^d$ components renormalize multiplicatively, while the lack of chiral symmetry allows $\mav$ to mix with the identity. Thus the singlet mass $\mav$ is also subject to additive renormalization. The bottom line is the following renormalization pattern:
\bea
\big [ {\widetilde m}^d (\gren) \big ]_{\rm R}  \,\,\, &=& \,\,\,  \lim_{a \rightarrow 0} \,\,\, Z_m(g_0,a\mu) \,\,\, {\widetilde m}^d(g_0) \,\,\, ,
\nonumber \\
\big [ \mav (\gren) \big ]_{\rm R} \,\,\, &=& \,\,\,  \lim_{a \rightarrow 0} \,\,\, Z_{m^0}(g_0,a\mu) \,\,\, [ \,\,\, \mav(g_0) \,\, - \,\, \mcrit \,\,\, ] \,\,\, .
\label{eq:m0tildmav}
\eea
We denote by $Z_m, Z_{m^0}$ the multiplicative renormalization constants and $\mu$ the renormalization scale; $g_{\rm R}(\mu)$ is the renormalized gauge coupling. One must also keep in mind the dependence of the bare coupling on the lattice spacing; $g_0^2(a) \sim 1/\ln(a)$. The additive renormalization factor $\mcrit = [C(g_0)/a]$, with $C(g_0)$ a coefficient bearing the dependence on the bare coupling, is a flavour independent, linearly divergent counter-term. At tree-level, $\mcrit = -4/a$. This counter-term is a well known feature of Wilson fermions. As already pointed out, it is a consequence of the loss of chiral symmetry by this regularization (even when the bare quark masses are switched off). The fact that there are two mass renormalization constants, $Z_m \ne Z_{m^0}$, albeit with the same scale-$\mu$ dependence, is also due to the loss of chiral symmetry. We will establish useful relations between them in sect~\ref{ward}.

It is common practice to omit, in expressions like (\ref{eq:m0tildmav}), the continuum limit on the r.h.s.. The resulting equations relate quantities at fixed lattice spacing and are true up to discretization effects.

Given that eq.~(\ref{eq:m0f}) relates the bare mass with flavour $f$ to the singlet and non-singlet bare masses, we {\it define} the
renormalized mass of this quark as the same combination of the renormalized singlet and non-singlet masses~\cite{Bhattacharya:2005rb}:
\bea
[ m_f ]_{\rm R} \,\,\, &=& \,\,\, [ \mav ]_{\rm R} \,\, + \,\,  {\widetilde \sum}_d \,\, [ {\widetilde m}^d ]_{\rm R}  \,\, \lambda^d_{ff}
\nonumber \\ 
&=& \,\,\,  Z_{m^0} [ \, \mav(g_0) \,\, - \,\, \mcrit \, ] \,\, + \,\,  Z_m  \,\,  {\widetilde \sum}_d \,\, \, {\widetilde m}^d \,\, \lambda^d_{ff}
\nonumber \\ 
&=& \,\,\,  Z_{m^0} [ \, \mav(g_0) \,\, - \,\, \mcrit \, ] \,\, + \,\,  Z_m [ \, m_{0f} \,\, - \,\, \mav \, ]  \,\,\, .
\eea
The second line is obtained by using eqs.~(\ref{eq:m0tildmav}). The third line is obtained from the second, by substituting
the sum in its last term from eq.~(\ref{eq:m0f}). This result can also be conveniently rewritten as
\be
[ m_f ]_{\rm R} \,\,\, = \,\,\,  Z_m \Big [ \, m_{0f} \,\, - \,\, \mcrit \,\, + \,\, \Big ( \dfrac{Z_{m^0}}{Z_m} \,\, - \,\, 1 \,\, \Big ) \,\, (\mav \,\, - \,\, \mcrit ) \,\, \Big ] \,\,\, .
\label{eq:mfRen}
\ee
Note that it is not enough to take $m_{0f} \rightarrow \mcrit$ in order to obtain the chiral limit $[ m_f ]_{\rm R} \rightarrow 0$; this is achieved only when {\it all bare quark masses} $m_{01}, \cdots m_{0N_F}$ tend to $\mcrit$. Also note that if the regularization scheme were chirally invariant (it isn't!) then $\mcrit$ would be absent and $Z_m = Z_{m^0}$; in this case the quark mass would be multiplicatively renormalizable ($[ m_f ]_{\rm R} =  Z_m  m_{0f}$), for each flavour.

It is instructive to consider the case of degenerate masses (${\widetilde m}^d =0$ and $\mav = m_{0f}$). Then eq.~(\ref{eq:mfRen}) reduces to the more familiar
\be
[\,\, m_f (\gren) \,\,]_{\rm R} \,\,\, = \,\,\, \lim_{a \rightarrow 0} [ \,\,\, Z_{m^0}(g_0,a\mu) \,\,\, m_f(g_0,m_{0f}) \,\,\, ]   \,\,\, ,
\label{eq:mrzm}   
\ee  
where the subtracted quark mass is defined as
\be
m_f (g_0, m_{0f}) = m_{0f}(g_0) - \mcrit  \,\,\, .
\label{eq:msub}
\ee
The chiral limit is now simply $m_f \rightarrow 0$; i.e. $m_{0f}  \rightarrow \mcrit$. 
In the mass degenerate case, or whenever flavour dependence is unimportant, the flavour index will be dropped, leaving us with $m_0, m, m_{\rm R}$ etc.

We also point out that combining eq.~(\ref{eq:mfRen}) for two flavours, say $f=1,2$ we obtain
\bea
[\,\, m_1 \,\,]_{\rm R} - [\,\, m_2  \,\,]_{\rm R} \,\,\, &=& \,\,\, Z_m \big [ \, m_{01} \,\, - \,\, m_{02} \, \big ]
\nonumber \\
&=& \,\,\, Z_m \big [ \, m_1 \,\, - \,\, m_2 \, \big ] \,\,\, ;
\label{eq:mdiff}
\eea
i.e. bare (and subtracted) mass differences renormalize with $Z_m$ rather than $Z_{m^0}$. As already anticipated, the two renormalization constants $Z_m$ and $Z_{m^0}$ have the same scale dependence. We will show this in sect.~\ref{ward}, and also prove that their ratio is a finite function of the bare coupling $g_0$, which tens to unity in the continuum limit.

\section{Quark propagator}

The bare quark propagator in coordinate space,  ${\cS}_f (x_1 - x_2;m) =  \langle \psi_f(x_1) \bar \psi_f(x_2) \rangle$, has
the Fourier Transform
\be   
\cS (p;m) \,\,\, = \,\,\, \int d^4x \,\,\, \exp(-i p x) \,\,\, \cS (x,m)   \,\,\, ,
\label{eq:Sft}
\ee
where the flavour index $f$ has been dropped (it will reappear wherever necessary). The renormalized propagator is  then given by
\be   
[ \,\,\ \cS (p; g_{\rm R}, m_{\rm R}, \mu) \,\, ]_{\rm R} \,\,\, = \,\,\, \lim_{a \rightarrow 0} \,\,\,
[\,\,\, Z_\psi(g_0,a\mu) \,\,\, \cS (p;g_0,m) \,\,\, ] \,\,\, ,
\label{eq:cp}   
\ee   
where $Z_\psi^{1/2}$ is the wave function renormalization; $\psi_{\rm R} = Z_\psi^{1/2} \psi$. 
Note that in the bare quantities, the bare mass $m_0$ has been traded off for the simpler
(from the point of view of the chiral limit) subtracted mass $m$. 
For later convenience, we define two ``projections'' (i.e. traces in spin and colour space)
of $\cS^{-1} (p,m)$ and its derivative with respect to momentum:
\bea
&& \Gamma_\Sigma(p;m) \,\,\, = \,\,\, \dfrac{-i}{48}  \,\,\,
\Tr \left[ \gamma_\mu \dfrac{\partial \cS^{-1} (p;m)}{\partial p_\mu} \right]  \,\,\, ,
\label{eq:gamwf}   
\\   
&& \Gamma_m(p;m) \,\,\, = \,\,\, \dfrac{1}{12} \,\,
\Tr \left[ \dfrac{\partial \cS^{-1} (p;m)}{\partial m} \right]   \,\,\, .
\label{eq:gamsm}   
\eea
These quantities have been defined so that their tree level values are equal to unity.
From eqs~(\ref{eq:mrzm}) and~(\ref{eq:mdiff}) we see that   
$\Gamma_\Sigma (p)$ and $\Gamma_m (p)$ renormalize like   
\bea   
\left[ \,\, \Gamma_\Sigma (p) \,\, \right]_{\rm R} \,\, &=& \,\, \lim_{a \rightarrow 0}  \,\,\,
 \left[ Z_\psi^{-1} \Gamma_\Sigma (p,m) \right]   \,\,\, ,
\label{eq:renwf}
 \\   
\left[ \,\, \Gamma_m (p) \,\, \right]_{\rm R} \,\,&=& \,\, \lim_{a \rightarrow 0}  \,\,\,
 \left[ Z_\psi^{-1} Z_m^{-1} \Gamma_m (p,m) \right]   \,\,\, .
\label{eq:rensig}   
\eea   
It is natural to fix the quark field and quark mass renormalization $Z_\psi$ and $Z_m$
by imposing renormalization conditions on $\left[ \Gamma_\Sigma (p) \right]_{\rm R}$ and
$\left [ \Gamma_m (p) \right]_{\rm R}$ at a given momentum $p=\mu$. Such a scheme will be the
subject of Section~\ref{sec:MOM}. It is clearly not a unique choice.

\section{Quark bilinear operators (composite fields)}

In these lectures we will make extensive use of local bilinear quark operators of the form:   
\be
O^f_\Gamma(x) \,\,\, = \,\,\, \bar \psi(x) \Gamma \dfrac{\lambda^f}{2} \psi(x)  \,\,\, ,
\label{eq:o}   
\ee   
where $\Gamma$ stands for a generic Dirac matrix and $f = 1, \cdots , N_F^2-1$ (flavour non-siglet case). Specific bilinear operators 
will be denoted according to their Lorentz group transformations: the scalar,    
pseudoscalar and tensor densities are 
\bea
S^f(x) \,\,\, = \,\,\, \bar \psi(x) \dfrac{\lambda^f}{2} \psi(x) \qquad & , & \qquad 
P^f(x) \,\,\, = \,\,\, \bar \psi(x) \dfrac{\lambda^f}{2} \gamma_5 \psi(x) \,\,\, , \nonumber \\
T^f_{\mu\nu}(x) \,\,\,& =& \,\,\, \bar \psi(x) \dfrac{\lambda^f}{2} \gamma_\mu \gamma_\nu \psi(x) \,\,\, ,
\label{eq:SPT}
\eea
respectively, whereas the vector and axial vector currents are
\be
V^f_\mu(x) \,\,\, = \,\,\, \bar \psi(x) \dfrac{\lambda^f}{2} \gamma_\mu \psi(x)  
\qquad  , \qquad
A^f_\mu(x) \,\,\, = \,\,\, \bar \psi(x) \dfrac{\lambda^f}{2} \gamma_\mu \gamma_5 \psi (x) \,\,\, . 
\label{eq:VA}
\ee
In the above, implicit colour and spin indices are contracted. Besides the non-singlet bilinear quark operators defined in eqs.~(\ref{eq:SPT}) and (\ref{eq:VA}), we will also be using the singlet ones
\bea
S^0(x) \,\,\, = \,\,\, \bar \psi(x) \dfrac{\lambda^0}{2} \psi(x) \qquad & , & \qquad 
P^0(x) \,\,\, = \,\,\, \bar \psi(x) \dfrac{\lambda^0}{2} \gamma_5 \psi(x) \,\,\, ,
\label{eq:SP0}
\eea
with $\lambda^0 \equiv \sqrt{2/N_F} I $. 

We will sometimes simplify the notation by using the symbols $S, V_\mu$ etc. (dropping the flavour superscript $f$). These are meant to be non-singlet operators of the form $O_\Gamma = \bar \psi_1 \Gamma \psi_2$. For example if we substitute $\lambda^f$ with $\lambda^+ = \lambda_1 + i \lambda_2$ in eqs.~(\ref{eq:SPT}) and~(\ref{eq:VA}), we end up with, say, $\bar u d$, $\bar u \gamma_\mu d$ etc.. The renormalization and improvement of four-fermion operators, although very important, is beyond the scope of the present lectures.
   
The insertion of the operator $O_\Gamma  = \bar \psi_1 \Gamma \psi_2$ in the $2$-point fermionic   
Green function gives the vertex function
\be   
G_O(x_1-x,x_2-x) \,\,\, = \,\,\, \langle \,\, \psi_1(x_1) \,\, O_\Gamma(x) \,\, \bar \psi_2(x_2) \,\, \rangle \,\,\, ,
\label{eq:gos}   
\ee
with translational invariance explicitly taken into account in the notation.
Placing, for simplicity, the operator at the origin $x=0$, the above correlation
function becomes, in momentum space, the vertex function
\be
G_O(p_1,p_2) \,\,\, = \,\,\, \int d^4x_1 d^4x_2 \,\, \exp(-ip_1x_1) \,\, \exp(ip_2x_2) \,\, G_O(x_1,x_2) \,\,\, .
\label{eq:gp}   
\ee   
The corresponding amputated Green function is given by   
\be   
\Lambda_O(p_1,p_2) \,\,\, = \,\,\, \cS_1 ^{-1}(p_1) \,\, G_O(p_1,p_2) \,\, \cS_2 ^{-1}(p_2)  \,\,\, ,
\label{eq:amp}   
\ee   
where the propagator subscripts denote flavour.
Note that $G_O$ and $\Lambda_O$ are rank-2 tensors in colour and spin space;
the colour and spin indices are suppressed in the notation.
Since it is preferable
to work with scalar, rather than tensor-like correlation functions, we define
the projected amputated Green function $\Gamma_O(p)$ as  follows:
\be   
\Gamma_O(p) \,\,\, = \,\,\, \dfrac{1}{12} \Tr \,\, \left[ \,\, P_O \,\, \Lambda_O(p,p) \,\, \right ] \,\,\, .
\label{eq:proj_GF}   
\ee   
The trace is over spin and colour indices. In some cases Lorentz indices need to be added; e.g.
the axial current $A_\mu$ has vertex functions $G_A^\mu$, $\Gamma_A^\mu$ etc. 
Thus, with the normalization factor 1/12,
$P_O$ is the Dirac matrix   
which renders the tree-level value of $\Gamma_O(p)$ equal to unity; i.e.   
it projects out the nominal Dirac structure of the Green function $\Lambda_O(p)$. It is
easy to work out these so-called ``projectors" for each of the operators
of eqs~(\ref{eq:SPT}) and (\ref{eq:VA}):
\bea   
P_S = I \qquad ; \qquad  P_P &=& \gamma_5 \qquad ; \qquad P_T^{\mu\nu} = \dfrac{1}{12} \gamma_\mu \gamma_\nu 
\nonumber \\   
P_V^\mu = \dfrac{1}{4} \gamma_\mu \qquad & ; & \qquad  
P_A^\mu = \dfrac{1}{4} \gamma_5 \gamma_\mu   \,\,\, .
\label{eq:proj}   
\eea   
Repeated Lorenz indices, appearing in the definition of these projectors and the corresponding
$\Lambda_O$'s are meant to be summed over. For simplicity of notation, specific Green functions will be denoted, for instance,
as $G_S, \Gamma_P, \Lambda_V$ etc.

The dimension-3 bilinear operators $S^f,P^f$ and $T^f_{\mu\nu}$ (with $a = f, \dots , N_F^2-1$), are non-singlet in flavour space.
As there are no other operators of equal or smaller dimension with the same quantum numbers, they
renormalize multiplicatively; their renormalization constants are denoted as $Z_S, Z_P$ and $Z_T$. Thus, the renormalized operator is formally given by   
\be   
[ \,\, O_\Gamma(\gren,\mren,\mu) \,\, ] _{\rm R} \,\,\, = \,\,\, \lim_{a \rightarrow 0}  \,\,
[ \,\, Z_O(g_0, a\mu) \,\, O_\Gamma(g_0,m) \,\, ]   \,\,\, .
\label{eq:zodef}   
\ee   
For the currents $V^f_\mu$ and $A^f_\mu$ there is a lattice version of the no-renormalization theorem,
which will be the object of the following sections on Ward identities. Essentially we will show in Sect.~\ref{ward} that 
these currents, once normalized by appropriate finite, scale independent
factors $Z_V$ and $Z_A$, go over to their continuum limit up to discretization effects which are proportional to positive powers of the lattice spacing. We then write the continuum currents as: 
\be   
[ \,\, O_{V,A}(\gren,\mren) \,\, ] _{\rm R} \,\,\, = \,\,\, \lim_{a \rightarrow 0}  \,\,
[ \,\, Z_{V,A}(g_0) \,\, O_{V,A}(g_0,m) \,\, ]   \,\,\, .
\label{eq:zodefcur}   
\ee   
Combining the above wih eqs.~(\ref{eq:cp}) and (\ref{eq:gos}) - (\ref{eq:amp}),   
we find for the renormalized Green functions:   
\bea   
&& [\,\, G_O(p_1,p_2) \,\, ]_{\rm R} \,\,\, = \,\,\, \lim_{a \rightarrow 0} \,\,
[ \,\, Z_\psi \,\, Z_O \,\, G_O (p_1,p_2) \,\, ]    \,\,\, ,
\label{eq:clG} \\   
&& [\,\, \Lambda_O (p_1,p_2)  \,\, ]_{\rm R} \,\,\, = \,\,\, \lim_{a \rightarrow 0} \,\, 
[ \,\, Z_\psi^{-1} \,\, Z_O \,\, \Lambda_O (p_1,p_2) \,\, ]   \,\,\, .
\label{eq:cl}   
\eea   
The renormalization of $\Gamma_O(p)$ is identical to that of  $\Lambda_O(p)$:   
\be   
[\,\,\, \Gamma_O (p)  \,\, ]_{\rm R} \,\,\, = \,\,\,  \lim_{a \rightarrow 0} \,\,\, 
\left[ \,\, Z_\psi^{-1} \,\, Z_O \,\, \Gamma_O (p) \,\, \right]   \,\,\, .
\label{eq:renG}   
\ee 
The singlet pseudoscalar density $P^0$ renormalizes multiplicatively in a similar fashion; its renormalization constant is denoted by $Z_{P^0}$. The singlet scalar density $S^0$, besides its multiplicative renormalization constant $Z_{S^0}$, also requires a power subtraction, as we will discuss in section~\ref{sec:chircond}.

The above definitions of renormalization constants are purely formal. By this we mean
that some arbitrary (still unspecified), choice of renormalization scheme and scale   
is implied. No reference to a specific scheme has been made so far.
However, it is implicit in the notation that only mass independent schemes are taken into consideration (otherwise $Z_m, Z_\psi$ and $Z_O$ would also depend on quark masses).
In sect.~\ref{sec:RIMOM} we will discuss a specific mass independent renormalization   
scheme, known as RI/MOM \cite{renorm_mom:paper1}.

\chapter{Lattice Ward identities}
\label{ward}

Ward identities are relations between Green functions; essentially they express
how a classical continuum symmetry is realized at the quantum level. Since lattice actions break
chiral symmetry\footnote{This statement is usually reserved for the Wilson fermion action only; 
staggered fermions are known to display a reduced chiral symmetry, while Ginsparg-Wilson 
ones preserve a lattice chiral symmetry. Yet, in practice even with the latter actions there
is some loss of chirality: for domain wall fermions the extension of the fifth dimension is never quite infinite.} 
the resulting Ward identities are not trivial discretized
transcriptions of the formal continuum ones. In fact they dictate the proper normalization of
(partially) conserved currents on the lattice and relate the renormalization parameters of
quantities which in the continuum belong to the same chiral multiplet, such as the scalar
and pseudoscalar densities. In this way the question of recovering chiral symmetry
in the continuum limit (up to discretization effects generically denoted here as $\cO(a)$-effects)
is linked to the appropriate (re)normalization of lattice operators in the lattice
regularization. In particular we will discuss the following topics in detail:
\begin{itemize}
\item the conservation of vector symmetry $SU(N_F)_V$ on the lattice with Wilson fermions;
\item the loss of chiral $SU(N_F)_L \otimes SU(N_F)_R$ symmetry on the lattice, even in the chiral limit, and its recovery in the continuum limit; 
\item the derivation of vector and axial lattice Ward identities and the finite normalization of the vector and axial currents, resulting from them;
\item the relations between quark mass renormalization and the renormalization of the scalar
and pseudoscalar densities, arising from lattice Ward identities.
\end{itemize}

Let us first recall that at the classical level, Ward identities connect the four-divergence of vector and axial currents to the scalar and pseudoscalar densities. The so-called ``partially conserved vector current" (PCVC) relation
\be
\partial_\mu V_\mu^f (x)  \,\, + \,\, \bar \psi(x) \bigg [ \dfrac{\lambda^f}{2} , M_0 \bigg ] \psi(x) \,\, = \,\, 0 \,\,\, ,
\label{eq:PCVC-class}
\ee
results, through the standard Noether construction, from the invariance of the QCD (Euclidean) fermionic action under $SU(N_F)_V$ transformations. Vector current conservation holds {\it in the degenerate mass case} ($M_0 \propto I$). The ``partially conserved axial current" (PCAC) relation
\be
\partial_\mu A_\mu^f (x)  \,\, - \,\, \bar \psi(x) \bigg \{ \dfrac{\lambda^f}{2} , M_0 \bigg \} \psi(x) \,\, = \,\, 0  \,\,\, ,
\label{eq:PCAC-class}
\ee
results from the invariance of the QCD action under axial transformations. Axial current conservation holds
{\it in the chiral limit} ($M_0 = 0$)\footnote{Note that unless otherwise stated, flavour indices run over $f = 1, \cdots, N_F^2-1$. The singlet case ($f=0)$, related to anomalous Ward identities, will not be discussed in these lectures.}.
Similar, possibly more familiar, expressions are obtained in the specific case with
$\lambda^a$ replaced by $\lambda^+ \equiv \lambda^1 + i \lambda^2$. This involves   
transformations on two flavours only, $\psi_1$ and $\psi_2$ (with   
corresponding masses $m_{1}$ and $m_{2}$). With this   
choice, the quantities $S, P, T_{\mu\nu}, V_\mu , A_\mu$ of eqs.~(\ref{eq:SPT}) and (\ref{eq:VA})
have the form $O_\Gamma = \bar \psi_1 \Gamma \psi_2$ and the above Ward identities become
\bea
\partial_\mu V_\mu (x)  \,\, &=& \,\, (m_1 - m_2) \,\, S(x) \,\,\, ,
\\
\partial_\mu A_\mu (x)  \,\, &=& \,\,  (m_1 + m_2) \,\, P(x) \,\,\, .
\eea

Beyond classical level, these Ward identities are to be understood as statements between operators.
More precisely, they are insertions in expectation values of multi-local operators $O(x_1,\dots,x_n)$,
consisting of a product of quark and gluon fields at different space-time points ($x_1 \neq x_2 \neq \dots \neq x_n$).
In the bare (lattice) theory these expectation values are defined in the path integral formalism as
\be
\langle O(x_1,\dots,x_n)\rangle \,\, = \,\, \dfrac{1}{Z} \,\, \int \,\, [\cD U] \,\, [\cD \psi] \,\, [\cD \bar\psi ] \,\,
O(x_1,\dots,x_n) \,\, \exp [ - S_G - S_F ] \,\,\, .
\label{eq:Ovev}
\ee
Lattice Ward identities are obtained in a way analogous to the continuum case. We recall that the basic trick
is to consider, instead of the usual  {\it global} $SU(N_F)_L\otimes SU(N_F)_R$ chiral transformations, infinitesimal
{\it local} vector and axial transformations of the fermionic fields; these are
\bea   
\delta \psi(x)  =  i \left[ \delta \alpha_V^a (x) \dfrac{\lambda^a}{2} \right] \psi (x)   
\qquad  ; \qquad 
\delta \bar \psi (x) = -i \bar \psi (x) \left[ \delta \alpha_V^a (x)  \dfrac{\lambda^a}{2}  \right]  
\label{eq:vtr}   
\eea   
and 
\bea   
\delta \psi(x)  =  i \left[ \delta \alpha_A^a (x) \dfrac{\lambda^a}{2} \gamma_5 \right] \psi (x)   
\qquad ; \qquad
\delta \bar \psi (x) = i \bar \psi (x) \left[ \delta \alpha_A^a (x) \dfrac{\lambda^a}{2} \gamma_5 \right]   
\label{eq:atr}   
\eea   
respectively. The operator expectation value of eq.~(\ref{eq:Ovev}) is
invariant under any change of the fermionic fields $\psi$ and $\bar \psi$, since they are integration variables. This implies that
\be
\dfrac{\delta }{\delta \alpha^a(x)} \big \langle O(x_1,\dots,x_n) \big \rangle \,\, = \,\, 0 \,\,\, ,
\ee
where $\delta \alpha^a$ may be either $\delta \alpha^a_V$ or $\delta \alpha^a_A$, depending on the
transformation under consideration (vector or axial). The above expression leads to the relation
\be
\Bigg \langle \dfrac{\delta O(x_1,\dots,x_n)}{\delta \alpha^a(x)} \Bigg \rangle \,\, = \,\, 
\Bigg \langle O(x_1,\dots,x_n) \,\,\, \dfrac{\delta S_F}{\delta \alpha^a(x)} \Bigg \rangle
 \,\,\, .
\label{eq:WIgen}
\ee
This is the quantum field theoretic expression leading to a Ward identity: for vector transformations $\delta \alpha^a_V$,
the variation of the fermionic action on the r.h.s. corresponds to the PCVC relation of eq.~({\ref{eq:PCVC-class}), while
for axial transformations $\delta \alpha^a_V$, it gives rise to the PCAC relation of eq.~({\ref{eq:PCAC-class}).
The exact meaning of this statement and the subtleties that accompany it in the case of Wilson fermions is the subject of the rest of
this section.

\section{Lattice Vector Ward identities}
\label{sec:VWI}

It is easy to check that the {\it global} vector transformations corresponding to those of eq.~(\ref{eq:vtr}) are
a symmetry of the action~(\ref{eq:sf}) when quarks are degenerate in mass (i.e. $M_0$ is proportional to
the unit matrix). In common lore, ``vector symmetry is a lattice symmetry". 
From the {\it local} transformations of eq.~(\ref{eq:vtr}) the following vector Ward identity can be derived \cite{Karsten:1980wd}:
\bea
i \Big \langle \dfrac{\delta O(x_1,\dots,x_n)}{\delta \alpha_V^a (x)} \Big \rangle \,\,\, & = & \,\,\,
a^4 \sum_\mu \nabla_x^\mu \big \langle \widetilde V^a_\mu(x) \,\, O(x_1,\dots,x_n) \big \rangle
\nonumber \\
\,\,\, & + & \,\,\, a^4 \big \langle \bar \psi(x) \bigg [ \dfrac{\lambda^a}{2},M_0 \bigg ] \psi(x) \,\, O(x_1,\dots,x_n) \big \rangle \,\,\, ,
\label{eq:vwi}
\eea
where $a \nabla_x^\mu f(x) = (f(x) - f(x-\mu))$ is an asymmetric lattice derivative, and
\bea
\widetilde V_\mu^a(x) = 
\dfrac{1}{2} \bigg [ && \bar \psi (x) ( \gamma_\mu - 1) U_\mu (x) \dfrac{\lambda^a}{2}
\psi(x +a\hat\mu) + \nonumber \\
&& \bar \psi (x+a\hat\mu) ( \gamma_\mu + 1) U_\mu^{\dagger}(x) \dfrac{\lambda^a}{2}
\psi(x) \,\, \bigg ]
\label{eq:vtilde}
\eea
is a point-split vector current. With $x \ne x_1,\dots, x_n$ and in the limit of degenerate bare 
quark masses ($M_0 \propto I$), eq.~(\ref{eq:vwi}) leads to the conservation of the 
point-split lattice vector current, $ \nabla_x^\mu \widetilde V^a_\mu(x) = 0$. 
The conservation of the standard local vector current $V^a_\mu(x) = \bar
\psi(x)\dfrac{\lambda^a}{2} \gamma_\mu \psi (x)$ on the lattice is somewhat more intricate, as we
will see below.

Without loss of generality, we again simplify matters by combining two versions of eq.~(\ref{eq:vwi}), one with flavour index $a=1$ and one with $a=2$, into a single Ward identity, involving the raising Gell-Mann matrix $\lambda^+$ of $SU(N_F)_V$ and the two flavours $\psi_1$ and $\psi_2$ (with corresponding bare masses $m_{01}$ and $m_{02}$). The vector current and scalar density are then bilinear operators of the form $O_\Gamma(x) = \bar \psi_1 \Gamma \psi_2$.  We also chose the specific operator $O(x_1,x_2) = \psi_1 (x_1) \bar \psi_2(x_2)$. In terms of these quantities the lattice vector Ward identity (\ref{eq:vwi}) becomes
\bea
&& \nabla^\mu_x \Big \langle \psi_1 (x_1) \,\, {\widetilde V}_\mu(x) \,\,
\bar \psi_2 (x_2) \Big \rangle = 
\big [\,\, m_{01}-m_{02} \,\, \big ] \,\,
\Big \langle \psi_1 (x_1) \,\, S (x) \,\, \bar \psi_2 (x_2) \Big \rangle
\nonumber \\
&& -\,\,\,  \delta (x_2-x) \,\, \big \langle \psi_1 (x_1) \bar \psi_1 (x_2)  \big \rangle
\,\,\, - \,\,\, \delta (x_1-x) \,\, \big \langle \psi_2 (x_1) \bar \psi_2 (x_2) \big \rangle \,\,\, .
\label{eq:zvqb}
\eea
Recalling the definition~(\ref{eq:gos}) of the vertex functions $G_O$, this is written as
\bea
&& \sum_\mu \nabla_x^\mu G_{\widetilde V}^\mu \left ( x_1-x,x_2-x;m_{01} ,m_{02} \right ) =
(m_{01} - m_{02}) \,\, G_S \left (x_1-x,x_2-x;m_{01} ,m_{02} \right ) \nonumber
\\   
&&+  \,\, \delta(x_2-x) \cS_1 \left (x_1-x_2;m_{01} \right )  \,\,
- \,\, \delta(x_1-x) \,\, \cS_2 \left ( x_1-x_2;m_{02} \right ) \,\,\, .
\label{eq:vwiud}    
\eea
This expression relates  bare quantities which are divergent in the continuum limit. Let us first work
in the degenerate mass limit ($m_{01} = m_{02} $), in which the first term on the r.h.s.
vanishes. A glance at eq.~(\ref{eq:cp}) shows that multiplying the resulting expression
by $Z_\psi$ renders the r.h.s. finite. Thus the l.h.s., which is now $ Z_\psi \sum_\mu \nabla_x^\mu G_{\widetilde V}^\mu$,
is also finite and does not require any further
renormalization. Consequently, the point-split conserved vector current (\ref{eq:vtilde}) satisfies the vector Ward identity
with its normalization given by the trivial factor
\be
Z_{\widetilde V}=1 \,\,\, .
\label{eq;consVC}
\ee
This is the no-renormalization theorem on the lattice with Wilson fermions \cite{Karsten:1980wd,Bochicchio:1985xa}.
It is an exact analogue of the familiar one in continuum QCD. The only difference
is that the lattice vector current satisfying it is not the familiar one $V_\mu = \bar \psi_1 \gamma_\mu \psi_2$, but the point-split one defined in
eq.~({\ref{eq:vtilde}). This result guarantees a proper 
definition of the vector charge and the validity of current algebra.

We will now show that the correlation functions of the local vector current $V_\mu(x) = \bar \psi_1 (x) \gamma_\mu \psi_2(x)$ differ from those of the conserved current by finite
contributions, which vanish in the continuum limit. The argument goes as follows: first we express the conserved 
current as
\bea
\label{eq:expv}
\widetilde V_\mu(x) &=& V_\mu(x) + 
\dfrac{a}{2}
\bigg [\bar \psi_1(x) (\gamma_\mu - 1) \dcr_\mu \psi_2(x)
+\bar \psi_1 (x) \dcl_\mu (\gamma_\mu + 1) \psi_2 (x) \bigg ]
\nonumber \\
&=& V_\mu(x) \,\,\, + \,\,\, a \,\, \Delta_\mu (x) \,\,\, ,
\eea
where we have used the lattice asymmetric covariant derivatives of eqs.~(\ref{eq:covder1}) and (\ref{eq:covderback}).
The second term on the r.h.s. of eq.~(\ref{eq:expv}) is a dimension-4 operator  $\Delta_\mu$. For definitiveness, 
we now consider the vertex function $G_{\widetilde V}(p)$; from eq.~(\ref{eq:expv})
it follows that
\be
G_{\widetilde V}(p) \,\, = \,\, G_V(p) \,\, + \,\, a \, G_\Delta (p) \,\,\, ,
\label{eq:expg}
\ee
which we have shown to be finite, up to an overall quark field renormalization $Z_\psi$; cf. eqs.~(\ref{eq:clG}) and (\ref{eq;consVC}).
The term $a G_\Delta (p)$ vanishes at tree-level in the continuum 
limit. Beyond tree-level however, this term contributes, due to the power 
divergence induced by mixing with a lower dimensional operator. To see this, write
the renormalized dimension-4 operator as
\be
[ \Delta_\mu ]_{\rm R} \,\, = \,\, Z_\Delta \,\, \Big [ \Delta_\mu \,\, + \,\, \dfrac{c(g_0,am)}{a} \,\, V_\mu \Big ] \,\,\, .
\ee
This renormalization pattern is dictated by dimensional arguments and the quantum numbers of
$ \Delta_\mu$. The dimensionless renormalization constant $Z_\Delta$ is at worst logarithmically
divergent. The dimensionless mixing coefficient $c(g_0,am)$ depends on the gauge coupling and the bare
quark masses (collectively denoted by $a m$). It has been shown that it cannot depend on a renormalization scale
$a\mu$~\cite{Testa:1998ez}. From the point of view of dimensional analysis, $c(g_0,am)$
could contain logarithms of the form $\ln(am)$. However, the absence of such a logarithmic dependence
has been explicitly shown for the axial current at all orders in perturbation theory; the situation is analogous for the vector current~\cite{Curci:1985se}. This is not unexpected, as $\ln(am)$-terms would lead to an ill-defined chiral limit at fixed lattice spacing. Thus we are left with a regular dependence of $c(g_0,am)$ on the quark mass
(i.e. positive powers of $a m$), which may be dropped in a mass independent scheme leaving us with $c(g_0)$.

From these considerations we deduce that, up to regular $\cO(a)$ terms (i.e. discretization effects), eq.~(\ref{eq:expg}) may be written as
\bea
G_{\widetilde V}(p) \,\, &=& \,\, G_V(p) \,\, + \,\, \dfrac{a}{Z_\Delta} \,\, [ \, G_\Delta (p) \, ]_{\rm R} 
\,\, - \,\, c(g_0) \,\, G_V(p)  
\nonumber \\
&=& \,\, [ \, 1 \,\, - \,\, c(g_0) \, ] G_V(p)  + \cdots \,\,\, .
\label{eq:expg2}
\eea
The term proportional to $[ G_\Delta ]_{\rm R}$ has been dropped in the last expression, being proportional to the lattice spacing
(the renormalized correlation is finite by construction, while $Z_\Delta$ is at most logarithmically divergent).
Consequently,  the local current $V_\mu$ has a finite normalization:
\be
Z_V (g_0) \,\, \equiv \,\, 1 -c(g_0) \,\, \ne 1 \,\,\, .
\label{eq:ZVloc}
\ee 
Note that eq.~(\ref{eq:vwiud}) (and any other vector Ward identity) can be expressed in terms of 
the local current, by substituting $\widetilde V_\mu$ with $[ Z_V \,V_\mu ]$.

Next we go back to eq.~(\ref{eq:vwiud}) and, keeping the two masses distinct,
integrate it w.r.t. $x$; the surface trerm on the l.h.s. vanishes and we obtain
\bea
&& \big (m_{02} - m_{01} \big ) \int d^4x \,\, G_S \left (x_1-x,x_2-x;m_{01} ,m_{02} \right ) =
\nonumber \\
&& \cS_1 \left (x_1-x_2;m_{01} \right ) - \cS_2 \left ( x_1-x_2;m_{02} \right ) \,\, .
\label{eq:vwiud2}
\eea
Again, the r.h.s. of above expression becomes finite (renormalized) once we
multiply through by $Z_\psi$. This means that, up to this quark field renormalization,
the l.h.s. is also finite. In other words, the product of the bare quark mass difference and the
integrated scalar operator $\int d^4 x S(x)$ is a scale independent (or renormalization group invariant) quantity.
As there is no operator of dimension $d \le 3$ that could mix with $S$ and vanish identically
under the integral $\int d^4 x$, also the product of the bare quark mass difference and the (unintegrated)
scalar density $S$ is scale independent. With the difference of renormalized quark masses formally given in eq.~(\ref{eq:mdiff}) and the operator renormalization given by eq.~{(\ref{eq:zodef}), this means that
\be   
Z_S(g_0, a\mu) = Z_m^{-1} (g_0,a\mu) \,\,\, .   
\label{eq:zsm}   
\ee   
Thus, vector Ward identities imply an exact relation between the quark mass renormalization $Z_m$ and that of the scalar density. In other words, once a renormalization condition is imposed on the quark mass in order to determine $Z_m$, the scalar density renormalization constant $Z_S$ is also known (or vice versa).  A corollary is that the mass anomalous dimension is the opposite of that of the scalar operator ($\gamma_m = - \gamma_S$).

\section{Lattice Axial Ward identities}
\label{sec:AWI}
   
Far less obvious are the consequences of axial transformations   
with Wilson fermions. This is because even at vanishing quark masses (i.e. $M_0 = 0$), the
Wilson term of the lattice action~(\ref{eq:sf}) is not invariant under the {\it global} version of the axial
transformations~(\ref{eq:atr}). However, by imposing suitable  renormalization conditions, 
PCAC is recovered in the continuum limit~\cite{Karsten:1980wd,Bochicchio:1985xa,Testa:1998ez}.  
The axial WI, namely eq.~(\ref{eq:WIgen}) for the axial transformations~(\ref{eq:atr}), is 
\bea
i \Big \langle \dfrac{\delta O(x_1,\dots,x_n)}{\delta \alpha_A^a (x)} \Big  \rangle =
a^4 \sum_\mu \nabla_x^\mu \Big \langle \widetilde A^a_\mu(x) \,\,  O(x_1,\dots,x_n)  \Big \rangle \nonumber \\
\hspace {-0.5 cm}   - a^4 \Big \langle \bar \psi(x)
\bigg \{ \dfrac{\lambda^a}{2}, M_0 \bigg \} \gamma_5 \psi(x) \,\,\, O(x_1,\dots,x_n) \Big \rangle
- a^4 \Big \langle X^a(x) \,\,\, O(x_1,\dots,x_n) \Big \rangle \,\, ,
\label{eq:awi}
\eea
where $\widetilde A_\mu^a(x)$ is a bilinear point-split axial current given by
\bea
\hspace {-0.5 cm} \widetilde A_\mu^a(x) =
\dfrac{1}{2} \Big [\bar \psi (x) \gamma_\mu \gamma_5 U_\mu (x) 
\dfrac{\lambda^a}{2} \psi(x +a\hat\mu) 
+ \bar \psi (x+a\hat\mu) \gamma_\mu \gamma_5 U_\mu^{\dagger}(x)
\dfrac{\lambda^a}{2} \psi(x) \Big ] \,\,\, .
\label{eq:atilde}
\eea
The term $X^a$ in the above Ward identity is the variation of the Wilson
term under axial transformations:
\be
X^a(x) = -\dfrac{1}{2} a 
\left[\bar \psi(x) \dfrac{\lambda^a}{2} \gamma_5 \dcr^2 \psi(x) +
\bar \psi(x) \dcl^2 \dfrac{\lambda^a}{2} \gamma_5 \psi(x) \right] \,\,\, ,
\ee
where
\bea
a^2 \dcr^2 \psi (x) &=& \sum_\mu \big [ U_\mu(x) \psi(x+a\hat\mu) +
U_\mu^\dagger(x-a\hat\mu) \psi(x-a\hat\mu) - 2 \psi(x) \big ]
\nonumber \\
a^2 \bar \psi (x) \dcl^2 &=& \sum_\mu \big [ \bar \psi(x+a\hat\mu) 
U_\mu^\dagger (x) + \bar \psi(x-a\hat\mu) U_\mu (x-a\hat\mu) - 2 \bar \psi(x) \big ] \,\,\, .
\eea
Since $X^a$ cannot be cast in the form of a four-divergence, it cannot be absorbed
in a redefinition of the axial current $\widetilde A_\mu^a$. This is of course a corollary of the fact
that the Wilson term in the fermionic action breaks chiral symmetry. It is a dimension-4 operator
which, in the naive continuum limit vanishes, being of the form $X^a = a O^a_5$,
with $O^a_5$ a dimension-5 operator. However, when inserted in correlation functions
such as (\ref{eq:awi}), $O^a_5$ will generate power divergences which
will cancel the prefactor $a$. It is therefore mandatory to understand its renormalization properties.
Its mixing with other dimension-5 operators may be ignored, since the 
corresponding mixing coefficients are at most logarithmically divergent and their contribution to
$X^a = a O^a_5$ vanishes in the continuum limit. Mixing with lower dimensional operators
is, on the other hand, relevant, as the corresponding coefficients are inverse powers of
the lattice spacing~\cite{Testa:1998ez}. The only possibility for such operator mixing, respecting
the symmetry properties of $O^a_5$ is:
\be
[\,\, O^a_5(x)\,\, ]_{\rm R} = Z_5 \bigg [O^a_5(x) +
\bar \psi(x) \bigg \{ \dfrac{\lambda^a}{2},\dfrac{\overline M}{a} \bigg \} \gamma_5 \psi(x)
+ \dfrac{(Z_{\widetilde A} -1)}{a} \nabla_x^\mu \widetilde A^a_\mu(x) \bigg ]  \,\, .
\label{eq:xbar}
\ee
The functional dependence of the mixing coefficients $Z_{\widetilde A}$ and $\overline M$
is established by arguments, very similar to the ones of sect.~\ref{sec:VWI}, concerning $Z_V$. 
These coefficients involve subtractions of operators of lower dimension than that of $O^a_5$, which
implies that they are independent of any renormalization scale $\mu$~\cite{Testa:1998ez}.  Any dependence on
$a M$ cannot be divergent (e.g. $\ln(a M)$) because this would compromise the chiral limit
of $[O^a_5(x)]_{\rm R}$-insertions in renormalized correlation functions. The
absence of $\ln(a M)$-dependence of $Z_{\widetilde A}$ has been explicitly proved in perturbation theory~\cite{Curci:1985se}.
Moreover, any regular dependence of $Z_{\widetilde A}$ on $aM$ will vanish in a mass independent renormalization scheme
and may therefore be considered a lattice artefact. We conclude that the dimensionless axial current coefficient only depends on the gauge coupling,
$Z_{\widetilde A}(g_0)$. Similarly, the dimension-1 mass coefficient has the functional form $\overline M = 
w(g_0, aM)/a$, with $w(g_0, aM)$ a regular function of $aM$.

Using eq.~(\ref{eq:xbar}), we substitute $X^a  = a \times O^a_5$ in eq.~(\ref{eq:awi}), obtaining
\bea
i \Big \langle \dfrac{\delta O(x_1,\dots,x_n)}{\delta \alpha_A^a (x)} \Big \rangle &=&
a^4 \sum_\mu \nabla_x^\mu  \Big \langle  Z_{\widetilde A} \widetilde A^a_\mu(x) O(x_1,\dots,x_n) \Big \rangle 
\nonumber \\
&-& a^4 \Big \langle  \bar \psi(x)  \bigg \{ \dfrac{\lambda^a}{2}, [M_0 - \overline M] \bigg \} \gamma_5 \psi(x) \,\, O(x_1,\dots,x_n) \Big \rangle
\nonumber \\
&-& a^4 \Big \langle \dfrac{X^a_{\rm R}(x)}{Z_5} \,\, O(x_1,\dots,x_n)  \Big \rangle \,\,\, .
\label{eq:awi2}
\eea
In order to simplify the presentation, we again specify the operator
$O(x_1,x_2) = \psi_1(x_1) \bar \psi_2(x_2)$, and the Gell-Mann matrix $\lambda^+$ of $SU(N_F)_A$, in place
of $\lambda^a$. The above expression becomes
\bea
&& \nabla^\mu_x \Big \langle \psi_1 (x_1) \,\, Z_{\widetilde A} {\widetilde A}_\mu(x) \,\,
\bar \psi_2 (x_2) \Big \rangle = 
\big [\,\, m_{01}+m_{02} - {\overline m_1} - {\overline m_2} \,\, \big ] \,\,
\Big \langle \psi_1 (x_1) \,\, P (x) \,\, \bar \psi_2 (x_2) \Big \rangle
\nonumber \\
&& -\,\,\,  \delta (x-x_2) \,\, \big \langle \psi_1 (x_1) \bar \psi_1 (x_2) \gamma_5  \big \rangle
\,\,\, - \,\,\, \delta (x-x_1) \,\, \big \langle \gamma_5 \psi_2 (x_1) \bar \psi_2 (x_2) \big \rangle \\
&& +\,\,\, \dfrac{a}{Z_5} \,\, \Big \langle \psi_1 (x_1) \,\, [ O_5(x) ]_{\rm R} \,\, \bar \psi_2 (x_2) \Big \rangle \,\,\, ,
\nonumber
\label{eq:zaqb}
\eea
with $\overline m_1, \overline m_2$ the elements of the diagonal\footnote{If the matrix $\overline M$ were not diagonal, quantum numbers like strangeness would be violated by the theory.} matrix $\overline M$. Note that these quantities depend on the bare coupling and all bare quark masses; i.e. $\overline m_f(g_0, m_{01}, \cdots, m_{0N_F})$. The last term on the r.h.s. is genuinely $\cO (a)$. Since it vanishes in the continuum limit, it may be safely dropped as a discretization effect. This leaves us with the axial Ward identity
\bea
&& Z_{\widetilde A} \sum_\mu \nabla_x^\mu G_{\widetilde A}^\mu \left ( x_1-x,x_2-x;m_{01} ,m_{02} \right ) 
\label{eq:awiud} \\
&& \,\,\, = \,\,\, 
\big (m_1^{\rm PCAC} + m_2^{\rm PCAC} \big ) \,\,\, G_P \left (x_1-x,x_2-x;m_{01} ,m_{02} \right ) \nonumber
\\   
&& \,\,\, - \,\,\, \delta(x_2-x) \,\, \cS_1 \left (x_1-x_2;m_{01} \right ) \gamma_5  \,\,
- \,\, \delta(x_1-x) \,\, \gamma_5 \,\, \cS_2 \left ( x_1-x_2;m_{02} \right ) \,\,\, + \,\,\, \cO(a) \,\,\, , \nonumber
\eea
where the PCAC (bare) quark mass is defined by
\be
m^{\rm PCAC}_f \,\,\, \equiv \,\,\, m_{0f} \,\, - \,\, {\overline m_f}(g_0, m_{01}, \cdots, m_{0N_F})
\,\,\, \qquad \,\,\, ( f \,\, = \,\, 1,2,3, \cdots , N_F ) \,\,\, .
\label{eq:mpcacren}
\ee
We may now define the chiral limit on the basis of Ward identity~(\ref{eq:awiud}): first we take for simplicity al masses $m_0$ to be degenerate, which allows us to simply the notation by writing ${\overline m_f}$ as ${\overline m}(g_0, m_0)$. Then the chiral limit is defined as the value $\mcrit$ of $m_0$
for which $\overline m (g_0,\mcrit) = \mcrit$ (i.e. $m^{\rm PCAC}$ vanishes). A relation between the PCAC quark mass 
defined in eq.~(\ref{eq:mpcacren}) and the subtracted bare mass $m = m_0 - \mcrit$ is readily obtained by expanding the former around the
chiral point $\mcrit$:
\bea
m^{\rm PCAC} \,\, &=& \,\, m_0 - {\overline m}(\mcrit) \,\, - \,\, \dfrac{\partial {\overline m}(m)}{\partial m} \Big \vert_{\mcrit} (m_0 - \mcrit) \,\, + \,\, \cdots
\nonumber \\
&=& \,\, (m_0 - \mcrit) \,\, \Big [ 1 \,\, - \,\, \dfrac{\partial {\overline m}(m)}{\partial m} \Big \vert_{\mcrit} \,\, + \,\, \cdots \Big ] \,\,\, .
\label{eq:mPCACmsub}
\eea

In the chiral limit, the first term on the r.h.s. of eq.~(\ref{eq:awiud}) vanishes. Multiplying the
resulting expression by $Z_\psi$ renders the r.h.s. finite. Thus, also the l.h.s. is finite and does not require any further
renormalization.  Now all terms of eq.~(\ref{eq:awiud}) differ from their continuum
expressions by $\cO(a)$ discretization effects. The point-split axial current (\ref{eq:atilde}) with the
normalization
\be
Z_{\widetilde A} \ne 1 \,\,\, ,
\ee
satisfies the continuum axial Ward identity up to $\cO(a)$.
In practice it turns out to be more convenient to work with the lattice
local axial current $A_\mu^a(x)$. We can show, in a fashion analogous to 
the case of the vector current (c.f. the power counting argument based on 
eqs.~(\ref{eq:expv})-(\ref{eq:ZVloc})), that $A_\mu^a(x)$ has a finite 
normalization constant $Z_A$. Thus we have $[ A^a_\mu ]_{\rm R}=
\lim_{a \rightarrow 0} [Z_{\widetilde A} \widetilde A^a_\mu] = 
\lim_{a \rightarrow 0} [Z_A A^a_\mu]$. From now on, 
the combination $Z_{\widetilde A} \widetilde A^a_\mu$ will always be 
substituted by $Z_A A^a_\mu$ wherever it appears in a WI. Also analogous to 
the vector current case is the lack of mass dependence of these constants.
We therefore have:
\begin{equation}
Z_{\widetilde A}(g_0^2) \,\, , \,\,  Z_A(g_0^2) \,\, \neq \,\, 1 \,\,\, .
\end{equation}
Note that the tree level value of both $Z_{\widetilde A}$ and $Z_A$ is unity. In perturbation theory $Z_A = 1 + z_1 g_0^2 + \cdots$
(and similarly for $Z_{\widetilde A}$). Thus in the continuum limit ($a \rightarrow 0$, $g_0^2(a) \rightarrow 0$) we have
$Z_A , Z_{\widetilde A} \rightarrow 1$.

From here on things proceed very much like the case of vector Ward identities.
Integrating eq.~(\ref{eq:zaqb}) over $x$ eliminates the surface term with the axial current divergence,
leaving us with
\bea
&& (m_1^{\rm PCAC} + m_2^{\rm PCAC}) \,\,\int d^4x \,\, G_P \left (x_1-x,x_2-x;m_{01} ,m_{02} \right ) \nonumber
 =  
 \nonumber \\
&&  \,\, \cS \left (x_1-x_2;m_{01} \right ) \gamma_5  \,\,
+ \,\, \gamma_5 \cS \left ( x_1-x_2;m_{02} \right ) \,\,\, .
\label{eq:awi-int}    
\eea
Upon multiplying through with the quark field renormalization constant $Z_\psi$, the r.h.s. becomes
finite. Thus also the l.h.s. is finite, which means that the product of the PCAC quark mass times
the pseudoscalar density is a renormalization group invariant quantity (i.e. it is scale independent).
The anomalous dimension of the the pseudoscalar density is the opposite to that of the quark mass
($\gamma_m = - \gamma_P$)\footnote{This is not unexpected: anomalous dimensions are continuum quantities and
since in the continuum scalar and pseudoscalar operators belong to the same chiral multiplet, they have the same anomalous dimension
$\gamma_S = \gamma_P$. We have already derived from vector Ward identities that $\gamma_S = -\gamma_m$.}.
Given that the pseudoscalar operator $P$ is multiplicatively renormalized by $Z_P$, this suggests
the following renormalization pattern for the PCAC quark mass:
\be
[ \,\, m_f (g_{\rm R}, \mu) \,\, ]_{\rm R} \,\,\, \equiv \,\,\, Z_P^{-1}(g_0,a\mu) \,\,\, m^{\rm PCAC}_f (g_0, m_{01}, \cdots m_{0N_F}) \,\,\, .
\label{eq:mpcacren2}
\ee
This expression amounts to a definition of the renormalized quark mass, in a scheme which fixes the renormalization of the pseudoscalar density  $Z_P$. Identifying this definition of $[m_f]_{\rm R}$ with the generic expression~(\ref{eq:mfRen}) gives a closed expression for $\overline m_f$.

In conclusion, lattice axial Ward identities lead to: (i) the definition of the chiral limit as the solution of the equation $m^{\rm PCAC}(\mcrit) = \mcrit$; (ii) the normalization of the axial current $A_\mu$ by $Z_A$. The last requirement may raise the following question: the properly normalized matrix element, from which the pion decay constant $f_\pi$ is computed, is given by
\be
Z_A \,\, \langle \pi \vert A_\mu \vert 0 \rangle \,\, = \,\, \langle \pi \vert [A_\mu]_{\rm R} \vert 0 \rangle \,\, + \,\, \cO(a) \,\,\, .
\label{eq:piA0}
\ee
Since we know that in the continuum limit $Z_A \rightarrow 1$, why bother to normalize the axial current by $Z_A$? The answer provided by axial Ward identities like eq.~(\ref{eq:awiud}) (and eq.~(\ref{eq:awi3}) below) is that close to the continuum limit the normalized matrix element in the above expression differs from its continuum limit by discretization effects which are positive powers of the lattice spacing. Consequently, omitting $Z_A$ (a regular function of the bare coupling $g_0^2$), implies that the bare matrix element converges to its continuum limit like $g_0^2 \sim 1/ \ln(a)$. This is much slower than the convergence of $Z_A A_\mu$, shown in eq.~(\ref{eq:piA0}). 

\section{Hadronic Ward identities for $Z_V$ and $Z_A$}
\label{sec:hadWIVA}

So far we have seen how Ward identities fix the normalization of lattice partially conserved currents and the ratio of the scalar and pseudoscalar renormalization constants. We have mostly used vector and axial Ward identities based on the variation of the operator $O(x_1,x_2) = \psi (x_1) \bar \psi (x_2)$, commonly referred to as ``Ward identities on quark states". This terminology refers to the fact that at large time separations, say $x^0_1 
\ll x^0 \ll x^0_2$, expressions like (\ref{eq:vwiud}) and (\ref{eq:awiud}) eventually produce PCVC and PCAC relations between matrix elements of quark states. These Ward identities can form a basis for the
determination of the current normalizations $Z_V, Z_A$ and the ratio $Z_S/Z_P$, but this is usually not very practical
(for instance they require gauge fixing, which is affected by the Gribov ambiguity etc.). It is therefore preferable to write down Ward identities involving correlation functions of gauge invariant operators, based on suitable choices of $O(x_1,x_2)$. These are called hadronic Ward identities because they give rise, in terms of the LSZ procedure, to relations between matrix elements of hadronic states. Here we will briefly describe some examples of Ward identities leading to non-perturbative computations of the scale independent parameters $Z_V, Z_A$. A similar discussion concerning  ratios of scalar and pseudoscalar renormalization constants is the subject of the next section.

The computation of $Z_V$ is based on the observation that, up to discretization effects, $\widetilde V_\mu = Z_V V_\mu + \cO(a)$. It is then straightforward to obtain, at fixed UV cutoff (i.e. fixed $g_0$) estimates of $Z_V$ by comparing the insertion of the point-split vector current in a correlation function, to that of the local current. For example,
equations
\bea
\label{eq:ZVcomp}
\sum_{k=1}^3 \,\, \int d^3 {\bf x}  \,\, \big \langle \,\, \widetilde V_k^a(x)  \,\, V_k^b(z) \,\, \big \rangle \,\,\, &=& \,\,\, Z_V
\sum_{k=1}^3 \,\, \int d^3 {\bf x}  \,\, \big \langle \,\, V_k^a(x) \,\, V_k^b(z) \,\, \big \rangle \,\,\, ,\\
\int d^3 {\bf x} \, \int d^3 {\bf y} \,\, \big \langle \,\, P^a(x) \,\, \widetilde V_0^b(z)  \,\, P^c(y) \,\, \big \rangle \,\,\, &=& \,\,\, Z_V
\int d^3 {\bf x} \, \int d^3 {\bf y} \,\, \big \langle \,\, P^a(x) \,\, V_0^b(z) \,\, P^c(y) \,\, \big \rangle \,\,\, , \nonumber
\eea
are valid up to $\cO(a)$. Thus they
may be solved for $Z_V$, providing two independent estimates for this quantity, which differ by discretization
effects. Strictly speaking, integration over space dimensions is not an essential feature, but it is useful in practice, as it averages out statistical fluctuations. In order to avoid singularities from contact terms, time-slices
are kept distinct ($x^0 \ne y^0 \ne z^0$). Note that the flavour superscripts $a,b,c$ must be chosen so as to give, through Wick contractions of the fermion fields, connected diagrams of valence quarks with non-vanishing expectation values.

A non-perturbative determination of $Z_A$ is based on two axial Ward identities. We write down eq.~(\ref{eq:awi2})
for the operator $O(y) = P^b(y)$; with $y \ne x$ the l.h.s. vanishes. As we have argued above, the term containing $X_{\rm R}^a$ may also be dropped, being a discretization effect. We also work with the local, rather than the point-split current. Thus, up to $\cO(a)$, we have 
\bea
\label{eq:awi3}
 \sum_\mu \nabla_x^\mu  \Big \langle A^a_\mu(x) \,\,\, P^b(y) \,\, \Big \rangle =
 \Big \langle  \,\, \bigg [ \bar \psi(x) \,\,
\bigg \{ \dfrac{\lambda^a}{2}, \dfrac{[M_0 - \overline M]}{Z_A}  \bigg \} \gamma_5 \psi(x) \bigg ] \,\, P^b(y)  \,\, \Big \rangle
 \,\,\, . 
\eea
Like before, flavour indices $a,b$ must be appropriately chosen. The above Ward identity may be solved
for the ratio $[M_0 - \overline M_0]/Z_A$; this is one of the most standard methods for the determination of
the PCAC quark mass, up to the axial current normalization factor $Z_A$. At fixed gauge coupling and degenerate
quark masses, the quantity
$[m_0 - \overline m_0]/Z_A$ is computed at several values of the bare quark mass $m_0$ and subsequently
extrapolated to zero, in order to obtain a non-perturbative estimate of $\mcrit$; cf. eq.~(\ref{eq:mpcacren}) and
the comment that follows it. Note that in order to increase
the signal stability, the above Ward identity is usually integrated over all space coordinates of $y$ (with $x^0 \ne y^0$).

The next step is to work out the axial Ward identity (\ref{eq:awi}), with
the operator choice $O(x,y) = A^b_\nu(x)  V^c_\rho(y)$ (for $N_F > 2$):
\bea
&& Z_A \,\, \nabla^\mu_z \,\,\big  \langle \,\, A^a_\mu(z) \,\, A^b_\nu(x) \,\, V^c_\rho(y) \,\, \big \rangle
\label{eq:aav} \\
&& \qquad = \big \langle \,\, \Big [ \bar \psi(z) \,\, \Big \{ \dfrac{1}{2} \lambda^a,(M_0 - \overline M) \Big \} \, \gamma_5 \,\, \psi(z) \Big ] 
\,\, A^b_\nu(x) \,\, V^c_\rho(y) \,\,\big \rangle \,\,\,
+ \,\,\, \big \langle \,\, \overline X^a(z) \,\, A^b_\nu(x) \,\, V^c_\rho(y) \,\, \big \rangle 
\nonumber \\
&& \qquad + \,\,\, i\,\, f^{abd} \,\, \delta (z-x) \,\, \big \langle \,\, V^d_\nu(x) \,\, V^c_\rho(y) \,\,
\big \rangle \,\,\, + \,\,\, i \,\, f^{acd} \,\, \delta (z-y) \,\,
\big \langle \,\, A^b_\nu(x) \,\, A^d_\rho(y) \,\, \big \rangle \,\,\, , \nonumber
\eea
where again use of eq.~(\ref{eq:xbar}) has been made and the shorthand notation $\overline X^a = a [O_5^a]_{\rm R} /Z_5$ has been
introduced. Once more, flavour superscripts $a,b,c$ are chosen so as
to give, upon Wick contraction,  non-vanishing expectation values of connected diagrams of valence quarks. The correlation function with the $\overline X^a$ operator gives rise to contact terms. Symmetry arguments impose that these contact terms must have the form
\bea
\big \langle \,\, \overline X^a(z) \,\, A^b_\nu(x) \,\, V^c_\rho(y) \,\, \big \rangle \,\,\, = \,\,\,
&-& \,\, i \,\, k_1(g_0^2) \,\, f^{abd} \,\, \delta (z-x) \,\, \big \langle \,\, V^d_\nu(x) \,\, V^c_\rho(y) \,\, \big \rangle \,\,\, 
\label{eq:xk1k2} \\
&+& \,\, i \,\, k_2(g_0^2) \,\, f^{acd} \,\, \delta (z-y) \,\,
\big \langle \,\, A^b_\nu(x) \,\, A^d_\rho(y) \,\, \big \rangle \,\,\, + \dots \,\,\, ,
\nonumber
\eea
where the ellipsis stands for localized Schwinger terms, which will vanish after
integration over $z$, to be performed below (we always keep $x \neq y$).  For $x,y \neq z$ the l.h.s. vanishes in the continuum limit
as discussed in section~\ref{sec:AWI}.

We now proceed as 
follows: (I) rewrite eq.~(\ref{eq:aav}), using eq.~(\ref{eq:xk1k2}) and 
expressing 
all bare quantities in terms of the renormalized ones; (II) recall that the product of the PCAC quark mass
and the pseudoscalar density in the first term of the r.h.s. of eq.~(\ref{eq:aav}) is renormalization group invariant
(c.f. eq.~(\ref{eq:mpcacren2})); (III) require that the renormalized 
quantities obey the corresponding continuum (nominal) axial Ward identity. Thus we obtain
\bea
&& k_1(g_0^2) = 1 -\dfrac{Z_V}{Z_A} \nonumber \\
&& k_2(g_0^2) = \dfrac{Z_A}{Z_V} - 1 \,\,\, ,
\label{eq:k1k2}
\eea
and the Ward identity in terms of properly normalized currents is:
\bea
&&Z_A^2 \,\, Z_V \,\,  \big \langle \,\, \Big [ \nabla^\mu_z A^a_\mu (z) - \bar \psi (z)
\Big \{ \dfrac{\lambda^a}{2} , \dfrac{\overline M - M_0}{Z_A} \Big \} \gamma_5 \psi (z) \Big ] \,\,
A^b_\nu(x) \,\, V^c_\rho(y) \,\, \big \rangle \,\,\, =
\nonumber \\
&&  +\,\, i \,\, f^{abd} \,\, \delta (z-x) \,\, Z_V^2 \,\,
\big \langle \,\, V^d_\nu(x) \,\, V^c_\rho(y) \,\, \big \rangle
\nonumber \\
&& + \,\, i \,\, f^{acd} \,\, \delta (z-y) \,\, Z_A^2 \,\,
\big \langle \,\, A^b_\nu(x) \,\, A^d_\rho(y) \,\, \big \rangle
\label{eq:xav}
\eea
Performing the integration over $z$ kills off the first term on the l.h.s.\footnote{
Note that eq.~(\ref{eq:pav}) is only valid away from the chiral limit, where the 
integral over $z$ of the total divergence $\nabla^\mu_z A^a_\mu (z)$ 
vanishes. At zero quark mass, the term containing the total divergence of the axial current
contributes, because of the presence of massless Goldstone bosons.}. 
We also integrate over ${\bf x}$ in order to improve the signal to noise ratio in practical computations (recall that $x \neq y$ 
is necessary in order to eliminate Schwinger terms; thus $x^0 \neq y^0$):
\bea
&& \int d^4z \,\, \int d^3 {\bf x} \,\,
\Big \langle \,\, 
\Big [ \bar \psi (z) \,\, \Big \{ \dfrac{1}{2}\lambda^a,  \dfrac{\overline M-M_0}{Z_A} \Big \} \gamma_5 \psi (z) \Big ] \,\,
A^b_\nu(x) \,\, V^c_\rho(y) \Big \rangle \,\, =
\label{eq:pav}  \\
&& \qquad -i \,\, \dfrac{Z_V}{Z_A^2} \,\, f^{abd} \,\, \int d^3 {\bf x} \,\,
\big \langle \,\, V^d_\nu (x) \,\, V^c_\rho (y) \,\, \big \rangle
-i \,\, \dfrac{1}{Z_V} \,\, f^{acd} \,\, \int d^3 {\bf x} \,\,
\big \langle \,\, A^b_\nu(x) \,\, A^d_\rho(y) \,\, \big \rangle
\nonumber
\eea
Since $Z_V$ is known from eqs.~(\ref{eq:ZVcomp}) and  $[\overline M-M_0]/Z_A$ is known from
eq.~(\ref{eq:awi3}), the above Ward identity may be solved for $Z_A$. For $N_F = 2$, this determination of $Z_A$ is clearly not viable. A method for obtaining $Z_A$ for any $N_F \ge 2$ is based on an axial Ward identity with quark external states~\cite{Martinelli:1993dq}.

\section{Hadronic Ward identity for the ratio $Z_S/Z_P$}
\label{sec:hadWISP}

A further application of hadronic lattice Ward identities concerns the renormalization of the scalar and pseudoscalar
densities. In a regularization which respects chiral symmetry, these operators, belonging to the same chiral multiplet,
are renormalized by the same parameters $Z_S = Z_P$. This is not the case of Wilson fermions, which break chiral symmetry.
Following a line of reasoning similar to the one of the previous section, applied to the operator $O(x,y)= S^g(x) P^h(y)$ (with $g \ne h$), we obtain the Ward identity (for $N_F > 2$)
\bea
&& \int d^4z \,\, \int d^3 {\bf x} \,\,
\Big \langle \,\, 
\Big [ \bar \psi (z) \,\, \Big \{ \dfrac{1}{2}\lambda^f ,  \dfrac{\overline M-M_0}{Z_A} \Big \} \gamma_5 \psi (z) \Big ] \,\,
S^g(x) \,\, P^h(y) \Big \rangle \,\, =
\label{eq:wisp}  \\
&& \qquad \,\, \dfrac{Z_P}{Z_A Z_S} \,\, d^{fgl} \,\, \int d^3 {\bf x} \,\,
\big \langle \,\, P^l(x) \,\, P^h(y) \,\, \big \rangle
+ \,\, \dfrac{Z_S}{Z_A Z_P} \,\, d^{fhl} \,\, \int d^3 {\bf x} \,\,
\big \langle \,\, S^g(x) \,\, S^l(y) \,\, \big \rangle
\nonumber
\eea
where the $d$'s are defined in eq.~(\ref{eq:ds}). Once $Z_A$ has been determined through, say, eq.~(\ref{eq:pav}), the above Ward identity may be used to compute the ratio $Z_S/Z_P$. 

Note that Ward identities such as~(\ref{eq:pav}) and (\ref{eq:wisp}) may be regarded as explicit  demonstrations of the fact that $Z_V, Z_A$ and $Z_S/Z_P$ are scale independent functions of the bare gauge coupling: upon solving them for these $Z$-factors, we obtain solutions which are combinations of {\it bare} correlation functions. These bear no dependence on renormalization schemes and scales and remain finite in the continuum limit. At fixed gauge coupling (i.e. fixed UV cutoff), they provide non-perturbative estimates of $Z_V, Z_A$ and $Z_S/Z_P$.

Practical simulations of these Ward identities are often performed on lattices with periodic and/or antiperiodic boundary conditions for the gauge and fermion fields. For a given gauge coupling, simulations are carried out at several non-zero quark masses and the resulting $Z$-factors are extrapolated to the chiral limit.  Alternatively, Schr\"odinger functional (Dirichlet) boundary conditions and suitably chosen correlation functions enable us to perform computations of $Z_V, Z_A$ and $Z_S/Z_P$ directly in the chiral limit\footnote{For scale dependent renormalization parameters,  such as $Z_P$, renormalization group running is naturally performed non-perturbatively in the  Schr\"odinger functional scheme, for a large range of renormalization scales.
As this topic is beyond the scope of the present lectures, the reader is advised to consult the literature on non-pertrubative renormalization and the Schr\"odinger Functional~\cite{Luscher:1998pe,Sommer:1997xw,Weisz:2010nr}.}. Possible differences among
numerical results  for the same $Z$-factors, obtained from different formulations and/or Ward identities, are due to discretization effects and provide an estimate of this source of systematic error.

A determination of the ratio $Z_S/Z_P$, valid also for $N_F = 2$, is based on the expression
\bea
[\,\, m_1 \,\,]_{\rm R} - [\,\, m_2  \,\,]_{\rm R} \,\,\, &=& \,\,\, Z_S^{-1} \big [ \, m_{01} \,\, - \,\, m_{02} \, \big ]
\nonumber \\
&=& \,\,\, Z_P^{-1} \big [ \, m^{\rm PCAC}_1 \,\, - \,\, m^{\rm PCAC}_2 \, \big ] \,\,\, .
\label{eq:mrendiff}
\eea
The first line is obtained from eqs.~(\ref{eq:mdiff}) and (\ref{eq:zsm}), while the second one from eq.~(\ref{eq:mpcacren2}). The ratio of bare to PCAC mass differences gives an estimate of $Z_S/Z_P$. Such a computation must be carried out for pairs of non-degenerate quark masses. Subsequently results are to be extrapolated to the chiral limit. To the best of our knowledge, this method has never been applied in practical computations.

\section{Singlet scalar and pseudoscalar operators}
\label{sec:hadWISPsing}

It is sometimes convenient to work with Ward identities involving matrix elements of gauge invariant operators between hadronic states, obtained from Ward identities between correlations functions such as eq.~(\ref{eq:aav}), and the use of the standard LSZ procedure. The reader should have no difficulty performing these steps. We focus on the following Ward identity, based on the axial variation of the pseudoscalar density $P^g$ amongst hadronic states $\vert h_1\rangle \ne \vert h_2 \rangle$:
\bea
&& \nabla^\mu_x \,\, \langle h_1 \vert \cT [ Z_A \,\, A^f_\mu(x) \,\, P^g(0) ] \,\, \vert h_2 \rangle \,\,\, - \,\,\, 2 \,\, [m_0 \,\, - \,\, \overline m] \,\, 
\langle h_1 \vert \cT [ P^f (x) \,\, P^g(0) ] \,\, \vert h_2 \rangle \nonumber \\
&&  = \langle h_1 \vert \cT [ \overline X^f (x) \,\, P^g(0) ] \,\, \vert h_2 \rangle \,\, + \,\, i \,\,
\Big \langle h_1 \Big \vert \dfrac{\delta P^g(0)}{\delta\alpha_A^f(x)} \,\, \Big \vert h_2 \Big \rangle \,\,\, .
\label{eq:hadWI}
\eea
The flavour index $f = 1, \cdots, N_F^2-1$ characterizes non-singlet currents, pseudoscalar densities etc., while  $g= 0, \cdots, N_F^2-1$ also covers the case of the flavour singlet pseudoscalar density $P^0$. For simplicity we assume exact flavour symmetry; i.e. $M_0 = m_0 I$ and thus $\{ \lambda^f,M_0\} = 2 m_0 \lambda^f$. The axial variation of the pseudoscalar density is
\be
\dfrac{\delta P^g(0)}{\delta\alpha_A^f(x)} \,\,\, = \,\,\, i \,\,
\delta(x) \,\, \Big [ \,\, d^{fgh}  \,\, S^h(0) \,\,
+ \,\, \delta^{fg} \sqrt{\dfrac{2}{N_F}} 
\,\, S^0(0) \,\, + \delta^{g0} \sqrt{\dfrac{2}{N_F}} 
\,\, S^f(0) \,\, \Big ]  \,\,\, .
\label{eq:Pvar}
\ee
The axial current normalization $Z_A$ and the mass subtraction $\overline m$ in eq.~(\ref{eq:hadWI}) are generated in standard fashion by the counter-terms of the operator $O^f_5 = X^f /a$; cf. eq.~(\ref{eq:xbar}). This is not, however the end of the story as we now allow for the case in which the space-time point $x$ comes close to the origin ($x \approx 0$). In this case, the insertion of $ \overline X^f$ in off shell green functions does not vanish as we approach the continuum limit. Rather, it generates local contact terms which are constrained, by flavour symmetry, to have the form:
\bea
&& \langle h_1 \vert \cT [ \overline X^f (x) \,\, P^g(0) ] \,\, \vert h_2 \rangle  \,\,\, = \,\,\, - \,\,
\delta(x) \,\, \Big [ \,\, c_1(g_0^2) \,\, d^{fgh}  \,\,\langle h_1 \vert  S^h(0) \vert h_2 \rangle 
\label{eq:hadWICT} \\
&& + \,\, c_2(g_0^2) \,\, \delta^{fg} \sqrt{\dfrac{2}{N_F}}  
\,\, \langle h_1 \vert S^0(0) \vert h_2 \rangle \,\, + \,\, c_3(g_0^2) \,\,\delta^{g0} \sqrt{\dfrac{2}{N_F}} 
\,\, \langle h_1 \vert S^f(0) \vert h_2 \rangle \,\, \Big ]  \, .
\nonumber 
\eea
Putting it all together, and going over to the chiral limit, yields the Ward identity
\bea
\nabla^\mu_x \,\, \langle h_1 \vert \cT [ && Z_A \,\, A^f_\mu(x) \,\, P^g(0) ] \,\, \vert h_2 \rangle \,\,\,
= \,\,\, - \,\, \delta(x) \,\, \Big \{ \,\, [1 + c_1(g_0^2)] \,\, d^{fgh}  \,\, \langle h_1 \vert S^h(0) \,\, \vert h_2 \rangle \nonumber \\
&& + \,\, \delta^{fg} \sqrt{\dfrac{2}{N_F}} [1 + c_2(g_0^2)] 
\,\, \langle h_1 \vert S^0(0) \,\, \vert h_2 \rangle \,\, 
\label{eq:hadWIfin}\\
&& + \,\, \delta^{g0} \sqrt{\dfrac{2}{N_F}}  [1 + c_3(g_0^2)] 
\,\, \langle h_1 \vert S^f(0) \,\, \vert h_2 \rangle \,\, \Big \} \,\,\, . \nonumber 
\eea

For a non-singlet pseudoscalar density $P^g$ (i.e. for $g \ne 0$) and for $f \ne g$, the last terms in the r.h.s. of the above expression vanish. Since $P^g$ is multiplicatively renormalizable, it is adequate to multiply the above expression by $Z_P$ to render it finite. This results to the following relation between the renormalization factor of the non-singlet scalar density and $Z_P$:
\bea
Z_S & \,\,\, = \,\,\, & [1 + c_1(g_0^2)] \,\, Z_P \,\,\, .
\eea
If, on the the hand,  $f=g\ne0$, the second term on the r.h.s. of eq.~(\ref{eq:hadWIfin}) survives and we obtain the relation:
\bea
Z_{S^0} & \,\,\, = \,\,\, & [1 + c_2(g_0^2)] \,\, Z_P \,\,\, .
\eea
Finally we consider the case in which $P^g$ is a singlet pseudoscalar density (i.e. $g=0$), for which only the last term  on the r.h.s. of eq.~(\ref{eq:hadWIfin}) survives. Multiplying through by the factor $Z_{P^0}$ we obtain the relation
\bea
Z_S & \,\,\, = \,\,\, & [1 + c_3(g_0^2)] \,\, Z_{P^0} \,\,\, .
\eea
Thus it has been established~\cite{Bochicchio:1985xa,RoMaTe} that the $[N_F,{\overline N_F} ] \oplus [{\overline N_F}, N_F ]$ representation of the chiral group $SU(N_F)_L \otimes SU(N_F)_R$ is formed by the rescaled  operators
\bea
P^f & \qquad \qquad & [1+c_1] \,\, S^f \qquad \qquad f \ne 0 \,\,\, ,
\\
\dfrac{1+c_1}{1+c_3} P^0 & \qquad \qquad & [1+c_2] \,\, S^0 \,\,\, ,
\eea
which renormalize by a common factor $Z_P$. Conversely, the renormalized operators $P_{\rm R}^f = Z_P P^f$, $S_{\rm R}^f = Z_S P^f$ (for $f \ne 0$) and $P_{\rm R}^0 = Z_{P^0} P^f$ have a common anomalous dimension (arising from, say, $Z_P$) but renormalization constants $Z_P$, $Z_S$ and $Z_{P^0}$ which differ by finite normalization factors. These factors tend to unity in the continuum limit. Their presence is yet another consequence of chiral symmetry breaking by the Wilson term. Analogous statements are true for the renormalization factor $Z_{S^0}$ of the singlet scalar density $S^0$, which however is also subject to power subtractions, as we will see below.

Another important result~~\cite{Maiani:1987by,RoMaTe} is the relation between the ratio $Z_P/Z_{S^0}$ and the ratio of PCAC to subtracted quark masses. The first step towards its derivation consists in a Ward identity analogous to that of eq.~(\ref{eq:hadWI}), but with $S^0$ in place of $P^g$:
\bea
&& \nabla^\mu_y \,\, \langle h_1 \vert \cT [ Z_A \,\, A^f_\mu(y) \,\, S^0(x) ] \,\, \vert h_2 \rangle \,\,\, - \,\,\, 2 \,\, [m_0 \,\, - \,\, \overline m] \,\, 
\langle h_1 \vert \cT [ P^f (y) \,\, S^0(x) ] \,\, \vert h_2 \rangle \nonumber \\
&&  = \langle h_1 \vert \cT [ \overline X^f (y) \,\, S^0(x) ] \,\, \vert h_2 \rangle \,\, + \,\, i \,\,
\Big \langle h_1 \Big \vert \dfrac{\delta S^0(x)}{\delta\alpha_A^f(y)} \,\, \Big \vert h_2 \Big \rangle 
\nonumber \\
&& = \,\, - \,\, \delta(x-y) \,\, \big [ 1 + t(g_0^2) \big ] \,\, \sqrt{\dfrac{2}{N_F}} \,\,\langle h_1 \vert P^f(y) \vert h_2 \rangle  \,\,\, .
\label{eq:hadWI2m}
\eea
The last equation is the result of performing the functional derivative $\delta S^0(x)/\delta\alpha_A^f(y)$ and establishing that the contact term, arising when $\overline  X^f (y)$ is in the vicinity of $S^0(x)$, is proportional to $P^f$, with $t(g_0^2)$ the proportionality factor.
Multiplying through by $Z_{S^0}$ gives the renormalized Ward identity
\bea
&& Z_{S^0} \,\, \nabla^\mu_y \,\, \langle h_1 \vert \cT [ Z_A \,\, A^f_\mu(y) \,\, S^0(x) ] \,\, \vert h_2 \rangle \,\,\, - \,\,\, 2 \,\, [m_0 \,\, - \,\, \overline m] \,\, Z_{S^0} \,\,  \langle h_1 \vert \cT [ P^f (y) \,\, S^0(x) ] \,\, \vert h_2 \rangle \nonumber \\
&& = \,\, - \,\, \delta(x-y) \,\, \sqrt{\dfrac{2}{N_F}} \,\,Z_P \,\, \langle h_1 \vert P^f(y) \vert h_2 \rangle  \,\,\, ,
\label{eq:hadWI3m}
\eea
with the identification $Z_P = [1 + t(g_0^2)] Z_{S^0}$.

The next step consists in differentiating the axial Ward identity between hadronic states with respect to the bare quark mass:
\bea
\dfrac{\partial}{\partial m_0} \nabla^\mu_y \,\, \langle h_1 \vert Z_A \,\, A^f_\mu(y) \,\, \vert h_2 \rangle \,\,\, =  \,\,\,
\dfrac{\partial}{\partial m_0} \,\,\, 2 \,\, [m_0 \,\, - \,\, \overline m] \,\, 
\langle h_1 \vert P^f (y) \vert h_2 \rangle \,\,\, .
\label{eq:derWI1}
\eea
The mass derivative on the r.h.s. gives
\bea
\dfrac{\partial}{\partial m_0} \,\,\, 2 \,\, [m_0 \,\, - \,\, \overline m] \,\, 
\langle h_1 \vert P^f (y) \vert h_2 \rangle
&=& \,\,\, 2 \Big [1 - \dfrac{\partial \overline m}{\partial m_0} \Big ] \,\,  \langle h_1 \vert P^f (y) \vert h_2 \rangle \nonumber \\
&+& \,\,\,  2 \,\, [m_0 \,\, - \,\, \overline m] \,\, \dfrac{\partial}{\partial m_0}
\langle h_1 \vert P^f (y) \vert h_2 \rangle  \,\,\, .
\label{eq:derWI2}
\eea
Now differentiating a Green's function with respect to the quark mass $m_0$ amounts to inserting the operator $\int d^4x \bar \psi(x) \psi(x) = \sqrt{2 N_F} \int d^4x S^0(x)$. Performing this insertion on the l.h.s. of eq.~(\ref{eq:derWI1}) and the last term on the  r.h.s. of eq.~(\ref{eq:derWI2}) results to the following identity
\bea
\nabla^\mu_y \,\, \langle h_1 \vert Z_A \,\, \cT \Big [ \,\,  A^f_\mu(y) && \int  d^4x S^0(x) \,\,  \Big ] \,\, \vert h_2 \rangle \,\,\,
= \,\,\, \sqrt{\dfrac{2}{N_F}} \Big [1 - \dfrac{\partial \overline m}{\partial m_0} \Big ] \,\,  \langle h_1 \vert P^f (y) \vert h_2 \rangle \nonumber \\
&& + \,\,\, 2 \,\, [m_0 \,\, - \,\, \overline m] \,\, \langle h_1 \vert \cT \Big [ \,\,  P^f(y) \,\, \int d^4x S^0(x) \,\, \Big ] \,\, \vert h_2 \rangle \,\,\, .
\eea
Multiplying both sides of the above by $Z_{S^0}$ gives a finite expression. Upon comparing it with eq.~(\ref{eq:hadWI3m}), and taking the chiral limit $m_0 \rightarrow \mcrit$, we obtain a relation between renormalization factors
\be
\dfrac{Z_P}{Z_{S^0}} \,\,\, = \,\,\, 1 \,\, - \,\, \dfrac{\partial \overline m}{\partial m_0} \Big \vert_{\mcrit} \,\,\, ,
\label{eqZPZS0}
\ee
Compared with~eqs.(\ref{eq:mpcacren}) and (\ref{eq:mPCACmsub}), this gives a relation between the PCAC and the subtracted quark masses:
\be
\lim_{m_0 \rightarrow \mcrit} \,\, \Big [ \dfrac{m^{\rm PCAC}}{m_0 - \mcrit} \Big ]  \,\,\, =  \,\,\, \dfrac{Z_P}{Z_{S^0}} \,\,\, .
\ee
\label{eq:mPCACmsub2}
Combining this expression with the renormalized quark mass definition~(\ref{eq:mpcacren2}), we obtain
\be
m_{\rm R} \,\,\, = \,\,\, Z^{-1}_{S^0} \,\, [ \,\, m_0 \,\, - \,\, \mcrit \,\, ] \,\,\, ,
\ee
which, upon comparison with eqs.~(\ref{eq:mrzm}) and (\ref{eq:msub}) leads to
\be
Z_{S^0}  \,\,\, = \,\,\, Z^{-1}_{m^0}
\ee
We see that the above expression, as well as eq.~(\ref{eq:zsm}), relate the singlet and non-siglet mass renormalization constants to the corresponding scalar density renormalization constants.

In perturbation theory, the difference between singlet and non-singlet operators arises from extra diagrams in the singlet case, involving fermion loops. As these drop out in the quenched approximation, the quenched renormalization constants obey $Z_S = Z_{S^0}$ and $Z_P = Z_{P^0}$.

\section{The chiral condensate}
\label{sec:chircond}

Our last topic related to lattice Ward identities is the derivation of the proper definition of the chiral condensate with Wilson 
fermions~\cite{Bochicchio:1985xa}. The starting point is the Ward identity~({\ref{eq:hadWI}) for the vacuum expectation values of
the non singlet axial current $A_\mu^f$ and the pseudoscalar density $P^g$ (i.e. $f,g \ne 0$ and $\vert h_1 \rangle = \vert h_2 \rangle = \vert 0 \rangle$):
\bea
&& \nabla^\mu_x \,\, \langle 0 \vert \,\, \cT [ Z_A \,\, A^f_\mu(x) \,\, P^g(0) ] \,\, \vert 0\rangle \,\,\, - \,\,\, 2 \,\, [m_0 \,\, - \,\, \overline m] \,\, 
\langle 0 \vert \,\, \cT [ P^f (x) \,\, P^g(0) ] \,\, \vert 0 \rangle \nonumber \\
&&  = \,\,\, \langle 0 \vert \,\, \cT [ \overline X^f (x) \,\, P^g(0) ] \,\, \vert 0  \rangle \,\, + \,\, \delta^{fg} \,\, \delta (x) \,\, \dfrac{1}{N_F} \,\, 
 \langle 0 \vert \bar \psi \psi \vert 0 \rangle \,\,\, .
\label{eq:hadWIcond}
\eea
Note that only the term proportional to $S^0$ survives in the vacuum expectation value of the axial variation of the pseudoscalar density~(\ref{eq:Pvar}). Moreover the contact terms arising in the limit of vanishing lattice spacing from the insertion of $\overline X^f(x)$ with $P^g(0)$ at $x \approx 0$ are not exactly those of eq.~(\ref{eq:hadWICT}): at first sight it appears that  the only contact term
surviving in eq.~(\ref{eq:hadWICT}), sandwiched between vacuum states is the one proportional to $\langle 0 \vert S^0 \vert 0 \rangle$.
But this is not the whole story, since
we should also take into account terms which vanish in eq.~(\ref{eq:hadWICT}), when hadronic states $\vert h_1 \rangle \ne \vert h_2 \rangle$, but survive when these states are equal (e.g. the vacuum). The reader can easily convince himself that in the chiral limit the contact term structure is
\be
\langle 0 \vert \,\, \cT [ \overline X^f (x) \,\, P^g(0) ] \,\, \vert 0  \rangle \,\,\, = \,\,\, \delta^{fg} \,\, \Big [ \dfrac{1}{N_F} \,\, \delta(x) \dfrac{b_0(g_0^2)}{a^3} \,\, + \,\, \Box \delta(x) \dfrac{b_1(g_0^2)}{a} \Big ] \,\,\, ,
\label{cont-terms-chcon}
\ee
where the term proportional to $\langle 0 \vert S^0 \vert 0 \rangle$ is subdominant (it may be considered as incorporated in $b_0$). Away from the chiral limit we also have terms proportional to the mass (e.g. $\propto m^{\rm PCAC}/a^2$) but these are subdominant compared to the cubic divergence and irrelevant for the present discussion.

Combining the last two equations in the chiral limit we obtain
\bea
&& \nabla^\mu_x \,\, \langle 0 \vert \,\, \cT [ Z_A \,\, A^f_\mu(x) \,\, P^g(0) ] \,\, \vert 0\rangle \,\,\, - \,\,\, \delta^{fg} \,\, \Box \delta(x) \dfrac{b_1(g_0^2)}{a} 
\nonumber \\
&&  = \,\,\, \delta^{fg} \,\, \delta (x) \,\, \dfrac{1}{N_F} \,\, \Big [
 \langle 0 \vert \bar \psi \psi \vert 0 \rangle  \,\, + \,\, \dfrac{b_0(g_0^2)}{a^3} \,\, \Big ] \,\,\, .
\label{eq:hadWIcond2}
\eea
The subtraction on the l.h.s. compensates a divergent term, arising in the expectation value
$\langle 0 \vert \,\, \cT [ A^f_\mu(x) \,\, P^g(0) ] \,\, \vert 0\rangle$ when $x \approx 0$. This term is proportional to $\partial_\mu \delta(x)$. Multiplying the Ward identity by $Z_P$ renders both sides finite. This leads to the following definition of the renormalized chiral condensate with Wilson fermions
\be
\langle \bar \psi \psi \rangle_{\rm R} \,\,\, \equiv \,\,\, Z_P\, \dfrac{1}{N_f} \,  \Big [ \langle 0 \vert \bar \psi \psi \vert 0 \rangle  \,\, + \,\, \dfrac{b_0(g_0^2)}{a^3} \,\, \Big ]
\,\,\, .
\label{eq:hadWIcond3}
\ee
Note that a chirally symmetric regularization would lead to the much simpler result $\langle \bar \psi \psi \rangle_{\rm R} = Z_{S^0} \langle 0 \vert \bar \psi \psi \vert 0 \rangle/N_f$. The more complicated renormalization pattern of the last expression is due to the loss of chiral symmetry
by Wilson fermions.

Useful information may be also be obtained from Ward identity~(\ref{eq:hadWIcond}), by integrating it over all space-time. We distinguish two cases. In the first case we take the chiral limit {\it before} performing the integration. Then the term proportional to $(m_0 - \overline m)$ vanishes. The integrated $\nabla^\mu_x \hat A^f_\mu$-term would also vanish, being the integral of a four-divergence, {\it provided chiral symmetry were not broken in QCD} (absence of Goldstone bosons). Since upon integration, the Schwinger $b_1$-term  of eq.~(\ref{cont-terms-chcon}) also vanishes, the integrated Ward identity~(\ref{eq:hadWIcond2}) would then imply a vanishing chiral condensate. However, in QCD the symmetry is broken, and the presence of Goldstone bosons guarantees a non-vanishing surface term upon integrating $\nabla^\mu_x \hat A^f_\mu$. Thus, also the r.h.s. of the Ward identity (i.e. the chiral condensate) is non-zero.

The second case of interest consists in integrating Ward identity~(\ref{eq:hadWIcond}) over all space-time, {\it before} going to the chiral limit. Now this integration will kill off the first term on the l.h.s. (it is an integral of the total derivative of the axial current in the presence of massive states) while the second term survives. Upon integration, the Schwinger $b_1$-term  of eq.~(\ref{cont-terms-chcon}) also vanishes. Taking the chiral limit {\it after} the integration, we obtain 
\be
\label{eq:wiccint}
\dfrac{\delta^{fg}}{N_f} \,\, \Big [ \langle 0 \vert \bar \psi \psi \vert 0 \rangle  \,\, + \,\, \dfrac{b_0(g_0^2)}{a^3} \,\, \Big ] \,\, = \,\, - \,
\lim_{m_0 \rightarrow m_{\rm cr}} 2 [ m_0 - \overline m] \, \int d^4 x\langle 0 \vert \,\, \cT [ P^f (x) \,\, P^g(0) ] \,\, \vert 0 \rangle  \,\, .
\ee
\par
From the above equation, we can derive two other expressions for the chiral condensate, by inserting, in standard fashion, a complete set of states in the time-ordered product of pseudoscalar densities. The spatial integration $\int d^3 x$ projects zero-momentum states. Contributions from higher mass states vanish in the chiral limit, leaving us with a zero-momentum pion state $\vert \pi(\vec 0) \rangle$. Upon performing the time integration $\int dx^0$, we find
\be
\label{eq:wip}
\dfrac{1}{N_f} \,\,  \,\, \Big [ \langle 0 \vert \bar \psi \psi \vert 0 \rangle  \,\, + \,\, \dfrac{b_0(g_0^2)}{a^3} \,\, \Big ] \,\, = \,\,
\lim_{m_0 \rightarrow m_{\rm cr}} \,\, 
\dfrac{(m_0 - \overline m)}{m_\pi^2} \,\,
\Big \vert \langle 0 \vert P(0) \vert \pi(\vec 0) \rangle \Big \vert^2\; ,
\ee
where $m_\pi$ is the mass of the pseudoscalar state $\vert \pi \rangle$ and the flavour indices $f,g$ have been suppressed for simplicity.

Next recall that the definition of the pion decay constant is given by
\begin{equation}
\langle 0 \vert [A_\mu(x)]_{\rm R} \vert \pi(\vec p) \rangle \,\,\, = \,\,\, i \,\, f_\pi \,\, p_\mu \exp(ipx) \,\,\, .
\label{eq:fpi}
\end{equation}
We also know that
\begin{equation}
Z_A \nabla^\mu_x \,\, \langle 0 \vert A_\mu(x) \vert  \pi(\vec p) \rangle \,\,\, = \,\,\, 2 [m_0  - \overline m] \langle 0 \vert P(x) \vert  \pi(\vec p) \rangle
\,\,\, = \,\,\, - f_\pi m_\pi^2 \exp(ipx) \,\,\, ,
\label{eq:AWIdervpfi}
\end{equation}
where the first equation is the axial Ward identity between states $\vert 0 \rangle$ and  $\vert \pi \rangle$, while the second one is obtained by derivation of eq.~(\ref{eq:fpi}). Taking the square of the second equation at $\vec p = \vec 0$ and combining it with eq.~(\ref{eq:wip}) we find
\be
\label{eq:wiaa}
\dfrac{1}{N_f}  \,\, \Big [ \langle 0 \vert \bar \psi \psi \vert 0 \rangle  \,\, + \,\, \dfrac{b_0(g_0^2)}{a^3} \,\, \Big ] \,\, = \,\,  \lim_{m_0 \rightarrow m_{\rm cr}} \,\, \dfrac{f_\pi^2 m_\pi^2}{4 (m_0 - \overline m)} \,\, .
\ee
Multiplying both sides by $Z_P$ yields the familiar Gell-Mann--Oakes--Renner relation \cite{GellMann:1968rz}:
\begin{equation}
\langle \bar \psi \psi \rangle_{\rm R} \,\,\, = \,\,\,  \lim_{m_{\rm R} \rightarrow 0} \dfrac{f_\pi^2 m_\pi^2}{4 m_{\rm R}} \,\,\, .
\label{eq:GMOR}
\end{equation}
Note that the non-vanishing of the chiral condensate in the last two equations implies the well known linear dependence of the pseudoscalar mass squared on the quark mass, close to the chiral limit. Provided that numerical simulations can be performed close to the chiral limit, eqs.~(\ref{eq:wip}) and (\ref{eq:wiaa}) may be used for the computation of the chiral condensate with Wilson fermions\footnote{Computing the chiral condensate directly form the trace of the quark propagator is not viable with Wilson fermions, due to the presence of the cubic divergence proportional to $b_0$.}.

This concludes our discussion of lattice Ward identities with Wilson fermions. A related subject is that of Ward identities of the singlet axial current, related to the question of $U(1)_A$ anomaly and the $\eta^\prime$ mass. Although extremely important, this topic is beyond the scope of the present lectures.

\chapter{Momentum subtraction (MOM) schemes}
\label{sec:MOM}

We have seen that Ward identities may be used in order to determine non-perturbatively the normalization of partially conserved currents and finite ratios of renormalization constants of operators of the same chiral multiplet. These quantities would be identically equal to unity if chiral symmetry were preserved by the lattice regularization. On the other hand, local operators like $S^a$, $P^a$ and $T^a_{\mu\nu}$ are subject to multiplicative renormalization by scale dependent renormalization parameters $Z_S$, $Z_P$ and $Z_T$, which must be fixed by a more or less arbitrary renormalization condition. We now present a renormalization scheme, known as the RI/MOM scheme, which is suitable for the non-perturbative evaluation of these parameters.

Before presenting the RI/MOM scheme explicitly, we will discuss a few general features of momentum subtraction schemes. For definitiveness, let us consider the dimension-3, multiplicatively renormalizable operator $O^a_\Gamma$ defined in eq.~(\ref{eq:o}). Mimicking what is usually done with operator renormalization in continuum momentum subtraction (MOM) schemes, we impose that a suitable renormalized vertex function between say, quark states, at a given momentum scale $\mu$, be equal to its tree level value. For instance,
\be
\Big [ \langle \,\, p^\prime \,\, \vert \,\, O^a_\Gamma \,\, \vert \,\, p \,\, \rangle \Big ]_{\rm R} {\Big \vert}_{p^{\prime 2} = p^2 = \mu^2} 
 \,\,\, = \,\,\, Z_O(a\mu) \,\, 
\langle \,\, p^\prime \,\, \vert \,\, O^a_\Gamma \,\, \vert \,\, p \,\, \rangle_{\rm bare} \,\,\, = \,\,\, 
\langle \,\, p^\prime \,\, \vert \,\, O^a_\Gamma \,\, \vert \,\, p \,\, \rangle_{\rm tree} \,\,\, ,
\label{eq:mom}
\ee
where $| p \rangle$,  $| p^\prime \rangle$ are single quark states of four-momenta $p, p^\prime$ respecively. This is a familiar example of a momentum subtraction scheme (MOM-scheme). Some important properties of such schemes are:
\begin{itemize}
\item The renormalization condition, imposed on quark states, is gauge dependent. Gauge fixing is required and one typically opts for the Landau gauge.  Care should be taken that no problems arise from such a choice. For instance, dependence of the results on Gribov copies of a given configuration ensemble is obviously undesirable. 
\item MOM is a mass independent, infinite volume renormalization scheme. This means that in principle the condition (\ref{eq:mom}) is written at infinite lattice volumes and vanishing quark masses. In practice numerical simulations are performed at large but finite volumes and non-zero quark masses. Results are then extrapolated to the chiral limit; their independence on the volume must be carefully checked.
\item The above renormalization condition is imposed in the chiral limit. By working at non-vanishing exceptional momenta, any infrared problems associated with vanishing quark masses are avoided~\cite{ItzyksonZuber}.
\item Once the bare matrix element in the above MOM condition is regularized on the lattice, the renormalization constant $Z_O(a\mu)$. computed non-perturbatively, is subject to systematic errors due to discretization effects which are typically  $\cO(a\mu)$, $\cO( a\Lambda_{\rm QCD})$ and $\cO(am)$, for non-improved Wilson fermions. Effects which are $\cO(am)$ are extrapolated away by going to the chiral limit, while those which are $\cO(a\mu)$ and $\cO(a\Lambda_{\rm QCD})$ are entangled with the non-perturbative definition of $Z_O$.
\end{itemize}

The reliability of non-perturbative $Z_O$ estimates, computed in a MOM scheme, depend on the existence of a so-called ``renormalization window" for the renormalization scale $\mu$, defined through the inequalities
\be
\Lambda_{\rm QCD} \,\, \ll \,\, \mu \,\, \ll \,\, \cO(a^{-1}) \,\,\, .
\label{eq:renwin}
\ee
The upper bound guarantees that $\cO(a\mu)$ discretization effects are under control (and so are  $\cO(a\Lambda_{\rm QCD})$ ones). The lower bound is imposed for two reasons. The first reson is that in many cases, related to operator weak matrix elements, the physical amplitude of interest $\cA$ is expressed in terms of an operator product expansion (OPE), schematically written as:
\be
\label{eq:OE}
\cA \,\, = \,\, \langle f | \cH_{\rm eff} | i \rangle \,\, = \,\, C_W \Big ( \dfrac{\mu}{M_W} \Big ) \,\,\,
 \langle f | O (\mu) | i \rangle_{\rm R} \,\,\, ,
\ee
where $|i\rangle$ and $|f\rangle$ are physical initial and final states. The scale dependent Wilson coefficient $C_W$ comprises all short distance effects of the physical process under consideration. It is known (typically to NLO) in perturbation theory. The renormalized weak matrix element 
\be
\label{eq:OEren}
 \langle f | O (\mu) | i \rangle_{\rm R} \,\,\, = \,\,\, \lim_{a \rightarrow 0} Z[g_0(a),a\mu] \,\,\, \langle f | O[g_0(a)] | i \rangle_{\rm bare}  \,\,\, ,
\ee
is a long distance quantity, computed non perturbatively. Its scale dependence cancels that of $C_W$ (to, say NLO), so as to have a scale independent physical amplitude $\cA$. Since $C_W$ is a perturbative quantity, it should be calculated at a scale $\mu$, well above $\Lambda_{\rm QCD}$. But $\mu$ is also the scale at which $Z_O$ must be computed non-perturbatively, which leads to the lower bound of the renormalization window.

The second reason for requiring that $\Lambda_{\rm QCD} \ll \mu$ is related to the existence of unwanted non-perturbative effects, which affect the determination of some $Z_O$'s at scales close to the infrared. This so-called ``Goldstone pole contamination" will be treated extensively below.

Moreover, the UV cutoff $a^{-1}$ must be well above the energy scales of the problem, such as $\Lambda_{\rm QCD}$, $\mu$ (and any quark masses of active flavours). Finally, all these scales must also be higher than the lattice extension $L$, which acts as an infrared cutoff. Thus we must tune the bare coupling $g_0(a)$ so that
\be
L^{-1} \,\, \ll \,\, \Lambda_{\rm QCD} \,\, \ll \mu \,\, \ll \,\, a^{-1} \,\, .
\label{eq:window}
\ee
With present-day computer resources, this hierarchy of scales is not always easy to satisfy in practice. One could then turn to finite-size scaling methods, similar to the ones based on the Schr\"odinger functional~\cite{Weisz:2010nr}. Although these methods have been described in general terms for the RI/MOM scheme some time ago~\cite{Donini:1999sf}, they have only been systematically formulated and developed very recently~\cite{Arthur:2010ht}, with extremely promising preliminary results.

\section{The RI/MOM scheme}
\label{sec:RIMOM}

A specific renormalization scheme, which is particularly well suited for non-perturbative computations of renormalization constants is the RI/MOM scheme. As the name betrays, it is a momentum-subtraction scheme,  similar to the one described above (the reason for the acronym ``RI" is described in Appendix~\ref{app:RI}. The difference lies in the fact that the renormalization condition is not imposed on operator matrix elements of single-quark states, but on the amputated projected correlation function defined in eq.~(\ref{eq:proj_GF}). Since the renormalization pattern of this correlation function is given by eq.~(\ref{eq:renG}), it immediately follows that the renormalization condition
\be   
\big[\,\Gamma_O (\mu,\gren,m_{\rm R}=0)  \, \big]_{\rm R} \, = \, \lim_{a \rightarrow 0} \,
\left[ \, Z_\psi^{-1}(a\mu,g_0) \,\, Z_O(a\mu,g_0) \,\, \Gamma_O (p,g_0,m) \,
\right]_{\begin{matrix} p^2 = \mu^2 \\ m \rightarrow 0 \end{matrix}}   \,\,\, = \, 1 \,\,\, ,
\label{eq:renGcond}   
\ee 
fixes the combination $Z_\psi^{-1} \,\, Z_O$. Note that unity on the r.h.s. stands for the tree level value of the
correlation function  $\Gamma_O$, which has been constructed so as to satisfy this property. 

There are several ways to disentangle the two renormalization constants from the product $ Z_\psi^{-1} \,\, Z_O$. A conceptually straightforward method is based on the conservation of the vector current $\tilde V_\mu$. We {\it assume}\footnote{This assumption will be proved in section~\ref{sect:RI-WI}.} that the RI/MOM condition for the conserved current is compatible with the Ward identity result of eq.(\ref{eq;consVC}), $Z_{\tilde V} =1$. This means that 
\be   
[\,\,\, \Gamma_{\tilde V} (\mu,\gren,m_{\rm R} =0)  \,\, ]_{\rm R} \,\,\, = \,\,\,  \lim_{a \rightarrow 0} \,\,\, 
\left[ \,\, Z_\psi^{-1}(a\mu,g_0) \,\, \Gamma_{\tilde V} (p,g_0,m) \,\, \right] _{\begin{matrix} p^2 = \mu^2 \\ m \rightarrow 0 \end{matrix}}    \,\,\, = \,\,\, 1 \,\,\, .
\label{eq:renVtilde}   
\ee
The quark field renormalization $Z_\psi$ cancels in the ratio $\Gamma_O (\mu) / \Gamma_{\tilde V} (\mu)$, which can be solved for $Z_O$. Alternatively, one often uses the product $Z_V V_\mu$ in place of $\tilde V_\mu$, with $Z_V$ known from some Ward identity. 

Another method is to first compute $Z_\psi$ from the RI/MOM condition of the quark propagator, based on eqs.~(\ref{eq:gamwf}) and (\ref{eq:renwf}):
\bea   
\left[ \,\, \Gamma_\Sigma (\mu) \,\, \right]_{\rm R} \,\, &=& \,\, \lim_{a \rightarrow 0}  \,\,\,
 \left[ Z_\psi^{-1}(a\mu,g_0) \,\,\, \Gamma_\Sigma (p,g_0,m) \right]_{\begin{matrix} p^2 = \mu^2 \\ m \rightarrow 0 \end{matrix}}  \,\, = \,\, 1 \,\,\, .
\label{eq:rensigRIMOM}   
\eea   
Once $Z_\psi$ is known, eq.~(\ref{eq:renGcond}) gives us $Z_O$. This method is avoided in practice, because the definition of $\Gamma_\Sigma$ involves derivatives, which look rather ugly with discretized momenta. A slightly different renormalization condition is preferred for the fermion field renormalization:
\be
\dfrac{i}{12} \,\, \Tr \left[ \dfrac{\pslash \,\, [ \cS^{-1}]_{\rm R}}{p^2} \right] \,\,\, = \,\,\,
\lim_{a \rightarrow 0}  \,\,\, ( Z_\psi')^{-1} \,\, \dfrac{i}{12}
\Tr \left[ \dfrac{\pslash\,\, \cS^{-1}}{p^2} \right]_{\begin{matrix} p^2 = \mu^2 \\ m \rightarrow 0 \end{matrix}}  \,\,\, = \,\,\, 1 \,\,\, .
\label{eq:rimom2}
\ee
Operator renormalization constants $Z_O$, computed with $Z_\psi'$ rather than $Z_\psi$, are said
to be in the RI'/MOM scheme. In general we expect that $Z_\psi' = Z_\psi + \cO(g_0^2)$. However, in the Landau gauge it is known that $Z_\psi'= Z_\psi + \cO(g_0^4)$. As this is the gauge of preference, differences between RI/MOM and RI'/MOM results are expected to be small.

For the quark mass, an RI/MOM renormalization condition, based on eqs.~(\ref{eq:gamsm}) and (\ref{eq:rensig}) would be:
\bea   
\hspace {-0.75 cm}  \left[ \,\, \Gamma_m (\mu) \,\, \right]_{\rm R} \, = \, \lim_{a \rightarrow 0}  \,
 \left[ Z_\psi^{-1}(a\mu,g_0) \,\,\, Z_m^{-1}(a\mu,g_0) \,\,\, \Gamma_m (p,g_0,m) \right]_{\begin{matrix} p^2 = \mu^2 \\ m \rightarrow 0 \end{matrix}}  \,\, = \,\, 1 \,\,\, .
\label{eq:renmass1RIMOM}   
\eea
In numerical simulations, this is also inconvenient, as $ \Gamma_m$ requires derivation of w.r.t the mass. The following condition is used in practice:
\bea   
\hspace {-0.75 cm}  \left[ \,\, \Gamma_m^\prime (\mu) \,\, \right]_{\rm R} \, = \, \lim_{a \rightarrow 0}  \,
 \left[ Z_\psi^{-1}(a\mu,g_0) \,\,\, Z_m^{-1}(a\mu,g_0) \,\,\, \Gamma_m^\prime (p,g_0,m) \right]_{\begin{matrix} p^2 = \mu^2 \\ m \rightarrow 0 \end{matrix}}  \,\, = \,\, 1 \,\,\, .
\label{eq:renmass2RIMOM}   
\eea
with
\be
\Gamma_m^\prime(p;m) \,\,\, = \,\,\, \dfrac{1}{12 m} \,\,
\Tr \left[ \cS^{-1} (p;m) \right]   \,\,\, .
\label{eq:gamsm-prim}   
\ee
Note that the renormalization pattern of $\Gamma_m^\prime$ is analogous to that of eq.~(\ref{eq:rensig}) for $\Gamma_m$.
The two conditions are equivalent, as differentiation of eq.~(\ref{eq:renmass2RIMOM}) w.r.t. the quark mass $m$ results to eq.~(\ref{eq:renmass1RIMOM}.)

In practical simulations, the RI/MOM' conditions are solved for $Z_\psi'$ and $Z_O$ at fixed bare coupling. The lattices are big with periodic and/or antiperiodic boundary conditions. At each $g_0$, one has to work at several small quark masses and then extrapolate the results to the chiral limit. The whole procedure is repeated for several lattice momenta $ap$ in order to empirically establish a range of renormalization scales $ap =a\mu$, within the renormalization window of eq.~(\ref{eq:window}).

\section{Goldstone pole contamination}
\label{sect:Goldstone}

It is clear from the previous section that RI/MOM is a mass independent renormalization scheme, i.e. the renormalization conditions are imposed in the chiral limit. For field theories with vertices with degree four (such as QCD), renormalized at some fixed Euclidean point, it is well known~\cite{ItzyksonZuber}, that if all masses of a Feynman diagram go to zero, infrared singularities may appear, unless the external momenta are non-exceptional\footnote{A correlation function has non-exceptional momenta if no partial sum of the incoming momenta $p_i$ vanishes.}. This is not our case:  the renormalization condition (\ref{eq:renGcond}) is imposed on the amputated-projected Green function $\Gamma_O(p)$, derived from the momentum-space correlation function $G_O(p_1,p_2)$ of eq.~(\ref{eq:gp}) for $p_1 = p_2$. In other words, the operator $O$ carries zero momentum ($q \equiv p_1-p_2 =0$) and quark-field external momenta are exceptional. In order to investigate which correlation functions are subject to dangerous infrared singularities, we apply the LSZ reduction formula~\cite{PeskinSchroeder} to the correlation functions $G_O(p_1,p_2)$~\cite{Papin:thesis}. Returning momentarily to Minkowski space-time, we obtain for the pseudoscalar operator $P$:
\begin{align}
G_P(p_1,p_2) \,\,\, &\sim \,\,\, \int d^4x_2\,\, \exp(ip_2 x_2) \,\, \Big [ \langle 0 \vert {\rm T} [ \psi(0) \bar \psi(x_2) ] \vert \pi(\vec q) \rangle \,\,\,
\dfrac{i \theta(q^0)}{q^2 - m_\pi^2} \,\,\, \langle \pi(\vec q) \vert [P(0)]_{\rm R} \vert 0 \rangle
\nonumber \\
&+ \,\,\, \langle 0 \vert [P(0)]_{\rm R} \vert \pi(-\vec q) \rangle \,\,\, \dfrac{i\theta(-q^0)}{q^2 - m_\pi^2} \,\,\,
\langle \pi(-\vec q) \vert {\rm T} [ \psi(0) \bar \psi(x_2) ] \vert 0 \rangle \Big ] \,\,\, + \,\,\, \dots  \,\,\, ,
\label{eq:LSZP}
\end{align}
where the ellipsis indicates terms without pion poles; isospin indices are implicit\footnote{Contrary to what we have done so far, we find it convenient to identify $x_1$ with the origin (i.e. $x_1=0$) in the correlation function $G_P(x_1-x,x_2-x)$ of eq.~(\ref{eq:gos}). Then the Fourier transform, analogous to that of eq.~(\ref{eq:gos}), is performed w.r.t to $q$ (conjugate variable of $x$) and $p_2$ (conjugate variable of $x_2$), in order to obtain $G(p_1,p_2)$, with $p_2 \equiv p_1-q$.}. 

Combining eqs.~(\ref{eq:AWIdervpfi}) and (\ref{eq:GMOR}) we obtain for the vacuum-to-pion pseudoscalar matrix element
\begin{equation}
\langle 0 \vert [P(0)]_{\rm R} \vert \pi(p) \rangle \,\,\, = \,\,- \, 2 \,\, \dfrac{\langle \bar \psi \psi \rangle_{\rm R}}{f_\pi} \exp(ipx) \,\,\, .
\end{equation}
Since $\langle \bar \psi \psi \rangle_{\rm R} \ne 0$ (spontaneous symmetry breaking), the above matrix element
survives the chiral limit. This last result, applied to eq.~(\ref{eq:LSZP}) in the limit $q_\mu \rightarrow 0$, implies that the correlation function $G_P(p,p)$ (and thus also $\Gamma_P(p)$) have a $1/m_\pi^2$ pole. The presence of this so-called Goldstone pole means that the RI/MOM renormalization condition, which strictly speaking is valid in the chiral limit, can only be applied to $Z_P$ with some precaution (see below for details).

The situation is less dramatic in the case of the axial current, for which the LSZ reduction formula is
\begin{align}
G^\mu_A(p_1,p_2) \,\,\, &\sim \,\,\, \int d^4x_2\,\, \exp(ip_2 x_2) \,\, \Big [ \langle 0 \vert {\rm T} [ \psi(0) \bar \psi(x_2) ] \vert \pi(\vec q) \rangle \,\,\,
\dfrac{i \theta(q^0)}{q^2 - m_\pi^2} \,\,\, \langle \pi(\vec q) \vert [A_\mu(0)]_{\rm R} \vert 0 \rangle
\nonumber \\
&+ \,\,\, \langle 0 \vert [A_\mu(0)]_{\rm R} \vert \pi(-\vec q) \rangle \,\,\, \dfrac{i \theta(-q^0)}{q^2 - m_\pi^2} \,\,\,
\langle \pi(-\vec q) \vert {\rm T} [ \psi(0) \bar \psi(x_2) ] \vert 0 \rangle \Big ] \,\,\, + \,\,\, \dots  \,\,\, .
\label{eq:LSZA}
\end{align}
Since the relevant matrix element behaves like $\langle 0 \vert [A_\mu(0)]_{\rm R} \vert \pi(q) \rangle \sim q_\mu$ (cf. eq~(\ref{eq:fpi})) in the
limit $q_\mu \rightarrow 0$ the Goldstone pole contribution of $G_A^\mu(p,p)$ vanishes. However, as we will see below, the pole of $G_A^\mu(p,p)$ still plays an important role in axial Ward identities, used for the determination of $Z_A$. Other correlation functions such as $G_V^\mu$, $G_S$ and $G_T$ (as well as $\Gamma_V$, $\Gamma_S$ and $\Gamma_T$) have analogous LSZ reduction formulae. In these cases however, the interpolating operators $V_0$, $S$ and $T_{\mu\nu}$ have Lorentz properties which ensure that the corresponding intermediate single-particle state is not a pion. As the mass of these particles does not vanish in the chiral limit, these correlation functions are free of Goldstone pole singularities\footnote{Intermediate two-pion states are of course allowed. These however give rise not to poles, but to branch cuts with less severe, logarithmic singularities in the chiral limit.}.

It is important to realize that the Goldstone pole term of $G_P(p,p)$ vanishes in the limit of large external momentum $p^2 \rightarrow \infty$. This has been discussed in general terms in the original RI/MOM paper~\cite{renorm_mom:paper1}, where the following correlation function was considered:
\begin{equation}
F_O(p) \,\,\, = \,\,\, \int d^4x \,\, d^4y \,\, \exp(-ipx) \,\, \langle \, [\bar \psi(0) \Gamma \psi(x) ] \,\, O_\Gamma(y) \, \rangle \,\,\, ,
\end{equation}
with $O_\Gamma(y)$ defined in eq.~(\ref{eq:o}). In the limit of large $p^2$, the dominant contribution to $F_O (p)$ comes from regions of integration where the integrand is singular; i.e. $x \sim 0$ and $x \sim y$. Based on the OPE and dimensional arguments, it is shown that~\cite{renorm_mom:paper1}
\begin{equation}
F_O(p) \,\,\, = \,\,\, c_\Gamma \,\, \dfrac{\ln^{\gamma_\Gamma} (p^2/\mu^2)}{p^2} \,\,\,
+ \,\,\, d_\Gamma \,\, \dfrac{\ln^{\delta_\Gamma} (p^2/\mu^2)}{p^4} \,\, \tilde \Delta_\Gamma(0) \,\,\, ,
\end{equation}
where $c_\Gamma$, $\gamma_\Gamma$, $d_\Gamma$ and $\delta_\Gamma$ can be calculated in perturbation theory, while
\begin{equation}
\tilde \Delta_\Gamma(0) \,\,\, = \,\,\, \int d^4y \,\, \langle \, O_\Gamma(0) \,\, O_\Gamma(y) \, \rangle 
\end{equation}
is a non-perturbative quantity. This means that, in the large $p^2$ limit, the perturbative contribution of  $F_O(p)$ (i.e. the first term, proportional to $c_\Gamma$) dominates  by one power of $p^2$ over the non-perturbative one (i.e. the term proportional to $d_\Gamma$). This is  why $F_O(p)$ is infrared safe in perturbation theory, when $p^2$ is large. In other words, in the large external momentum limit, the contamination by the Goldstone pole residing in  $\tilde \Delta_\Gamma(0)$, as well as other infrared chiral effects, vanishes. At finite momenta however, the Green function
$\Gamma_P$, defined in eq.~(\ref{eq:proj_GF}), is not regular in the chiral limit. By requiring that $\Lambda_{\rm QCD} << \mu$ (the lower bound of the renormalization window (\ref{eq:window})) one is attempting to reduce, in practical simulations, the contamination due infrared chiral effects, such as the Goldstone pole.

\section{RI/MOM scheme and Ward Identities}
\label{sect:RI-WI}

Once the wave function renormalization $Z_\psi$ is computed, the RI/MOM renormalization condition (\ref{eq:renGcond})  fixes the renormalization parameter $Z_O$; examples are $Z_S$, $Z_P$, $Z_T$ but also $Z_V$ and $Z_A$. However, $Z_V, Z_A$ and the ratio of $Z_S/Z_P$ do not depend on any renormalization scheme (or renormalization scale), as they are fixed by Ward identities. It is then important that their determination through the RI/MOM renormalization condition be compatible with such identities\footnote{Strictly speaking, it is of course conceivable that one renormalizes, say $S$ and $P$, without taking Ward identities into account. This is acceptable from the point of view of renormalization (i.e. removal of divergences) but distorts, by terms finite in the bare coupling $g_0$, the recovery of symmetry as the regularization is removed.}. In this section we will establish that the RI/MOM determination of  $Z_S/Z_P$ agrees with that from Ward identities only in the limit of large renormalization scales $\mu >> \Lambda_{\rm QCD}$, where Goldstone pole contaminations die off\footnote{For simplicity we limit our discussion to the renormalization properties of quark bilinear operators. Analogous conclusions may be drawn for other more complicated cases. For example, the scale independent mixing coefficients of the four-fermion operators, such as the ones involved in $K \rightarrow \pi\pi$ decays, are also fixed by Ward identities~\cite{Bochicchio:1985xa,Donini:1999sf}. When computed in the RI/MOM scheme~\cite{Donini:1999sf}, they are also subject to Goldstone pole contamination~\cite{Zrimom:clov}.}. On the other hand, the RI/MOM determination of $Z_V$ is compatible with Ward identities at all momentum scales. Finally, Goldstone poles do not affect the RI/MOM determination of $Z_A$, provided the chiral limit is approached with care.

Let us start from the vector Ward identity (\ref{eq:vwiud}), with $\tilde V_\mu$ replaced by $Z_V V_\mu$.
We Fourier-transform and amputate eq.~ (\ref{eq:vwiud}), obtaining (at small $a q_\mu$)
\begin{eqnarray}
\nonumber   
\sum_\mu iq_\mu &Z_V& \Lambda_V^\mu    
\left( p+\frac{q}{2}, p-\frac{q}{2};  m_{01}, m_{02} \right)    
- (m_{02} - m_{01})  \Lambda_S   
\left( p+\frac{q}{2}, p-\frac{q}{2}; m_{01},  m_{02} \right) \\   
&=&   
\cS^{-1} \biggl (p+\frac{q}{2};  m_{01} \biggr)   
-  \cS^{-1} \biggl (p-\frac{q}{2}; m_{02} \biggr)   \,\,\, .
\label{eq:vwift2}
\end{eqnarray}   
We next go through the following steps: (i) derive the above expression w.r.t. $q_\mu$; (ii) let $m_{01} = m_{02}$, so as to dispose of the term with the scalar vertex function $\Lambda_S$; (iii)  trace the resulting equation with $\gamma_\mu$, (iv) take the limit $q_\mu \rightarrow 0$. This gives
\begin{equation}
Z_V \, \Gamma_V(p) \,\, + \,\, Z_V \lim_{q_\mu \rightarrow 0}
\dfrac{q_\mu}{48} {\rm Tr} \Big [ \gamma_\rho \dfrac{\partial \Lambda_V(p+q/2,p-q/2)}{\partial q_\rho}
\Big ] \,\,\, = \,\,\,  \Gamma_\Sigma(p)
\label{eq:vwift3}
\end{equation}
The LSZ reduction formula for the vector current is analogous to eq.~(\ref{eq:LSZA}), but the denominator does not involve a Goldstone pole $m_\pi$. This implies that, upon taking its derivative w.r.t. $q_\rho$, and in the limit $q_\mu \rightarrow 0$, the second term on the l.h.s. of the above Ward identity vanishes (there are no infrared singularities).  Thus we obtain
\begin{equation}
Z_V \,\, \Gamma_V(p) \,\,\, = \,\,\, \Gamma_\Sigma (p)
\end{equation}
Solved for $Z_V$ this is yet another Ward identity determination of the vector current normalization. On the other hand, if the above expression is multiplied by $Z_\psi^{-1}$, determined in the RI/MOM scheme through eq.~(\ref{eq:rensigRIMOM}), we obtain at momenta $p = \mu$
\begin{equation}
Z_\psi^{-1} \,\, Z_V \,\, \Gamma_V(\mu) \,\,\, = \,\,\, 1 \,\,\, ,
\end{equation}
which is the RI/MOM condition for $Z_V$. We have shown that for the vector current the RI/MOM scheme is consistent with Ward identities.

The situation is not quite the same for axial Ward identities. Starting from eq.~(\ref{eq:awiud}), written for degenerate quark masses,
we perform the same steps which led from eq.~(\ref{eq:vwift2}) to eq.~(\ref{eq:vwift3}) and obtain
\begin{eqnarray}
Z_A \, \Gamma_A(p) \,\, &+& \,\, Z_A \lim_{q_\mu \rightarrow 0}
\dfrac{q_\mu}{48} {\rm Tr} \Big [ \gamma_5 \gamma_\rho \dfrac{\partial \Lambda_A^\mu(p+q/2,p-q/2)}{\partial q_\rho} \Big ]
\\
& + & \,\, \dfrac{2 m^{\rm PCAC}}{48} \lim_{q_\mu \rightarrow 0}
{\rm Tr} \Big [ \gamma_5 \gamma_\rho \dfrac{\partial \Lambda_P(p+q/2,p-q/2)}{\partial q_\rho}
\Big ] \,\,\, = \,\,\,  \Gamma_\Sigma(p) \,\,\, .
\nonumber
\label{eq:awift3}
\end{eqnarray}
The LSZ reduction formulae (i.e. eq.~(\ref{eq:LSZA}) for the axial current  and eq.~(\ref{eq:LSZP}) for the pseudoscalar density) may now be used in order to study the Goldstone pole contributions to the second and third term on the l.h.s.. It can be easily shown that, away from the chiral limit,  as $q_\mu \rightarrow 0$, these terms vanish. Approaching subsequently the chiral limit, just like in the vector current case, the axial Ward identity determination of $Z_A$ is equivalent to the RI/MOM condition. If however, the chiral limit $m^{\rm PCAC} \rightarrow 0$ is taken {\it before} $q_\mu \rightarrow 0$, the third term vanishes but the second one survives. This term is essential to the saturation of the Ward identity, in the  presence of a massless Goldstone boson (i.e. any divergences due to the presence of massless pions in the first and second term will cancel). In this regime the determination of $Z_A$ from the RI/MOM condition is not feasible, unless one considers the $p \rightarrow \infty$ case. As we have seen in sect.~\ref{sect:Goldstone}, in this limit these Goldstone pole terms become negligible.

The simplest way to see a discrepancy between Ward identities and RI/MOM is through $Z_S/Z_P$~\cite{zp:givl}. We start from the Ward identities (\ref{eq:vwiud2}) and (\ref{eq:awi-int}), which we Fourier-transform to momentum space; cf. eqs.~(\ref{eq:Sft}) and (\ref{eq:gp}).
We next amputate the resulting correlation functions as in eq.~(\ref{eq:amp}) and project them as in eqs.~(\ref{eq:proj_GF}) and (\ref{eq:proj}), obtaining the following vector and axial Ward identities:
\begin{eqnarray}
\hspace {-0.75 cm}
(m_{02} - m_{01}) \Gamma_S (p; m_{01}, m_{02}) \, &=& \, \dfrac{1}{12} {\rm Tr} \Big [ {\cS}^{-1}(p; m_{02}) - {\cS}^{-1}(p; m_{01}) \Big ] \,\,\, ,
\\
\hspace {-0.75 cm}
(m_{2}^{\rm PCAC} + m_{1}^{\rm PCAC}) \Gamma_P (p; m_{01}, m_{02}) \, &=& \, \dfrac{1}{12} {\rm Tr} \Big [ {\cS}^{-1}(p; m_{01}) +  {\cS}^{-1}(p; m_{02}) \Big ] \, .
\end{eqnarray}
In the mass degenerate limit these become
\begin{eqnarray}
\Gamma_S \left( p \right)  \,\,\, &=& \,\,\, \dfrac{1}{12} {\rm Tr} \Bigg [ \dfrac{\partial {\cS}^{-1} \left (p; m_0 \right) }{\partial m_0} \Bigg ] \,\,\, ,
\label{eq:GS-WI-mom}
\\
m^{\rm PCAC} \,\, \Gamma_P \left( p \right) \,\,\, &=& \,\,\, \dfrac{1}{12} {\rm Tr} \Bigg [ {\cS}^{-1} \left (p; m_0 \right) \Bigg ] \,\,\, .
\label{eq:GP-WI-mom}
\end{eqnarray}
The vector Ward identity (\ref{eq:GS-WI-mom}) and the definition~(\ref{eq:gamsm}) imply that $\Gamma_S \left( p \right) = \Gamma_m \left( p \right)$. Once again we see from this and eq.~(\ref{eq:rensig}) that if $Z_S$ is determined by the RI/MOM condition, then compatibility with this Ward identity requires that the mass renormalization satisfy $Z_m = Z_S^{-1}$.

We now turn to the axial Ward identity (\ref{eq:GP-WI-mom}).
The inverse quark propagator may be considered a function of either bare quark mass, $m_0$ or $m^{\rm PCAC}$.  We differentiate the
above axial Ward identity w.r.t. $m^{\rm PCAC}$, obtaining
\bea
\Gamma_P \left( p \right) \,\, + \,\, m^{\rm PCAC} \dfrac{\partial \Gamma_P}{\partial m^{\rm PCAC}} \,\,\,
&=& \,\,\, \dfrac{1}{12} {\rm Tr} \Bigg [ \dfrac{\partial {\cS}^{-1} \left (p; m^{\rm PCAC} \right) }{\partial m^{\rm PCAC}} \Bigg ]
\nonumber \\
&=& \,\,\, \dfrac{\partial m_0}{\partial m^{\rm PCAC}} \,\,\, \dfrac{1}{12} {\rm Tr} \Bigg [ \dfrac{\partial {\cS}^{-1} \left (p; m_0 \right) }{\partial m_0} \Bigg ] 
\nonumber \\
&=& \,\,\, \dfrac{Z_S}{Z_P} \,\,  \dfrac{1}{12} {\rm Tr} \Bigg [ \frac{\partial {\cS}^{-1} \left (p; m_0 \right) }{\partial m_0}  \Bigg ]  \,\,\, .
\label{eq:GP-WI-mom2} 
\eea
In the last step, the substitution of the derivative $\partial m_0 / \partial m^{\rm PCAC}$ by the ratio $Z_S/Z_P$ is a consequence of eq.~(\ref{eq:mrendiff}). With the aid of the LSZ expression~(\ref{eq:LSZP}) at zero momentum transfer ($q_\mu = 0$) we easily confirm that the second term of the l.h.s. 
of~eq.(\ref{eq:GP-WI-mom2}) diverges in the chiral limit, due to the presence of a Goldstone pole. This term is essential for the cancellation of a similar contribution of the first term on the l.h.s.. Again as $p \rightarrow \infty$ these contributions become negligible.

The ratio of the above vector and axial Ward identities (\ref{eq:GS-WI-mom}) and (\ref{eq:GP-WI-mom2}) gives the following result
for $Z_P/Z_S$:
\begin{equation}
\frac{Z_P}{Z_S} \,\,\, = \,\,\, \dfrac{\dfrac{\Gamma_S}{\Gamma_P}}
{1 \, + \, \dfrac{m^{\rm PCAC}}{\Gamma_P} \dfrac{\partial \Gamma_P}{\partial m^{\rm PCAC}} } \,\,\, .
\label{eq:WIGPGS}
\end{equation}
But this is not what the RI/MOM renormalization condition~(\ref{eq:renGcond}) gives for the same ratio:
\begin{equation}
\dfrac{Z_P}{Z_S} \,\,\, = \,\,\, \dfrac{\Gamma_S}{\Gamma_P} \,\,\,.
\end{equation}
The two determinations differ by a factor which, as we have already discussed, is necessary in order to cancel Goldstone pole divergences in $\Gamma_P$ and which becomes negligible in the limit $p^2 \rightarrow  \infty$; in practice this limit is realized as $p^2 >> \Lambda_{\rm QCD}$. The absence of this factor from the RI/MOM determination is commonly referred to as Goldstone pole contamination.

It is also instructive to identify the origin of the Goldstone pole in the non-perturbative part of the quark propagator:
\begin{equation}
\def\slashchar#1{\setbox0=\hbox{$#1$}           
   \dimen0=\wd0                                 
   \setbox1=\hbox{/} \dimen1=\wd1               
   \ifdim\dimen0>\dimen1                        
      \rlap{\hbox to \dimen0{\hfil/\hfil}}      
      #1                                        
   \else                                        
      \rlap{\hbox to \dimen1{\hfil$#1$\hfil}}   
      /                                         
   \fi}                                         %
{\cS}^{-1}(p; m) \,\,\, = \,\,\, i {\slashchar{p}} \,\, \Sigma_1(p^2;m;\mu^2) + m \,\, \Sigma_2(p^2;m;\mu^2)  
\,\, + \,\,\Sigma_3(p;m;\mu^2)  
\label{eq:q-prop}
\end{equation}
where
$\Sigma_k$ (with $k=1, \cdots, 3)$ are form factors. The functional form of the first two terms is dictated by general symmetry arguments; the form factors $\Sigma_1$ and $\Sigma_2$ may be calculated in perturbation theory. The last term is a non-perturbative form factor, known to several  orders in the OPE; to $\cO (1/p^2)$ it is given by~\cite{Politzer:1976tv,Pascual:1981jr}:
\begin{equation}
\Sigma_3 (p;\mu;m) = g_0 ^2 \,\, K \,\, \langle \bar \psi \psi \rangle \dfrac{1}{p^2} \,\, +{\cO}\left(p^{-4} \right) \,\,\, ,
\label{eq:qsigma3}
\end{equation}
where $K$ is a mass-independent, gauge-dependent factor, in which logarithmic divergences have also been absorbed. Note the non-perturbative nature of this term: it is proportional to the chiral condensate $\langle \bar \psi \psi \rangle$ and vanishes in the large scale limit $p^2 \rightarrow \infty$. Upon inserting the quark propagator expressions (\ref{eq:q-prop}) and (\ref{eq:qsigma3}) into Ward identities (\ref{eq:GS-WI-mom}) and (\ref{eq:GP-WI-mom}), we find\footnote{We have swapped, in 
eq.~(\ref{eq:GS-WI-mom}) the dependence from the bare quark mass $m_0$ with that of the subtracted bare mass $m$.}
\begin{eqnarray}
\Gamma_S(p;m) \,\, &=& \,\, \Sigma_2(p;m) \,\, + \,\, 
m \dfrac{\partial\Sigma_2(p;m)}{\partial m} \,\,
+ \,\, {\cO}\left(p^{-4} \right)
\\
\Gamma_P(p;m) &=& \dfrac{m}{m^{\rm PCAC}} \,\, \Sigma_2 \left(ap,am\right)
\,\, + \,\, g_0^2 \,\, K
\dfrac{\langle \bar \psi \psi \rangle}{m^{\rm PCAC}\,\, p^2} \,\, + \,\, {\cO}\left(p^{-4} \right) \,\,\, = \,\,\, 
\nonumber
\\
&=& \dfrac{Z_{S^0}}{Z_P} \Big [ \Sigma_2 \left(ap,am\right)
\,\, + \,\, g_0^2 \,\, K
\dfrac{\langle \bar \psi \psi \rangle}{m \,\, p^2} \Big ]\,\, + \,\, {\cO}\left(p^{-4} \right)
\end{eqnarray}
Thus, with the aid of an axial Ward identity, we see that, to $\cO (1/p^2)$ in the OPE, the correlation function $\Gamma_P$ has a pole in the quark mass, proportional to the chiral symmetry breaking order parameter $\langle \bar \psi \psi \rangle$.
On the contrary, using a vector Ward identity we see that such a pole is absent, to $\cO (1/p^2)$, from the scalar
correlation function $\Gamma_S$\footnote{An instructive exercise consists in plugging-in the above OPE expressions for $\Gamma_S$ and $\Gamma_P$ on the r.h.s. of eq.~(\ref{eq:WIGPGS}) in order to explicitly confirm that they combine to give the ratio $Z_P/Z_S$. Contributions proportional to the chiral condensate cancel out.}.

As stated repeatedly, this non-perturbative contribution to $\Gamma_P$ vanishes like $1/p^2$ at large momenta. At finite momenta however, $\Gamma_P$ is not regular in the chiral limit and $p^2 m^{\rm PCAC} >> \Lambda_{\rm QCD}^3$ must be enforced. In practical simulations it is usually difficult to satisfy this requirement. A remedy consists in fitting
$\Gamma_P$, at fixed momenta $p^2$, by $A(p^2) + B(p^2)/m$. Once the fit parameters $A(p^2)$ and $B(p^2)$ are determined, the Goldstone pole contamination is removed from $Z_P$ by imposing the RI/MOM condition on the subtracted correlation function
$A(p^2)$~\cite{Cudell:1998ic,Cudell:2001ny}.

As we have shown to $\cO(p^{-2})$, $\Gamma_S$ does not suffer from Goldstone pole contaminations and therefore the RI/MOM determination of $Z_S$ should be reliable even at relatively low renormalization scales. The same is true for the ratio $Z_P/Z_S$, computed from Ward identities, which are completely free of these problems. Thus, a reliable evaluation of $Z_P$ consists in multiplying the ratio of $Z_P/Z_S$, obtained from a Ward identity, by $Z_S$ in the RI/MOM scheme. The ratio $Z_P/Z_S$ may be computed from hadronic Ward identities, cf. eq.~(\ref{eq:wisp}). Alternatively, Ward identities in momentum space, based on the correlation functions $\Gamma_S$ and
$\Gamma_P$ have also been used with satisfactory results~\cite{zp:givl}.

Besides the single Goldstone pole, other, more complicated non-perturbative effects are of course also be present. Some of these could be revealed by higher orders in the OPE, which involve higher-dimension condensates~\cite{Pascual:1981jr,Lavelle:1991vr}. In any case, all these contributions ought to disappear as $p^2 >> \Lambda_{\rm QCD}$.

\section{The RI/SMOM scheme}
\label{sec:RI-SMOM}

In practice it is not easy to satisfy the bounds imposed by the renormalization window of eq.~(\ref{eq:renwin}). At least one fine example where everything falls in place has been provided in the literature~\cite{Zrimom:clov}. However, several cases exist in which the renormalization window turns out to be quite narrow: upon attempting to compute RI/MOM renormalization parameters at very high momenta, so as to reduce contamination due to the infrared chiral-symmetry breaking effects, one is faced with the problem of discretization errors, which start flawing the data, even in an $\cO(a)$-improved setup. The current precision of simulations is such that the infrared non-perturbatve effects are a significant source of error. Clearly it would be advantageous to renormalize the operators in a scheme which is free of this problem, suppressing these effects by choosing kinematics without channels of exceptional momenta. One such choice, which for quark bilinears is characterized by very simple and convenient kinematics is the recently proposed variant of the RI/MOM scheme, called RI/SMOM (SMOM stands for {\it symmetric} momentum subtraction scheme). Although the new scheme has been introduced in the framework of domain wall fermions~\cite{Aoki:2007xm,Sturm:2009kb}, its definition and the advantages derived from it are independent of the regularization details.

Essentially the scheme consists in the choice of new kinematics for the vertex function $G_O(p_1,p_2)$ of eq.~(\ref{eq:gp}). 
In the standard RI/MOM scheme the renormalization conditions for quark bilinear operators are imposed on Green functions with the operator inserted between equal incoming and outgoing momenta, satisfying (in Euclidean space-time)
\begin{equation}
p_1^2 \,\, = \,\, p_2^2 \,\, = \,\, \mu^2  \qquad ;  \qquad q  = p_1 \,\, - \,\, p_2 \,\, = \,\, 0 \,\,\, .
\end{equation}
In the above momentum configuration, the momentum inserted at the 
operator is therefore $q=0$ so that there is an {\it exceptional} channel, i.e. one in which the squared 
momentum transfer $q^2$ is much smaller than the typical large scale $\mu^2$. 
These kinematics define an asymmetric subtraction point.
The renormalization procedure for RI/SMOM is very similar, but with the incoming and outgoing quarks 
having different momenta, $p_1$ and $p_2$, satisfying
\begin{eqnarray}
p_1^2 \,\, = \,\, p_2^2 \,\, = q^2 \,\, = \,\, \mu^2 \qquad ; \qquad & q\,\, = \,\, p_1 - p_2 \ne 0 \,\,\, .
\label{eq:symmom}   
\end{eqnarray}
There are now no exceptional channels. An example of such a symmetric momentum configuration
on a lattice of linear extension $L$ is 
\bea
p_1 \,\, = \,\, \dfrac{2\pi}{L} (0,1,1,0) \qquad; \qquad p_2 \,\, = \,\, \dfrac{2\pi}{L} (1,1,0,0) \,\,\, ,
\eea
where $p = 2\pi/L (n_x,n_y,n_z,n_t)$. With this choice of a symmetric subtraction point, renormalized quantities such as $[G_O]_{\rm R}$,
$[\Lambda_O]_{\rm R}$ and $[\Gamma_O]_{\rm R}$ depend only on a single scale $\mu^2$.
The amputated Green function $\Lambda_O$ is obtained once more as in eq.~(\ref{eq:amp}), whereas for the projected amputated vertex function $\Gamma_O$ we introduce the following set of projectors:
\bea   
P_S = I \qquad ; \qquad  P_P &=& \gamma_5 \qquad ; \qquad P_T = \dfrac{1}{12} \gamma_\mu \gamma_\nu 
\nonumber \\   
P_V = \dfrac{1}{q^2} q_\mu  \qslash \qquad  & ; & \qquad  
P_A = \dfrac{1}{q^2} q_\mu \gamma_5 \qslash   \,\,\, .
\label{eq:proj2}   
\eea   
The RI/SMOM renormalization condition is that of eq.~(\ref {eq:renGcond}), imposed on the vertex functions $[\Gamma_O]_{\rm R}$ (implicitly defined though the above projectors), in the symmetric momentum configuration of  eq.~(\ref{eq:symmom}). We will use the shorthand notation $\{\mu^2\}$ for these kinematics.

Comparing with eq.~(\ref{eq:proj}), we see that the RI/SMOM definitions of $\Gamma_S$,  $\Gamma_P$ and  $\Gamma_T$ are identical to those of the standard RI/MOM scheme, while  $\Gamma_V$  and $\Gamma_A$ have been redefined through new projectors. The reason behind these changes is that we must ensure that  the new scheme is consistent with Ward identities. If we wish to maintain condition (\ref{eq:rimom2}) for the wave function renormalization $Z_\psi$, we must 
modify the definitions of $\Gamma_V$ and $\Gamma_A$, as implied by eq.~(\ref{eq:proj2}). These modifications must be accompanied by a redefinition of the quark mass renormalization: instead of condition (\ref{eq:rensig}), defined through (\ref{eq:gamsm}), we must impose,
in a symmetric momentum configuration, that
\be
\lim_{a \rightarrow 0} \,\, \Big \{  [\Gamma_m^\prime(p)]_{\rm R} \,\, - \,\, \dfrac{1}{2} \Tr \Big [ q_\mu [\Lambda^\mu_A]_{\rm R} \gamma_5 \Big ] \Big \}
{\Big \vert}_{\{\mu^2\}, m_{\rm R} \rightarrow 0} \,\,\, = 1 \,\,\, ,
\ee
where $\Gamma_m^\prime(p)$ is defined in eq.~(\ref{eq:renmass2RIMOM}) and $\Lambda^\mu_A$ in eq.~(\ref{eq:amp}). Following a line of reasoning similar to that of sect.~\ref{sect:RI-WI}, it has been shown that this RI/SMOM scheme is consistent with Ward identities~\cite{Sturm:2009kb}. Clearly these choices are not unique. It is also possible to maintain the standard RI/MOM definitions for $\Gamma_V$ and $\Gamma_A$ but then the wave function renormalization condition must be modified in order to ensure compatibility with Ward identities~\cite{Sturm:2009kb}. This is a different RI/SMOM scheme.

Compared to the standard RI/MOM, the RI/SMOM scheme displays an improved infrared behaviour. We have explicitly shown in
sects.~\ref {sect:Goldstone} and \ref{sect:RI-WI} how the Goldstone pole contaminates the chiral limit of certain correlation functions as their quark external momenta become exceptional ($p_1 = p_2$) and the vertex operator does not inject any momentum ($q=0$). In the 
RI/SMOM scheme the kinematics are arranged so that this situation is avoided. In fact, using Weinberg' s theorem it has been shown that 
for the asymmetric subtraction point, chiral symmetry breaking effects vanish like $1/p^2$ for large external momenta $p^2$. On the
other hand, with bilinear operator  renormalization constants defined at a symmetric subtraction point (with non-exceptional kinematics), unwanted infrared effects are better behaved and vanish with larger asymptotic powers which are $\cO(1/p^6)$~\cite{Aoki:2007xm}. Numerical evidence, in the framework of domain wall fermion discretization, provide strong support to these arguments~\cite{Aoki:2007xm}.

\chapter{Twisted mass QCD (tmQCD), renormalization  and improvement}
\label{sec:tmQCD}

In recent years, there has been a newcomer to the family of lattice actions, known as lattice QCD with a chirally twisted mass term (tmQCD for short). It consists in a modification of the mass term of the lattice Wilson fermion action. The new regularization has several advantages, but these come at a price. On the positive side, the tmQCD fermion matrix is free of the spurious zero modes which plague quenched simulations with Wilson fermions at small quark mass, as well as dynamical fermion algorithms. Moreover, the renormalization pattern of certain physical quantities is much simpler with tmQCD. Finally, the bare mass parameters of the theory may be tuned in such a way that improvement is ``automatic"; i.e. physical quantities such as masses and matrix elements are free of $\cO(a)$-effects, without having to intorduce Symanzik counter-terms. The price to pay is loss of symmetry: parity and time reversal are only recovered in the continuum limit. Moreover, flavour symmetry is also affected and this leads to loss of degeneracy between some hadrons. An important example concerns pions: the neutral pion mass differs from that of the two charged pions. Since this is due to discretization effects, degeneracy is restored in the continuum. A complete presentation of tmQCD is beyond our scope. The interested reader may consult existing review  articles~\cite{Sint:2007ug,Shindler:2007vp}. Here we will concentrate on the renormalization and improvement properties of the theory.

\section{Classical tmQCD}

It is instructive to begin by writing down the tmQCD classical Lagrangean density in the continuum. For simplicity we will consider QCD with two degenerate flavours (called up and down quarks). The Dirac spinor in flavour space is then given by the doublet $\bar \chi = (\bar u~~\bar d)$ and the fermionic Lagrangean density is defined as
\be
{\cL}_{\rm tm}  \,\, = \,\, \bar\chi \, \Big [ \Dslash \,\, + \,\, m_0 \,\, + \,\, i \mu_q \tau^3 \gamma_5 \Big ] \, \chi \,\,\, .
\label{eq:tmQCDclass}
\ee
Compared to the familiar standard QCD Lagrangean density, we now have an additional mass term which, being a pseudoscalar, is ``twisted" in chiral space. Apparently, this is not classical QCD, as the extra twisted mass term breaks parity and $SU(2)$ flavour symmetry. However, this is illusory. To see this, we redefine fermionic fields through chiral transformations (chiral rotations) in the third direction of flavour (isospin) space:
\begin{eqnarray}
\chi \,\, & \rightarrow & \chi^\prime \,\, = \,\, \exp \big [ i \, \dfrac{\alpha}{2} \, \gamma_5 \, \tau^3 \big ] \,\, \chi
\nonumber \\
\bar  \chi \,\, & \rightarrow & \bar \chi^\prime \,\, = \,\, \bar \chi \,\, \exp \big [ i \, \dfrac{\alpha}{2} \, \gamma_5 \, \tau^3 \big ] \,\,\, .
\label{eq:chirot}
\end{eqnarray}
We also redefine the two mass parameters through spurionic transformations
\begin{eqnarray}
m_0 \,\,  \rightarrow  \,\, m_0^\prime \,\,  & =  &\,\,  \cos(\alpha) \, m_0 \,\,  + \,\, \sin(\alpha) \,\, \mu_q
\nonumber \\
\mu_q \,\,  \rightarrow  \,\, \mu_q^\prime \,\,   & = & \,\,  \cos(\alpha) \, \mu_q \,\,  - \,\, \sin(\alpha) \,\, m_0 \,\,\, .
\label{eq:spur}
\end{eqnarray}
Under these transformations, the Lagrangean density transforms as follows:
\be
{\cL}_{\rm tm}  \,\, \rightarrow \,\, {\cL}_{\rm tm}^\prime \,\, = \,\,  \bar\chi^\prime \, \Big [ \Dslash \,\, + \,\, m_0^\prime \,\, + \,\, i \mu_q^\prime \tau^3 \gamma_5 \Big ] \, \chi^\prime  \,\,\, .
\label{eq:Lrot}
\ee
This means that the theory is form invariant under the chiral rotations~(\ref{eq:chirot}), provided they are accompanied by the mass spurionic transfrormations~(\ref{eq:spur}). In other words, these changes of field variables and mass definitions do not change the content of the theory. In fact tmQCD is a family of equivalent Lagrangean densities, connected by chiral and spurionic transformations. A member of this family of equivalent theories is standard QCD, obtained through a specific transformation angle $\alpha$, chosen so that $\mu_q \rightarrow \mu_q^\prime = 0$.

The above argument may be reformulated by defining an {\it invariant mass} $M_{\rm inv}$ and a {\it twist angle} $\omega$:
\be
M_{\rm inv} \,\, = \,\, \sqrt{m_0^2 \,\, + \,\, \mu_q^2} \qquad ; \qquad
\tan(\omega) \,\, = \,\, \dfrac{\mu_q}{m_0} \,\,\, .
\label{eq:omega}
\ee
The reason for this terminology will be clarified shortly. The Lagrangean density may be rewritten as
\be
{\cL}_{\rm tm}  \,\, = \,\, \bar\chi \, \Big [ \Dslash  \,\, + \,\, M_{\rm inv} \,\, \exp \big ( i \omega \tau^3 \gamma_5 \big ) \Big ] \, \chi \,\,\, ;
\label{eq;tmQCDclass2}
\ee
i.e. it is a function of a single mass parameter $M_{\rm inv}$ and a dimensionless angle $\omega$.
The transformation~(\ref{eq:Lrot}) of the Lagrangean density may also be written as
\begin{eqnarray}
{\cL}_{\rm tm}  \,\, \rightarrow \,\, {\cL}_{\rm tm}^\prime \,\, = \,\, \bar\chi^\prime \, \Big [ \Dslash  \,\, + \,\, M^\prime_{\rm inv} \,\, \exp \big ( i \omega^\prime \tau^3 \gamma_5 \big ) \Big ] \, \chi^\prime \,\,\, ,
\end{eqnarray}
with the invariant mass and twist angle defined as follows:
\be
M^\prime_{\rm inv} \,\, = \,\, \sqrt{[m_0^\prime]^2 \,\, + \,\, [\mu_q^\prime]^2} \qquad ; \qquad
\tan(\omega^\prime) \,\, = \,\, \dfrac{\mu_q^\prime}{m_0^\prime} \,\,\, .
\ee
Using eqs.~(\ref{eq:spur}), it is easy to show that $M^\prime_{\rm inv} = M_{\rm inv}$; i.e. the invariant mass is indeed invariant under spurionic transformations. In physical terms, this is the quark mass, which in tmQCD is seen to be a combination of both the standard mass $m_0$ and the twisted mass $\mu_q$. Morevoer, the new twist angle $\omega^\prime$ can easily be expressed in terms of the old twist angle $\omega$ and the rotation angle $\alpha$:
\be
\tan(\omega^\prime) \,\, = \,\, \tan(\omega \,\, - \,\, \alpha) \,\, .
\ee
Again, we see that tmQCD  may be regarded as a family of equivalent theories, parametrized by an invariant mass $M_{\rm inv}$ and a twist angle $\omega$.
Starting with a specific tmQCD theory (i.e. a given value of $\omega$, defined through $\tan(\omega) = \mu_q/m_0$), we obtain standard QCD by performing chiral rotations with a transformation angle $\alpha = \omega$, which brings us to $\omega^\prime = 0$ (i.e. $\mu_q^\prime = 0$).

It is worth noting two values of the twist angle which are of special interest to us. First, as already explained, for chiral rotations with $\alpha = \omega$, we obtain standard QCD (i.e. $\omega^\prime = 0$, $\mu_q^\prime = 0$ and $m_0^\prime = M_{\rm inv}$). In this case, the quark mass is carried entirely by the standard mass $m_0^\prime$. The second case of interest is when the chiral rotations are $\alpha = \omega - \pi/2$ (i.e. $\omega^\prime = \pi/2$, $\mu_q^\prime = M_{\rm inv}$ and $m_0^\prime = 0$). In this case, the quark mass is carried entirely by the twisted mass $\mu_q^\prime$. This is known as the {\it maximally twisted} theory.

Since QCD and tmQCD at the classical level are equivalent theories, they should share the same symmetries. This means, for example, that loss of parity due to the presence of the twisted mass term in the tmQCD Lagrangean density is only apparent. It is indeed straightforward to confirm that Lagrangean density~(\ref{eq:tmQCDclass}) is invariant under the parity transformations :
\begin{eqnarray}
x = (x^0,{\bf x}) \,\, & \rightarrow & \,\, x^\cP = (x^0,-{\bf x}) \,\,\, ,
\nonumber \\
U_0(x) \,\, & \rightarrow & \,\, U_0(x^\cP) \,\,\, ,
\nonumber \\
U_k(x) \,\, & \rightarrow & \,\, U_k(x^\cP - \hat k)^\dagger \,\,\,  ,
\nonumber \\
\chi(x) \,\, & \rightarrow & \,\, \gamma_0 \,\, \exp[ i \omega \gamma_5 \tau^3] \,\, \chi(x^\cP) \,\,\, ,
\nonumber \\
\bar \chi(x) \,\, & \rightarrow & \,\, \bar \chi(x^\cP) \,\, \exp[ i \omega \gamma_5 \tau^3]  \,\, \gamma_0 \,\,\, .
\label{eq:tmparity}
\end{eqnarray}
This means that in the classical tmQCD framework, parity is still a symmetry, albeit with modified transformations of the fermion fields (cf. eq~(\ref{eq:parity})). It should come as no surprise that the new parity transformations of $\chi$ and $\bar \chi$ involve the very same chiral rotations that connect tmQCD to standard QCD.

Analogous results may be obtained for the vector (isospin) symmetry. It is not really lost, as the twisted mass term of eq.~(\ref{eq:tmQCDclass})
may lead us to believe at first sight. Rather, it is simply transcribed as follows:
\begin{eqnarray}
\chi(x) \,\, & \rightarrow & \,\, \exp \big [ - i \dfrac{\omega}{2} \gamma_5 \tau^3 \big ] \exp \big [ i \dfrac{\theta^a}{2} \tau^a \big ] 
\exp \big [ i \dfrac{\omega}{2} \gamma_5 \tau^3 \big ] \,\, \chi(x)
\nonumber \\
\bar \chi(x) \,\, & \rightarrow & \,\, \bar \chi(x) \exp \big [  i \dfrac{\omega}{2} \gamma_5 \tau^3 \big ] \exp \big [- i \dfrac{\theta^a}{2} \tau^a \big ] 
\exp \big [ - i \dfrac{\omega}{2} \gamma_5 \tau^3 \big ] \,\,\, ,
\label{eq:tmvector}
\end{eqnarray}
with $\theta^a~~(a=1,2,3)$ the three rotation angles. We denote this symmetry group as $SU(2)_V^\omega$. Recall that from the beginning tmQCD has been formulated for a mass degenerate isospin doublet.

Axial transformations are transcribed in the following form:
\begin{eqnarray}
\chi(x) \,\, & \rightarrow & \,\, \exp \big [ - i \dfrac{\omega}{2} \gamma_5 \tau^3 \big ] \exp \big [ i \dfrac{\theta^a}{2} \tau^a \gamma_5 \big ] 
\exp \big [ i \dfrac{\omega}{2} \gamma_5 \tau^3 \big ] \,\, \chi(x)
\nonumber \\
\bar \chi(x) \,\, & \rightarrow & \,\, \bar \chi(x) \exp \big [  i \dfrac{\omega}{2} \gamma_5 \tau^3 \big ] \exp \big [ i \dfrac{\theta^a}{2} \tau^a \gamma_5 \big ] 
\exp \big [ - i \dfrac{\omega}{2} \gamma_5 \tau^3 \big ] \,\,\, .
\label{eq:tmaxial}
\end{eqnarray}
It is easy to verify that these transformations are a symmetry of the Lagrangean density~(\ref{eq;tmQCDclass2}), when $M_{\rm inv} =0$.

The field rotations, relating standard QCD and tmQCD also relate composite field operators, defined in the two theories. Since tmQCD (with twist angle $\omega$) and standard QCD (with twist angle $\omega^\prime=0$) are related by chiral rotations~(\ref{eq:chirot}) with $\alpha = \omega$,  the following relations hold between the vector and axial currents of the two theories:
\be
[ V^a_\mu ]_{\rm tmQCD} \,\, = \,\, \cos(\omega) \,\, [ V^a_\mu ]_{\rm QCD} \,\, - \epsilon^{3ab} \,\, \sin(\omega) \,\, [ A^b_\mu ]_{\rm QCD} 
\qquad a,b =1,2 \,\,\, ,
\nonumber
\ee
\be
[ A^a_\mu ]_{\rm tmQCD} \,\, = \,\, \cos(\omega) \,\, [ A^a_\mu ]_{\rm QCD} \,\, - \epsilon^{3ab} \,\, \sin(\omega) \,\,   [ V^b_\mu ]_{\rm QCD} 
\qquad a,b =1,2 \,\,\, ,
\nonumber
\ee
\be
[ V^3_\mu ]_{\rm tmQCD} \,\, = \,\, [ V^3_\mu ]_{\rm QCD} \,\,\, ,
\nonumber
\ee
\be
[ A^3_\mu ]_{\rm tmQCD}  \,\, = \,\, [ A^3_\mu ]_{\rm QCD} \,\,\, .
\label{eq:currents-class}
\ee
Recall that $a, b$ are flavour indices. Similarly, for the pseudoscalar density and the isospin singlet scalar density $S^0 \equiv \bar \chi \chi$ we obtain
\be
[ P^a ]_{\rm tmQCD} \,\, = \,\, [ P^a ]_{\rm QCD}  \qquad a,b =1,2 \,\,\, ,
\nonumber
\ee
\be
[ P^3 ]_{\rm tmQCD} \,\, = \,\, \cos(\omega) \,\, [ P^3 ]_{\rm QCD} \,\, - \dfrac{i}{2} \,\, \sin(\omega) \,\, [ S^0 ]_{\rm QCD}  \,\,\, ,
\nonumber
\ee
\be
[ S^0 ]_{\rm tmQCD} \,\, = \,\, \cos(\omega) \,\, [ S^0 ]_{\rm QCD} \,\, - 2i \,\, \,\, \sin(\omega) \,\,  [ P^3 ]_{\rm QCD}  \,\,\, .
\label{eq:S0-tmqCD}
\ee

Finally, the familiar PCVC and PCAC Ward identities, have the following form in the tmQCD formulation:
\begin{eqnarray}
\partial_\mu \, [ V_\mu^a ]_{\rm tmQCD} \,\,\, & = & \,\,\, - 2 \mu_q \epsilon^{3ab} \,\, [ P^b ]_{\rm tmQCD} 
\label{eq:tmQCD-PCVC}
\\
\partial_\mu \, [ A_\mu^a ]_{\rm tmQCD} \,\,\, & = & \,\,\, 2 m_0 \,\, [ P^a ]_{\rm tmQCD} \,\, + \,\, i \mu_q \delta^{3a} \,\, [ S^0 ]_{\rm tmQCD} 
\label{eq:tmQCD-PCAC}
\end{eqnarray}

\section{Lattice tmQCD}

So far we have achieved precious little! The relation between classical QCD and tmQCD is simply a change of fermionic field variables, accompanied by a redefinition of the masses. A change of variables cannot bring about new Physics. Thus, classical tmQCD is simply an intricate way of writing down QCD. It is upon passing over to the field theoretic formulation that the equivalence between the two theories is less trivial and this has important consequences, both in renormalization and improvement properties of many Physical quantities.

The proof that standard Wilson fermion QCD and tmQCD are equivalent Field Theories will only be shown schematically here. For a detailed demonstration, the reader is advised to consult the original reference \cite{Frezzotti:2000nk}. The starting point is that the equivalence in question proceeds through linear relations between renormalized Green functions of the two theories. As a first step, we consider QCD and tmQCD, regularized on the lattice with Ginsparg-Wilson fermions. The exact form of the lattice actions is not important in this discussion; for example in both lattice actions  the discretization of $\Dslash$ may be the Neuberger operator~\cite{exactchi:neub,Neuberger:1998wv,Neuberger:1997bg}. What is important is that chiral symmetry is not broken, owing to the Ginsparg-Wilson relation obeyed by the regularized Dirac operator. This implies that the chiral rotations~(\ref{eq:chirot}) map the lattice tmQCD bare action to the lattice QCD one, just like in the classical case. On the other hand, the variation of composite operators  transform members of a given multiplet among each other, plus some extra terms. It has been shown that the presence of these extra terms does not affect the main line of reasoning \cite{Frezzotti:2000nk}. Thus they will be ignored in this discussion. The fact that chiral rotations transform both the tmQCD action and the composite fields into those of standard QCD implies that bare Green functions of the two theories are related. For example, the insertion of the scalar operator $S^0$ in a Green function gives rise to the following relation 
\be
\langle \cdots S^0 \cdots \rangle_{\rm QCD}^{\rm GW} \,\,\, = \,\,\, \cos(\omega) \langle \cdots S^0 \cdots \rangle_{\rm tmQCD}^{\rm GW} \,\, + \,\, i \sin(\omega) \,\, \langle \cdots P^3 \cdots \rangle_{\rm tmQCD}^{\rm GW} \,\,\, ,
\ee
where the ellipses stand for other operators, which have the same form in QCD and tmQCD (e.g. $V^3_\mu$, $P^1$ etc.) and the superscript ${\rm GW}$ indicates bare Green functions, regularized in a Ginsparg-Wilson lattice framework. The subscripts QCD and tmQCD denote the mass term regularization (standard or twisted) in which these Green functions are defined.

The next step is to pass from bare to renormalized Green functions. Since chiral symmetry is preserved by the Ginsparg-Wilson regularization, members of the same chiral multiplet, such as $S^0$ and $P^3$, renormalize with the same renormalization factor $Z_{S^0} = Z_P$. In a mass independent renormalization scheme, this factor is the same both for standard QCD and tmQCD. This is easy to understand, since the two theories only differ in the way mass terms are introduced in the action. The bottom line is that when both sides of the last equation are multiplied by the same factor $Z_{S^0} = Z_P$, we obtain a linear mapping between renormalized quantities, calculated in two frameworks (QCD and tmQCD):
\be
\langle \cdots [ S^0 ]_{\rm R}  \cdots \rangle_{\rm QCD} \,\,\, = \,\,\, \cos(\omega) \langle \cdots [S^0]_{\rm R} \cdots \rangle_{\rm tmQCD} \,\, + \,\, i \sin(\omega) \,\, \langle \cdots [ P^3 ]_{\rm R} \cdots \rangle_{\rm tmQCD} \,\,\, .
\label{eq:tmQCD-QCD-Green}
\ee
Thus, based on a specific lattice regularization which respects chiral symmetry, we have shown the equivalence of QCD and tmQCD beyond the classical level. So far the line of reasoning has been similar to the one of the classical theories: a change of variables and mass definitions in the actions and the path integrals maps QCD lattice (bare) expectation values into those of tmQCD. Multiplicative renormalization subsequently ensures that the same relations hold for the continuum (renormalized) Green functions. Just like in the classical case, we do not expect anything interesting out of these changes of variables.

Things become non-trivial however, once we realize that the last expression, being true in the continuum, does not depend on the regularization details. This is a consequence of the principle of universality, which is a generally accepted  assumption. In particular, universality implies that eq.~(\ref{eq:tmQCD-QCD-Green}) is also true, up to discretization effects, for renormalized (continuum) Green functions, obtained from any other regularization; e.g. lattice Wilson fermions. Since chiral symmetry is broken by the Wilson term, the renormalization patterns of bare operators (which in the continuum belong to the same chiral multiplet) are now very different. This has important consequences. For example, as will be discussed below, the renormalization of $S^0$ with Wilson fermions is fairly complicated, and this renders the direct computation of the l.h.s. of eq.~(\ref{eq:tmQCD-QCD-Green}) rather cumbersome. On the other hand, the renormalization of $P^3$ with Wilson fermions is much simpler (be it in standard QCD or in tmQCD). One may tune the mass parameters of tmQCD so that the twist angle is $\omega = \pi/2$ and the first term in the r.h.s. of eq.~(\ref{eq:tmQCD-QCD-Green}) vanishes. Rather than computing the l.h.s. using standard lattice QCD with Wilson fermions, one may then compute the r.h.s. with the tmQCD lattice regularization 
of Wilson fermions, which has simpler renormalization properties. There are several interesting cases, besides the one sketched above, in which it is preferable, from the renormalization point of view, to work in a tmQCD framework.

Having explained the main idea, we now present Wilson fermion tmQCD in some detail. The standard Wilson action  (\ref{eq:sf})  is modified by the addition of the twisted mass term:
\be
S_F^{\rm tm} \,\, = \,\, a^4 \sum_x \bar \chi(x) \Big [ \dfrac{1}{2} \sum_\mu \big \{ \gamma_\mu ( \dcr^*_\mu + \dcr_\mu ) - a \dcr^*_\mu \dcr_\mu \}
+ m_0 + i \mu_q \tau^3 \gamma_5 \Big ] \chi(x) \,\,\, .
\label{eq:lattW-tmQCD}
\ee
The difference between the classical case and the present formulation is that the Wilson term breaks several symmetries. We saw, for example, that in the classical case parity is preserved in the modified form of eqs.~(\ref{eq:tmparity}). This symmetry is now broken by the Wilson term of the action. It does however survive, when combined with flavour exchange
\begin{eqnarray}
\chi(x) \,\, & \rightarrow & i \,\, \gamma_0 \,\, \tau^1 \,\, \chi(x^\cP) \,\,\, ,
\nonumber \\
\bar \chi(x) \,\, & \rightarrow & -i \,\, \bar \chi(x^\cP) \,\, \tau^1  \,\, \gamma_0 \,\,\, ,
\label{eq:tmparity-fl}
\end{eqnarray}
where the gauge-field transformations are the usual ones\footnote{The transformations of eq.(\ref{eq:tmparity-fl}), with $\tau^1$ replaced by $\tau^2$, are also a symmetry.}. Equivalently, instead of flavour exchange we can combine parity with a flip of the twisted mass sign,
\begin{eqnarray}
\chi(x) \,\, & \rightarrow & i \,\, \gamma_0 \,\, \chi(x^\cP) \,\,\, ,
\nonumber \\
\bar \chi(x) \,\, & \rightarrow & -i \,\, \bar \chi(x^\cP) \,\, \gamma_0 \,\,\, ,
\nonumber \\
\mu_q  \,\, & \rightarrow & - \,\, \mu_q \,\,\, ,
\label{eq:tmparity-fl-mass}
\end{eqnarray}
leaving the action $S_F^{\rm tm}$ invariant. Analogous considerations can be made for time-reversal. The consequence of this loss of symmetry is that matrix elements such as $\langle 0 \vert V_0^3(0) \vert \pi^0 \rangle$, which, due to parity, vanish in the standard Wilson theory, are non-zero in tmQCD. Of course, since we have argued in the previous section that tmQCD is a legitimate regularization of continuum QCD, such matrix elements will vanish in the continuum limit. They are lattice artefacts proportional to (some power of) the lattice spacing. Also note that often we study the asymptotic behaviour of correlation functions such as $\langle P(x) \,\, P^\dagger(0) \rangle$ at large time-separations,
by  introducing  a complete set of states between the operators. In the standard QCD case these would be pseudoscalar states; in tmQCD loss of parity implies that more states (e.g. scalars) are also allowed.

We now proceed to examine continuous symmetries.  The twisted vector symmetry of eq.~(\ref{eq:tmvector}) is hard-broken by the Wilson term. Axial symmetry  (cf. eq.~(\ref{eq:tmaxial})) is softly broken by the mass term $M_{\rm inv}$ as expected, but also by the Wilson term; the latter breaking is hard. Once the continuum and chiral limits are taken, these symmetries are expected to be restored. It is important to note
however, that the subgroup of transformations 
\begin{eqnarray}
\chi(x) \,\, & \rightarrow & \,\, \exp \big [ i \dfrac{\theta^3}{2} \tau^3 \big ] \,\, \chi(x)
\nonumber \\
\bar \chi(x) \,\, & \rightarrow & \,\, \bar \chi(x) \exp \big [- i \dfrac{\theta^3}{2} \tau^3 \big ] \,\,\, ,
\label{eq:tmvectorU1}
\end{eqnarray}
obtained from eq.~(\ref{eq:tmvector}) by setting $\theta^1 = \theta^2 = 0$,
remains a vector symmetry of the Wilson tmQCD action. We denote this group as $U(1)_V^3$. Thus, due to the Wilson term we have
the symmetry breaking pattern $SU(2)_V^\omega \rightarrow U(1)_V^3$. The reduced symmetry causes a lack of degeneracy between the neutral pion $\pi^0$ and the two degenerate charged pions $\pi^\pm$. The mass difference between charged and neutral pions is a discretization effect, proportional to (some power of) the lattice spacing, which vanishes in the continuum limit, where the full symmetry is restored.

Another interesting symmetry is derived from the axial transformations~(\ref{eq:tmaxial}) which, upon setting $\omega = \pi/2$ and $\theta^2 = \theta^3 = 0$, reduce to
\begin{eqnarray}
\chi(x) \,\, & \rightarrow & \,\, \exp \big [ i \dfrac{\theta^1}{2} \tau^2 \big ] \,\, \chi(x)
\nonumber \\
\bar \chi(x) \,\, & \rightarrow & \,\, \bar \chi(x) \exp \big [ - i \dfrac{\theta^1}{2} \tau^2 \big ] \,\,\, .
\label{eq:tmaxialU1}
\end{eqnarray}
This group of transformations is denoted by $U(1)_A^1$, though it has the appearance of the usual vector symmetry. Analogously, for $\theta^1 = \theta^3 = 0$ we obtain a similar group of transformations, called $U(1)_A^2$. Indeed these are axial symmetries, being symmetries of the massless Wilson action, softly broken by the twisted mass term. Thus, in the chiral limit the twisted theory is symmetric under $U(1)_V^3 \otimes U(1)_A^1 \otimes U(1)_A^2$, which amounts to an $SU(2)$ group, with one ``vector"and two ``axial" generators. It is not surprising that in the chiral limit, where the action~(\ref{eq:lattW-tmQCD}) reduces to the standard Wilson one, the full $SU(2)$ symmetry is recovered. The interpretation of this symmetries as vector (in the standard case) or vector/axial (in tmQCD), depends on how the soft mass term is introduced in the action~\cite{Sint:2007ug}.

\section{Renormalization with tmQCD}

So far we have seen that the introduction of a twisted mass term in the Wilson fermion regularization breaks some discrete and continuous symmetries. We have also argued that this lattice theory goes over to QCD as the lattice cutoff is removed, so any effects due to loss of symmetry by the twisted lattice action vanish in the continuum limit. Having seen the shortcomings of tmQCD, it is high time we discuss some of the advantages. First of all, it is straightforward to see that the fermion determinant corresponding to the tmQCD action~(\ref{eq:lattW-tmQCD}) is positive definite, as long as $\mu_q^2 \ne 0$. This is an important advantage for lattice simulations close to the chiral regime, but will not be further discussed here. A second advantage concerns renormalization. We have already argued, in rather general terms, that some operator renormalization is simpler in tmQCD than in standard Wilson fermion lattice theory. Here we will discuss some important examples in detail.

Before we do so, we should understand the renormalization properties of the two bare mass parameters of tmQCD, $m_0$ and $\mu_q$ (or equvalently, $M_{\rm inv}$ and $\omega$). The symmetries of the tmQCD action suggest that, for degenerate flavours, the standard quark mass $m_0$ renormalizes as in eqs.~(\ref{eq:mrzm}) and (\ref{eq:msub}), while the twisted mass $\mu_q$ renormalizes multiplicatively.
Moreover, the PCVC relation~(\ref{eq:tmQCD-PCVC}) is exact in lattice tmQCD, with the local vector current replaced by the point-split one (\ref{eq:vtilde}). This in turn implies that the product of twisted mass $\mu_q$ and pseudoscalar density $P^b$ (i.e. the r.h.s. of eq.~(\ref{eq:tmQCD-PCVC})) is renormalization group invariant. Thus $Z_P^{-1}$ renormalizes $\mu_q$:
\be
[ \mu_q ]_{\rm R} \,\,\, = \,\,\, Z_P^{-1} \,\, \mu_q \,\,\, .
\ee
The above statements are generalizations of the arguments of Section~\ref{ward}, concerning standard lattice QCD with Wilson quarks. The reader should have no problem in convincing himself of their validity for tmQCD.

Given these mass renormalizations, we define the twist angle through the following ratio of renormalized quantities:
\be
\tan(\omega) \,\,\, = \,\,\, \dfrac{[ \mu_q ]_{\rm R}}{m_{\rm R}} \,\,\, = \,\,\, \dfrac{Z_P^{-1} \mu_q}{Z_{S^0}^{-1} [ m_0 - \mcrit]} \,\,\, .
\label{eq:omegaW}
\ee
Note that with Ginsparg-Wilson fermions, chiral symmetry ensures that the two bare masses $m_0$ and $\mu_q$ renormalize multiplicatively with the same renormalization factor, and thus it makes no difference whether the twist angle is defined through a ratio of bare or renormalized masses. It is the loss of chiral symmetry with Wilson fermions that leads to the above redefinition of the twist angle w.r.t. eq.~(\ref{eq:omega}).

We stress that for mass independent renormalization, carried out in the chiral limit, there is no distinction between standard QCD and tmQCD, since the mass terms are absent from the action. Therefore, all renormalization constants computed in standard QCD simulations (e.g. $Z_m$, $Z_P$) may also be used for tmQCD quantities. For instance, the renormalization factors  of $m$ and $\mu_q$ in the last equation are the very same $Z_{S^0}$ and $Z_P$ that renormalize the densities $S^0$ and $P$ in the standard Wilson theory\footnote{The same is true of the power subtraction $\mcrit$, being the counter-term of the standard bare mass $m_0$. However, non-perturbatove determinations of $\mcrit$, computed in a tmQCD framework, require extra care \cite{Frezzotti:2005gi}.}.

With the above observations in mind, we can immediately generalize the tree level expressions (\ref{eq:currents-class}) for the full theory. In particular, the continuum axial current $A^a$ (for $a=1,2$) is expressed as linear combinations of renormalized tmQCD ones:
\be
[ A^a_\mu ]_{\rm cont} \,\, = \,\, Z_A \,\, [ A^a_\mu ]_{\rm QCD} \,\, = \,\,
\cos(\omega) \,\, Z_A \,\, [ A^a_\mu ]_{\rm tmQCD} \,\, + \epsilon^{3ab} \,\, \sin(\omega) \,\, [ \tilde V^b_\mu ]_{\rm tmQCD}  \,\,\, .
\label{eq:Acont}
\ee
The subscripts QCD and tmQCD indicate bare quantities in the respective lattice regularizations. The first equation in the above expression links the continuum axial current to the lattice one in the standard formulation, as detailed in Section~\ref{sec:AWI}. The second equation does the same job in the tmQCD framework. As previously explained, the normalization factor $Z_A$ is the same in both regularizations. The last
term of the second equation contains the exactly conserved point-split vector current; thus the absence of a normalization factor. We may of course use, instead of $[  \tilde V^a_\mu ]_{\rm tmQCD}$, the local current $[  V^a_\mu ]_{\rm tmQCD}$, multiplied by $Z_V$. Similarly for the pseudoscalar density $P^a$ (for $a=1,2$) we have
\be
[ P^a ]_{\rm cont} \,\, = \,\, Z_P \,\, [ P^a ]_{\rm QCD} \,\, = \,\, Z_P \,\, [ P^a ]_{\rm tmQCD}  \,\,\, .
\label{eq:Pcont}
\ee
Equalities (\ref{eq:Acont}) and (\ref{eq:Pcont}) are valid up to discretization errors. 

We are finally ready to see a first advantage of maximally twisted tmQCD. To ensure maximal twist (i.e. $\omega  = \pi/2$), the standard bare mass $m_0$ of the tmQCD regularization is tuned to its critical value $\mcrit$; cf. eq.~(\ref{eq:omegaW}). Note that knowledge the ratio $Z_P/Z_{S^0}$ is not required for this tuning. Then, according to eq.~(\ref{eq:Acont}) the axial current in the continuum limit is the vector current in tmQCD, which requires no normalization. Using eqs.~(\ref{eq:Acont}) and (\ref{eq:Pcont}), we see that the standard PCAC Ward identity in the continuum corresponds to the PCVC relation 
\be
\sum_{\vec x} \,\, \nabla_x^\mu \,\, \langle \tilde V_\mu^1(x) \,\, P^2(0) \rangle_{\rm tmQCD}  \,\,\, = \,\,\, - \,\, 2 \mu_q \,\, \sum_{\vec x} \,\, \langle P^1(x) \,\, P^2(0) \rangle_{\rm tmQCD} \,\,\, .
\ee
This is an exact Ward identity in lattice tmQCD. Inserting, in standard fashion, a complete set of states between operators in the above expectations values, we obtain, in the limit of large time separations
\be
m_\pi \,\, \langle 0 \vert \,\, [ \tilde V_0(0) ]_{\rm tmQCD} \,\, \vert \pi \rangle  \,\,\, = \,\,\,  2 \mu_q \,\, \langle 0 \vert \,\, [ P(0) ]_{\rm tmQCD} \,\,\vert \pi \rangle \,\,\, .
\ee
Flavour indices have been dropped for simplicity. Combining these expressions and eq.~(\ref{eq:Acont}) we get, for the pion decay constant $f_\pi$ 
\begin{eqnarray}
f_\pi \,\,\, &\equiv& \,\,\, \dfrac{1}{m_\pi} \,\, \langle 0 \vert \,\, [  A_0(0) ]_{\rm cont} \,\, \vert \pi \rangle
\nonumber \\
&=& \,\,\, \lim_{a\rightarrow 0} \,\, \dfrac{1}{m_\pi} \,\, \langle 0 \vert \,\, [ \tilde V_0(0) ]_{\rm tmQCD} \,\, \vert \pi \rangle
\nonumber \\
&=& \,\,\, \lim_{a\rightarrow 0} \,\, \dfrac{2\mu_q}{m_\pi^2} \,\, \langle 0 \vert \,\, [  P(0) ]_{\rm tmQCD} \,\, \vert \pi \rangle \,\,\, .
\end{eqnarray}
The last expression provides, within the tmQCD framework, a definition of $f_\pi$ which is free of any normalization or renormalization factors. Recall that in the standard Wilson fermion case, the computation of $f_\pi$ from the matrix element $\langle 0 \vert A_0 \vert \pi\rangle$ requires knowledge of the axial current normalization $Z_A$. Thus, the systematic error due to $Z_A$ has been eliminated in tmQCD.

Another example of the advantages of tmQCD is provided by the renormalization of the chiral condensate operator $S^0$. In the standard Wilson quark case,  the symmetries of the theory imply the renormalization pattern (cf.~eq.~(\ref{eq:hadWIcond3})):
\be
[ \, S^0 \, ]_{\rm R} \,\,\, = \,\,\, Z_{S^0} \,\, \Big [ [S^0]_{\rm QCD} + \dfrac{c_S(g_0^2)}{a^3} \Big ] \,\, + \,\, \cdots \,\,\, ,
\ee
where the ellipsis stands for less vigorous power divergences (quadratic and linear) which, for dimensional reasons, are proportional to powers of the quark mass $m$ (i.e. they are absent in the chiral limit).  In tmQCD with maximal twist ($\omega = \pi/2$), this operator is mapped on $P^3$ (cf. eq.~(\ref{eq:S0-tmqCD})). The symmetries of lattice tmQCD with Wilson quarks imply the renormalization pattern
\be
[ \, S^0 \, ]_{\rm R} \,\,\, = \,\,\, Z_P \,\, \Big [ [P^3]_{\rm tmQCD} + \dfrac{\mu_q c_P(g_0^2)}{a^2} \Big ] \,\, + \,\, \cdots \,\,\, ,
\label{eq:tmQCDS0}
\ee
where the ellipsis again stands for less vigorous linear power divergences, which depend on the quark mass $\mu_q$. Thus, in the chiral limit the chiral condensate, regularized in tmQCD with maximal twist, is multiplicatively renormalizable. It is well known that power divergences are hard to control, whether calculated perturbatively or computed non-perturbatively (i.e. numerically). Suppose that the dimensionless coefficients $c_S$ and $c_P$ are calculated perturbatively. Such a calculation misses out dimensionless non-perturbative contributions, which are $\cO(a \Lambda_{\rm QCD})$. Since these contributions are to be multiplied by the power divergences $1/a^3$ and $1/a^2$ of the power subtractions, the perturbative calculation of  $c_S$ and $c_P$ is inaccurate by terms which diverge in the continuum limit! On the other hand, if the coefficients $c_S$ and $c_P$ are calculated non-perturbatively, they are bound to be subject to discretization effects. These are expected to be $\cO(a)$\footnote{If Symanzik-improvement is used, these discretization effects are $\cO(a^2)$.}. 
So again the error of the power subtractions diverges in the continuum limit. Nevertheless, it is remarkable that in tmQCD the chiral condensate suffers from less severe power divergences. If somehow one could work in the chiral limit of maximally tmQCD, determined non-perturbatively at very high accuracy (i.e. $\mcrit \sim \cO(a^3)$), then the power subtraction of eq.~({\ref{eq:tmQCDS0}) could be ignored altogether and the chiral condensate computation would be greatly simplified.

These considerations have been extended to the calculation of matrix elements of dimension-6 four fermion operators, which determine the non-perturbative contributions in say, neutral Kaon oscillations and Kaon non-leptonic decays into two pions. In lattice QCD with standard Wilson quarks, the renormalization pattern of these operators is very complicated, due to the introduction of counter-terms which would have been absent, if chiral symmetry were preserved. In tmQCD it is possible to map these operators into their partners of opposite parity, which have a much simpler renormalization pattern, in spite of the loss of chiral symmetry. Explaining this in detail is beyond the scope of the present lectures. The interested reader is advised to consult the relevant literature, where these results are presented in considerable detail~\cite{Frezzotti:2000nk,Dimopoulos:2006dm,tmqcd:DIrule,Frezzotti:2004wz}.

\section{``Automatic" improvement in maximally twisted QCD}

Besides simplified renormalization patterns, tmQCD in its maximally twisted version has another nice property, commonly known as ``automatic improvement". By this we mean that physical quantities (e.g. masses, matrix elements etc.) have discretization effects which are sub-leading; i.e. $\cO(a^2)$. Improvement of these quantities comes about by carefully tuning the twist angle $\omega = \pi/2$, without the need of introducing Symanzik
counter-terms\footnote{In the realistic case of QCD with non-degenerate quark masses, the non-pertubative determination of Symanzik counter-terms is a non-trivial and intricate task~\cite{Bhattacharya:2005rb}.}. Following the original proof of this statement~\cite{FrezzoRoss1}, several simplified versions have appeared in the literature. In order to understand the line of reasoning of this section, the reader is strongly advised to acquaint himself with the relevant Sections of Chapter~2 of the present volume~\cite{Weisz:2010nr}.

The study of discretization effects of lattice observables is based on the Symanzik expansion, in the light of which the lattice  action, close to the continuum, is described in terms of an effective theory
\be
S^{\rm tm}_F \,\,\, = \,\,\, \int d^4y \,\, \cL_0 \,\, + \,\, a \,\, \int d^4y \,\, \cL_1 \,\,\, + \,\,\, \cdots \,\,\, .
\label{eq:Sym-act}
\ee
As the notation in the above expression implies, our starting point is the tmQCD Wilson fermion action~(\ref{eq:lattW-tmQCD}) at maximal twist (i.e. with $m_0 = \mcrit$). Thus the Symanzik counter-terms reflect the symmetries of this action:
\begin{eqnarray}
\cL_0 \,\,\, &=& \,\,\, \bar \chi \,\, \Big [ \,\, \Dslash \,\, + \,\, i \,\, [\mu_q]_{\rm R} \,\, \gamma_5 \,\, \tau^3 \,\, \Big ] \,\, \chi
\label{eq:cL0} \\
\cL_1 \,\,\, &=& \,\,\, i \,\, b_{\rm sw} \,\, \big [ \,\, \bar \chi \,\, \sigma \cdot F \chi \,\, \big ] \,\,\, + \,\,\, b_\mu \,\, [\mu_q]_{\rm R}^2 \,\, \big [ \,\, \bar \chi \chi \,\, \big ] \,\,\, .
\label{eq:cL1}
\end{eqnarray}
Besides the above expansion for the action, Symanzik-improvement implies that $d$-dimensional composite lattice fileds $\Phi_{\rm latt}$ are also subject to an effective theory description of the form
\be
\Phi_{\rm latt} \,\,\, = \,\,\, \Phi_0 \,\, + \,\, a \,\, \Phi_1 \,\,\, ,
\label{eq:Sym-op}
\ee
where $\Phi_0, \Phi_1$ are continuum composite fields of dimensions $d$ and $d+1$ respectively. Actually, $\Phi_{\rm latt}$ is a lattice transcription of $\Phi_0$. Combining eqs.~(\ref{eq:Sym-act}) and (\ref{eq:Sym-op}) gives, the lowest-order Symanzik expansion  for the vacuum expectation value of $\Phi_{\rm latt}$
\be
\langle \,\, \Phi_{\rm latt} \rangle_{\rm tm} \,\,\, = \,\,\, \langle \,\, \Phi_0 \,\, \rangle_0 \,\, + \,\, a \,\, \langle \,\, \Phi_1 \,\, \rangle_0 \,\,\, - \,\,\,
a \,\, \int d^4y  \,\, \langle \,\, \Phi_0 \,\, \cL_1 \,\, \rangle_0 \,\, + \,\, \cO(a^2) \,\,\, .
\label{eq:Sym-opvev}
\ee
The l.h.s. is the expectation value of the lattice operator in maximally twisted lattice QCD. The expectation values of the r.h.s. are continuum quantities; their subscript $\langle \cdots \rangle_0$ indicates that thet are defined in terms of the continuum tree level action $S_0 = \int d^4y \cL_0$. Automatic improvement means that the last two terms on the r.h.s. vanish identically, due to some symmetries of the maximally twisted continuum theory.
\begin{center}
\begin{table}
\begin{tabular}{l | ccc}
\hline
$\cS_F^{\rm tm}$ & $\cR_5^1$ &  $\cD$ & $[ \mu_q \rightarrow - \mu_q ]$ \\ \\
\hline
$\sum_x \sum_\mu \bar \chi  \gamma_\mu ( \dcr^*_\mu + \dcr_\mu ) \chi$ & + & + & + \\ \\
$\sum_x \sum_\mu \bar \chi \dcr^*_\mu \dcr_\mu \chi$ & $-$ & $-$ & + \\ \\
$ \mcrit \,\,  \sum_x \bar \chi \chi$ & $-$ & $-$ & + \\ \\
$i \,\, \sum_x \bar \chi \,\, \mu_q \tau^3 \gamma_5 \,\, \chi$ & + & $-$ & $-$ \\ \\
\hline
\end{tabular}
\caption{``Parity" of the various terms of the fermion tmQCD action under discrete symmetries.}
\label{tab:discr-transf} 
\end{table}
\end{center}

To prove this, we first explain what the relevant lattice symmetries are. We define the {\it discrete chiral transformations in the first isospin direction} $\cR^1_5$ as:
\begin{eqnarray}
\chi \,\,\, &\rightarrow& \,\,\, i \,\, \gamma_5 \,\, \tau^1 \,\, \chi \,\,\, ,
\nonumber \\
\bar \chi \,\,\, &\rightarrow& \,\,\, i \,\, \bar  \chi \,\, \gamma_5 \,\, \tau^1 \,\,\, .
\label{eq:R5parity}
\end{eqnarray}
Next we define the {\it operator dimensionality transformations} $\cD$:
\begin{eqnarray}
\chi(x) \,\,\, &\rightarrow& \,\,\,  \exp [ \, \dfrac{3i\pi}{2} \, ]  \,\, \chi(-x) \,\,\, ,
\nonumber \\
\bar \chi(x) \,\,\, &\rightarrow& \,\,\, \bar  \chi(-x) \,\, \exp [ \, \dfrac{3i\pi}{2} \, ]  
\nonumber \\  
U_\mu(x) \,\,\, &\rightarrow& \,\,\, U_\mu^\dagger (-x - a \hat \mu) \,\,\, .
\end{eqnarray}
The gauge lattice action is invariant under the above transformations of the link fields $U_\mu(x)$. Composite operators of even (odd) dimension $d$ are even (odd) under these transformations, up to the flip of sign of the space-time argument $x$. Finally we define the {\it twisted mass sign flip transformation}
\be
\mu_q \,\,\, \rightarrow \,\,\, - \,\, \mu_q
\ee
Each term of the lattice tmQCD action is even or odd under these transformations; i.e. it has definitive ``parity" with respect to these symmetries. In Table~\ref{tab:discr-transf} we list these properties explicitly. It is then clear that the tmQCD fermion action is invariant under the combined transformations $\cR_5^1 \otimes \cD \otimes [ \mu_q \rightarrow - \mu_q ]$, which is therefore a symmetry of the lattice theory. On the other hand, the continuum tmQCD action $\cS_0 = \int d^4y \cL_0$ has positive  $\cR_5^1$-parity (cf. eq.~(\ref{eq:cL0}) and the first and last rows of Table~\ref{tab:discr-transf}). Thus $\cR_5^1$ is a symmetry of the continuum theory.

Let us now examine each term of eq.~(\ref{eq:Sym-opvev}) in the light of these symmetries. It is important to keep in mind that the three vacuum expectation values on the r.h.s. are continuum quantities, determined by the continuum tmQCD action $\cS_0  =  \int d^4y \cL_0$, which is symmetric under $\cR_5^1$.
\begin{enumerate}
\item $\Phi_{\rm latt} $: Without loss of generality, we assume this operator to be even under $\cR_5^1$ and to have even dimension $d$. Then its vacuum expectation value $\langle \,\, \Phi_{\rm latt} \rangle_{\rm tm}$ is invariant under the combined transformations $\cR_5^1 \otimes \cD \otimes [ \mu_q \rightarrow - \mu_q ]$, which is a symmetry of the lattice theory. This implies that the terms on the r.h.s.  of eq.~(\ref{eq:Sym-opvev}) are also invariant under the same combined transformations. 
\item $\Phi_0$: This operator has positive $\cR_5^1$-parity and even dimension $d$, being the continuum counterpart of the lattice operator $\Phi_{\rm latt} $.
\item $\Phi_1$: This operator has dimension $d+1$, so it is odd under $\cD$. Since its expectation value  $\langle \,\, \Phi_1 \rangle_0$ is even under $\cR_5^1 \otimes \cD \otimes [ \mu_q \rightarrow - \mu_q ]$, the operator $\Phi_1$ must be odd under $\cR_5^1$. This implies that $\langle \,\, \Phi_1 \,\, \rangle_0$ vanishes, being the expectation value of an $\cR_5^1$-odd observable $\Phi_1$, weighted by an $\cR_5^1$-even action $\cS_0$.
\item $\cL_1$: This is the $\cO(a)$-counter-term of the continuum Lagrangean $\cL_0$. It is a continuum dimension-5 operator. From eqs.~(\ref{eq:cL1}) and (\ref{eq:R5parity}) we deduce that $\cL_1$ is odd under $\cR_5^1$ transformations.
\item $\Phi_0 \,\, \cL_1$: This is a product of two composite operators. The operator $\Phi_0$ is $\cR_5^1$-even while the operator $\cL_1$  is $\cR_5^1$-odd. It follows that the $\langle \,\, \Phi_0 \,\, \,\, \cL_1 \,\, \rangle_0$ vanishes, being the expectation value of an $\cR_5^1$-odd observable $\Phi_1$, weighted by an $\cR_5^1$-even action $\cS_0$.
\end{enumerate}
We have shown that the lattice expectation value $\langle \,\, \Phi_{\rm latt} \rangle_{\rm tm}$ of eq.~(\ref{eq:Sym-opvev}) is equal to the leading order term of the Symanzik expansion $\langle \,\, \Phi_0 \rangle_0$ up to $\cO(a^2)$ counter-terms. The $\cO(a)$ counter-terms of the expansion are identically zero. Therefore, without introducing Symanzik counter-terms, we find that expectation values of lattice operators in the maximally twisted theory are $\cO(a)$-improved. Note that it is the vacuum expectation values, rather than the maximally twisted action itself, which are automatically improved.

The previous arguments have overlooked the following subtlety. The proof rests on the vanishing of the continuum vacuum expectation values $\langle \,\, \Phi_1 \,\, \rangle_0$ and $\langle \,\, \Phi_0 \,\, \,\, \cL_1 \,\, \rangle_0$, due to their breaking of the discrete ``chiral" symmetry $\cR_5^1$, which is a  symmetry of the continuum tmQCD action $\cS_0$. But do these ``chiral condensates" vanish in a theory which, being a regularization of QCD, is characterized by spontaneous symmetry breaking? The answer is affirmative, because the term generating spontaneous symmetry breaking is the twisted mass term, while possible $\cO(a)$ counter-tems are generated by the ``chirally orthogonal" Wilson term, which in tmQCD breaks vector symmetry. In other words, the $\cR_5^1$-symmetry on which the automatic improvement is based is not really a chiral symmetry. It is a discrete subgroup of the flavour symmetry $SU(2)_V$ of eq.~(\ref{eq:tmvector}), for $\omega = \pi/2$ (i.e. maximal twist) and transformation angles $(\alpha_1,\alpha_2,\alpha_3) = (0,\pi,0)$. Flavour symmetry is not spontaneously broken in the continuum, and thus the quantities $\langle \,\, \Phi_1 \,\, \rangle_0$ and $\langle \,\, \Phi_0 \,\, \,\, \cL_1 \,\, \rangle_0$ do indeed vanish by symmetry arguments. The situation is not altered by simulations, provided we take extra care that the continuum limit is approached before the chiral limit. In this way the chiral phase of the vacuum is driven by the mass term and not by the Wilson term. This is ensured by imposing the condition $\mu_q >> a \Lambda_{\rm QCD}^2$.

We have shown that tmQCD is a variant of Wilson fermion regularization which, at the price of sacrificing some symmetries, enjoys simpler renormalization properties of composite operators and automatic improvement. The advantages on renormalization patterns are typically obtained by tuning the twist angle to a value which maps the original operator to its counterpart of opposite parity (e.g. axial current to vector current, scalar density to pseudoscalar density etc.). In some cases, this twist angle turns out to be different than $\pi/2$, the maximal twist value which ensures automatic improvement. Thus the renromalization and improvement advantages are not always satisfied simultaneously. Such difficulties arise with the four-fermion operators related to neutral Kaon oscillations and $K \rightarrow \pi\pi$ decays~\cite{tmqcd:DIrule}. A way out of these problems has been proposed~\cite{Frezzotti:2004wz}, based on mixed actions with maximally twisted valence quarks of the so-called Osterwalder-Seiler variety. Once again, the gains are accompanied by certain disadvantages, but these are issues beyond the scope of these lectures.

\chapter{Conclusions}

The present lectures focus on three specific topics: (i) lattice Ward identities; (ii) the non perturbative RI/MOM renormalization scheme; (iii)
renormalization and improvement of twisted mass QCD. From the study of these topics, significant theoretical and practical advantages may be obtained for the renormalization and improvement of lattice composite operators. Throughout we have used the Wilson regularization of the fermionic action. However, at least as far as the first two topics are concerned, the issues discussed go beyond a specific regularization.

Once chiral symmetry is lost by the regularization, Ward identities ensure its recovery in the renormalized theory, as the UV cutoff is removed. In the Wilson fermion case chiral symmetry is broken by an irrelevant operator. The consequences are immediately seen in the power subtraction of the quark mass, the finite normalization of the axial current, the different renormalization constants of operators belonging to the same chiral multiplet etc. In other regularizations, such as domain wall fermions, the loss of chirality only appears at the ``practical" level (i.e. when the fifth dimension of the domain wall is not infinite in simulations). The resulting loss of chiral symmetry is less manifest in such cases, but its effects are analogous to those of Wilson fermions (albeit quantitatively less important). 

The RI/MOM scheme has been presented in considerable detail. Emphasis has been laid on the theoretical aspects of the scheme, the conceptual problems and their resolution, without presenting any results, which may be easily dug out in the literature. Again, the language used was that of Wilson fermions, but it is quite clear that non-perturbative operator renormalization is nowadays a requisite for all lattice regularization schemes.

Finally, the variant of Wilson fermions known as twisted mass QCD has been discussed. Only selected aspects of this formulation have been exposed. The fact that the tmQCD fermion determinant is positive definite at non-zero twisted mass $\mu_q$ has only been touched upon, although it is of great importance to simulations close to the chiral regime. Moreover, we have limited our discussion to tmQCD with two light degenerate quarks, although it is possible to generalize the theory so as to include heavy flavours \cite{Frezzotti:2003xj}.
This subject has many faces, but the most important properties of  tmQCD are arguably those connected to renormalization and improvement. Significant simplifications of the standard Wilson formulation are obtained, which are however accompanied by loss of symmetry, recoverable only in the continuum limit. This is hardly surprising, being yet another example of the well known fact that most significant theoretical gains in Quantum Field Theory come at a price.

\chapter {Appendix: Spurionic chiral symmetry}
\label{app:spurionic}

Fermion fields are decomposed into left- and right-components $\psi = \psi_L + \psi_R$, with
\bea
\psi_L \,\,\, = \,\,\, \dfrac{1 - \gamma_5}{2} \,\, \psi
&\qquad , \qquad&
\bar \psi_L \,\,\, = \,\,\, \bar \psi \,\, \dfrac{1 + \gamma_5}{2}  \,\,\, ,
\nonumber \\
\psi_R \,\,\, = \,\,\, \dfrac{1 + \gamma_5}{2} \,\, \psi
&\qquad , \qquad&
\bar \psi_R \,\,\, = \,\,\, \bar \psi \,\, \dfrac{1 - \gamma_5}{2}  \,\,\, .
\eea
They transform under the chiral group $SU(N_F)_L \otimes SU(N_F)_R$ as follows:
\bea
\psi_L \rightarrow \psi_L^\prime \,\, = \,\, U_L \,\, \psi_L
&\qquad , \qquad&
\bar \psi_L \rightarrow \bar \psi_L^\prime \,\, = \,\, \bar \psi_L \,\, U_L^\dagger \,\,\, ,
\nonumber \\
\psi_R \rightarrow \psi_R^\prime \,\, = \,\, U_R \,\, \psi_R
&\qquad , \qquad&
\bar \psi_R \rightarrow \bar \psi_R^\prime \,\, = \,\, \bar \psi_R \,\, U_R^\dagger \,\,\, ,
\label{eq:chritransf}
\eea
with
\be
U_{L,R} \,\,\, = \,\,\, \exp \Big[ i \alpha^a_{L,R} \, \dfrac{\lambda^a}{2} \Big ] \qquad a = 1, \cdots , N_F^2-1
\ee
The vector transformations~(\ref{eq:glob-vtransf}) are recovered for $U_L = U_R$ (i.e. $\alpha^a_L = \alpha^a_R$), while the axial transformations~(\ref{eq:glob-atransf}) are recovered for $U_L = U_R^\dagger$ (i.e. $\alpha^a_L = - \alpha^a_R$).

The kinetic term of the lattice action~(\ref{eq:sf}) is invariant under these transformations, while the mass- and Wilson-terms break chiral symmetry. However, the symmetry may be enforced on the mass term, once the mass matrix $M_0$ is generalized, so that it is neither real nor diagonal (the physically relevant case corresponds to $M_0 = M_0^\dagger$ and diagonal). In order for the action to remain Hermitean, the mass term must be modified as follows: 
\be
\bar \psi_L M_0 \psi_R \,\, + \,\, \bar \psi_R M_0 \psi_L \,\,\, \rightarrow \,\,\,
\bar \psi_L M_0 \psi_R \,\, + \,\, \bar \psi_R M_0^\dagger \psi_L  \,\,\, .
\ee
We also impose the following chiral transformation of the mass matrix:
\be
M_0 \,\,\, \rightarrow \,\,\, M_0^\prime \,\, = \,\, U_L \,\, M_0 \,\, U_R^\dagger \,\,\, .
\label{eq:spchritransf}
\ee
Transformations of the mass parameters are called spurionic. Under the transformations~(\ref{eq:chritransf}) and (\ref{eq:spchritransf}) the modified mass term of the action is chirally invariant. Note that the Wilson term in the action remains a (spurionic) chiral symmetry-breaking term.

The generalized, non-diagonal bare mass matrix $M_0$ may be decomposed into non-singlet and singlet quark mass contributions (this terminology will become clear shortly):
\be
M_0 \,\,\, = \,\,\, \sum_{a=1}^{N_F^2-1} \, \widetilde m^a  \lambda^a \,\, + \,\, \mav \, I \,\,\, .
\label{eq:mdecomp-gen}
\ee
Using the trace property of eq.~(\ref{eq:ds}) for the group generators $\lambda^a$ we see that
\bea
\widetilde m^a \,\,\, = \,\,\, \dfrac{1}{2} \, \Tr [ M_0 \, \lambda^a \, ]  \qquad  , \qquad
\mav  \,\,\, = \,\,\, \dfrac{1}{N_F} \, \Tr [ M_0 ]  \,\,\, .
\label{eq:mass-comp}
\eea
Clearly $\mav$ is the average of the diagonal elements of $M_0$. It is straightforward to show that under infinitesimal chiral transformations these mass components transform as follows:
\bea
\hspace {-0.3 cm}
\widetilde m^a \,\,&\rightarrow& \,\, \big [ \widetilde m^a \big ]^\prime \,\, = \,\, \widetilde m^a \,\, - \,\, f^{abc} \,\, \dfrac{\alpha_L^b + \alpha_R^b}{2} \widetilde m^c \,\, + \,\, i d^{abc} \,\, \dfrac{\alpha_L^b - \alpha_R^b}{2}   \widetilde m^c \,\, + \,\, i \,\, \dfrac{\alpha_L^a - \alpha_R^a}{2} \mav  \,\,\, ,
\nonumber \\
\mav  &\rightarrow& \,\, \big [ \mav \big ]^\prime \,\, = \,\, \mav \,\,\, + \,\,\, \dfrac{i}{N_F} \,\, (\alpha_L^a - \alpha_R^a) \,\, \widetilde m^a \,\,\, .
\label{eq:mchirtran}
\eea
In the special case of vector transformations ($\alpha^a_L = \alpha^a_R = \alpha^a$), the above simplify to:
\bea
\widetilde m^a \,\,\, &\rightarrow& \,\,\, \big [ \widetilde m^a \big ]^\prime \,\,\, = \,\,\, \widetilde m^a \,\,\, - \,\,\, f^{abc} \,\, \alpha^b  \widetilde m^c \,\,\, ;
\nonumber \\
\mav  &\rightarrow& \,\, \big [ \mav \big ]^\prime \,\, = \,\, \mav \,\,\, .
\label{eq:mnsvtran}
\eea
Thus, the non-singlet mass components $\widetilde m^a$ are indeed $SU(N_F)_V$ multiplets in the adjoint representation (triplets for $N_F=2$, octets for $N_F=3$ etc.). On the other hand, $\mav$ is a flavour singlet. Under infinitesimal axial transformations ($\alpha^a_L = -\alpha^a_R = \alpha^a$),  eqs.~(\ref{eq:mchirtran}) reduce to
\bea
\widetilde m^a \,\,\, &\rightarrow& \,\,\, \big [ \widetilde m^a \big ]^\prime \,\,\, = \,\,\, \widetilde m^a \,\,\, + \,\,\, d^{abc} \,\, \alpha^b  \widetilde m^c \,\,\, + \,\,\, i \alpha^a \,\, \mav  \,\,\, ,
\nonumber \\
\mav \,\,\, &\rightarrow& \,\,\, \big [ \mav \big ]^\prime \,\,\, = \,\,\, \mav \,\,\, + \,\,\, \dfrac{2i}{N_F} \,\, \alpha^a \,\, \widetilde m^a \,\,\, .
\label{eq:mnsatran}
\eea
Therefore, flavour non-singlets $\widetilde m^a$ and the singlet $\mav$ transform into each other under axial transformations; i.e. they belong to the same chiral multiplet.

\chapter {Appendix: Lattice discrete symmetries}
\label{app:disc-symm}

In this Appendix we gather the discrete symmetry transformations of the Wilson action~(\ref{eq:sg}), (\ref{eq:sf})~\cite{Bernard:1989nb}. Although of rather limited value to the present set of lecture notes, they are generally useful for the classification of physical states, the mixing of operators under renormalization etc. We start with parity, which is defined as follows:
\begin{eqnarray}
x = (x^0,{\bf x}) \,\, & \rightarrow & \,\, x^\cP = (x^0,-{\bf x}) \,\,\, ,
\nonumber \\
U_0(x) \,\, & \rightarrow & \,\, U_0(x^\cP) \,\,\, ,
\nonumber \\
U_k(x) \,\, & \rightarrow & \,\, U_k(x^\cP - \hat k)^\dagger \,\,\,  ,
\nonumber \\
\psi(x) \,\, & \rightarrow & \,\, \gamma_0 \,\, \psi(x^\cP) \,\,\, ,
\nonumber \\
\bar \psi(x) \,\, & \rightarrow & \,\, \bar \psi(x^\cP) \,\, \gamma_0 \,\,\, .
\label{eq:parity}
\end{eqnarray}
Time reversal is given by
\begin{eqnarray}
x = (x^0,{\bf x}) \,\, & \rightarrow & \,\, x^\cT = (-x^0,{\bf x}) \,\,\, ,
\nonumber \\
U_0(x) \,\, & \rightarrow & \,\, U_0(x^\cT - \hat 0)^\dagger \,\,\, ,
\nonumber \\
U_k(x) \,\, & \rightarrow & \,\, U_k(x^\cT) \,\,\,  ,
\nonumber \\
\psi(x) \,\, & \rightarrow & \,\, \gamma_0 \,\, \gamma_5 \,\, \psi(x^\cT) \,\,\, ,
\nonumber \\
\bar \psi(x) \,\, & \rightarrow & \,\, \bar \psi(x^\cT) \,\, \gamma_5  \,\, \gamma_0 \,\,\, .
\label{eq:t-rev}
\end{eqnarray}
Charge conjugation is defined as
\begin{eqnarray}
U_\mu(x) \,\, & \rightarrow & \,\, U_\mu(x)^\ast \,\,\, ,
\nonumber \\
\psi(x) \,\, & \rightarrow & \,\, \cC \,\, \bar \psi^T(x) \,\,\, ,
\nonumber \\
\bar \psi(x) \,\, & \rightarrow & \,\, - \psi^T(x) \,\,\cC^{-1} \,\,\, .
\label{eq:CC}
\end{eqnarray}
The $T$ superscript denotes transpose vectors and matrices. The charge conjugation matrix satisfies
\be
\cC \,\, \gamma_\mu \,\, \cC \,\,\, = \,\,\, - \gamma_\mu^\ast \,\,\, = \,\,\, - \gamma^T_\mu \,\,\, .
\ee
A realization of $\cC$ is $\cC = \gamma_0 \gamma_2$.

\chapter {Appendix: Regularization dependent scheme}
\label{app:RI}

The acronym ``RI" stands for ``regularization independent" and the reason for this is mostly historical: before the advent of lattice non-perturbative renormalization, $Z_O$-factors were calculated in lattice perturbation theory, with  ${\overline{\rm MS}}$ as the scheme of preference. This means that two perturbative calculations had to be carried out; one in the continuum and one on the lattice. Schematically, in dimensional regularization (DR) the continuum perturbative expression for the bare correlation function $\Gamma_O$, at one-loop is
\be
\Gamma_O^{\rm DR}(p,g_0,\epsilon) \,\,\, = \,\,\, \Big [ 1 \,\, + \,\, \dfrac{g_0^2(\mu)}{(4\pi)^2}
\,\, \Big ( \gamma_\Gamma^{(0)} \dfrac{1}{\hat\epsilon} - \gamma_\Gamma^{(0)} \ln(\dfrac{p^2}{\mu^2}) + C_\Gamma^{\rm DR}
\,\, \Big ) \,\, \Big ] \,\,\, ,
\label{eq:dr}
\ee
where $\gamma_\Gamma^{(0)}$ is the LO anomalous dimension of the correlation function $\Gamma_O$. To this order it is a universal, scheme-independent quantity. As implied by the notation, the finite constant $C_\Gamma^{\rm DR}$  depends on the regularization scheme and the chosen gauge. The factor $1 / \hat\epsilon$ is an abbreviation
for
\be
\frac{1}{\hat\epsilon} \,\,\, = \,\,\, \frac{1}{\epsilon} \,\, + \,\, \ln(4\pi) \,\, - \,\, \gamma_E \,\,\, ,
\ee
where $\epsilon=(4-D)/2$ and $\gamma_E$ stands for Euler's constant. In DR
we work in $D$ dimensions, where the original bare coupling $g_0$ has
dimension $\epsilon$. Here the scale $\mu$ is introduced to render the bare
coupling $g_0(\mu)$ dimensionless. The $\mu$-dependence of the r.h.s. of
eq.~(\ref{eq:dr}) is only apparent.

Imposing the $\msb$ renormalization condition amounts to removing
the $1/ \hat\epsilon$ divergence. Since the renormalization constant of the projected Green function 
$\Gamma_O$ is given by ${Z_\Gamma} = Z_\psi^{-1} Z_O$ (see
eq.~(\ref{eq:renG})), this implies for ${Z_\Gamma}$ the value
\be
Z_\Gamma^{\msb,{\rm DR}} (g_0(\mu),\epsilon) \,\,\,= \,\,\, 1 \,\,  -  \,\,
\frac{g_0^2(\mu)}{(4\pi)^2} \,\, \gamma_\Gamma^{(0)} \,\, \frac{1}{\hat\epsilon} \,\,\, .
\label{eq:rmsb}
\ee
Consequently, the renormalized Green function is given by
\bea
& & \Big [ \Gamma_O^{\msb} (p,g_{\msb}(\mu),\mu) \Big ]_{\rm R} \,\,\, = \,\,\, 
\lim_{\epsilon \rightarrow 0} \,\,\, \left[ \,\, Z_\Gamma^{\msb,\DR}(g_0(\mu),\epsilon) \,\, 
\Gamma_O^\DR (p,g_0,\epsilon) \,\,\, \right] 
\nonumber \\
& & \qquad =  \,\,\, 1 \,\, + \,\, \frac{g_{\msb}^2(\mu)}{(4\pi)^2} \,\,
\Bigl[ \,\, - \gamma_\Gamma^{(0)} \,\, \ln(p/\mu)^2 \,\, + \,\, C^{\DR}_\Gamma \,\, \Bigr] \,\, \,\,\, ,
\label{eq:gren_msb}
\eea
where, to this order, we are free to replace $g_0$ by
$g_{\msb}(\mu)$, the $\msb$ renormalized coupling constant.

The same calculation can be repeated on the lattice. Now the UV cutoff
is provided by the inverse finite lattice spacing $a^{-1}$ and thus
the 1-loop calculation yields:
\be
\Gamma_O^{\LAT}(p,g_0(a),a) \,\,\, = \,\,\, 
1\,\, + \,\, \frac{g_0(a)^2}{(4\pi)^2} \,\, \Big( \,- \, \gamma_\Gamma^{(0)} \,\, \ln(pa)^2 \,\, +\,\, C^{\LAT}_\Gamma 
\Big) \,\,\, + \,\,\, \cO(a) \,\,\, ,
\label{eq:lat}
\ee
where $g_0(a)$ is the bare coupling of the lattice action.  The renormalization scheme can again be chosen at will; as previously stated, the
$\msb$ is often chosen also on the lattice\footnote{This seemingly unnatural
choice (the $\msb$ is closely linked to continuum DR) has a few
advantages. For example, matrix elements of effective Hamiltonians, once calculated
non-perturbatively on the lattice, must be renormalized and combined
with perturbatively calculated Wilson coefficients, in order to obtain
physical amplitudes; cf. eqs.~(\ref{eq:OE}) and (\ref{eq:OEren}). The renormalization-group invariance of these
amplitudes is guaranteed only if the Wilson coefficients and the
renormalization constants are calculated in the same renormalization scheme. 
Since the former are often known in the $\msb$ scheme,
this scheme is also preferred for the calculation of the latter.}.
Thus, we must satisfy the renormalization condition
\be
\Big [ \Gamma_O^{\msb} (p,g_{\msb}(\mu),\mu) \Big ]_{\rm R} \,\,\, = \,\,\, \lim_{a \rightarrow 0} \left[
Z_\Gamma^{\msb,\LAT}(\mu a, g_0(a)) \,\,\, \Gamma_O^{\LAT} (p,g_0(a),a)\right] \,\,\, ,
\label{eq:rgf_l}
\ee
where again the lattice coupling $g_0(a)$ should be traded for the $\msb$
renormalized coupling constant $g_{\msb}(\mu)$. This point of principle is of
limited relevance for a 1-loop calculation.
From eqs.~(\ref{eq:gren_msb}), (\ref{eq:lat}) and (\ref{eq:rgf_l}), the 
following renormalization constant is obtained:
\be
Z_\Gamma^{\msb,\LAT} (\mu a,g_0(a))  \,\,\, = \,\,\,
1 \,\, + \,\, \frac{g_0(a)^2}{(4\pi)^2} \,\, \left[ \gamma_\Gamma^{(0)} \ln(\mu a)^2 \,\,
+ \,\, C^{\DR}_\Gamma - C^{\LAT}_\Gamma \right] \,\,\, .
\label{eq:rlat}
\ee
We now recall that the renormalization constant of the amputated vertex $\Gamma_O$ is 
$Z_\Gamma = Z_\psi^{-1} Z_O$. The quark field renormalization $Z_\psi$ can 
be calculated, from the quark propagator $\cS (p)$,
with an analogous procedure; c.f. eq.~(\ref{eq:rensig}). The result is
\be
Z_\psi^{\msb,\LAT} (\mu a,g_0(a))  \,\,\, = \,\,\, 
1 \,\, + \,\, \frac{g_0(a)^2}{(4\pi)^2} \,\, \left[ \gamma_{\Sigma}^{(0)} \ln(\mu a)^2
\,\, + \,\, C^{DR}_\Sigma \,\, - \,\, C^{\LAT}_\Sigma \right] \,\,\, .
\label{eq:rlatSig}
\ee
Combining eqs.~(\ref{eq:rlat}) and (\ref{eq:rlatSig}) we obtain
\be
Z_O^{\msb,\LAT} (\mu a,g_0(a)) \,\,\, = \,\,\,
1 \,\, + \,\, \frac{g_0(a)^2}{(4\pi)^2} \,\, \left[ \gamma_O^{(0)} \ln(\mu a)^2
\,\, + \,\, \Delta_\Gamma + \Delta_\Sigma \right] \,\,\, ,
\label{eq:rlatZ}
\ee
where
\bea
\gamma_O &=& \gamma_\Gamma \,\, + \,\, \gamma_\Sigma \,\,\, ,\nonumber \\
\Delta_\Gamma &=& C^{\DR}_\Gamma \,\, - \,\, C^{\LAT}_\Gamma \,\,\, ,\\
\Delta_\Sigma &=& C^{\DR}_\Sigma \,\, -\,\, C^{\LAT}_\Sigma \,\,\, .\nonumber
\eea
It is this renormalization constant  (with this choice of renormalization condition)
which is usually denoted by $Z_O$ in lattice perturbation theory calculations. 
The dependence of $Z_O^{\msb,\LAT}$ on the coefficients
$C^{\DR}_\Gamma$ and $C^{\DR}_\Sigma$ comes from the choice of the $\msb$
renormalization condition (see eqs.~(\ref{eq:gren_msb}) and (\ref{eq:rgf_l}))
whereas its dependence on $C^{\LAT}_\Gamma$ and $C^{\LAT}_\Sigma$ from 
the lattice regularization (see eq.~(\ref{eq:lat})). Two perturbative
calculations are thus necessary, one in the continuum for the $C^{\DR}$'s and
one on the lattice for $C^{\LAT}$'s. The presence of the $C^{\DR}$'s on the r.h.s.
of eq.~(\ref{eq:rlat}) is sometimes referred to as the ``regularization dependence" of 
the renormalization scheme.

On the other hand, the RI/MOM scheme is obtained by imposing condition~(\ref{eq:renGcond}) on the lattice
correlation function~(\ref{eq:lat}). It follows that the renormalization constant of interest, in one-loop perturbation theory, is
given by
\be
Z_\Gamma^{\rimom} (\mu a,g_0(a))  \,\,\, = \,\,\,
1 \,\, + \,\, \frac{g_0(a)^2}{(4\pi)^2} \,\, \left[ \gamma_\Gamma^{(0)} \ln(\mu a)^2 \,\,
- \,\, C^{\LAT}_\Gamma \right] \,\,\, .
\label{eq:rimomPT}
\ee
Similarly, condition~(\ref{eq:rensig}), related to the quark field renormalization, gives
\be
Z_\psi^{\rimom} (\mu a,g_0(a))  \,\,\, = \,\,\, 
1 \,\, + \,\, \frac{g_0(a)^2}{(4\pi)^2} \,\, \left[ \gamma_{\Sigma}^{(0)} \ln(\mu a)^2
 - \,\, C^{\LAT}_\Sigma \right] \,\,\, .
\label{eq:rlatSigrimom}
\ee
The last two expressions combine to give
\be
Z_O^{\rimom} (\mu a,g_0(a)) \,\,\, = \,\,\,
1 \,\, + \,\, \frac{g_0(a)^2}{(4\pi)^2} \,\, \left[ \gamma_O^{(0)} \ln(\mu a)^2
\,\, - \,\, C^{\LAT}_\Gamma \,\, - \,\, C_\Sigma^\LAT \right] \,\,\, ,
\label{eq:rlatZ2}
\ee
Clearly, the above RI/MOM result does not depend on a continuum regularization (such as DR), but only on the lattice. This is referred to as ``regularization independence".

\bibliographystyle{OUPnamed_notitle}
\bibliography{vladikas_refs}

\end{document}